\documentclass[twocolumn,twocolappendix]{aastex701}

\usepackage{graphicx}
\usepackage{xcolor}
\usepackage{amsmath}
\usepackage[version=4]{mhchem}
\usepackage{multirow}

\usepackage{siunitx}
\DeclareSIUnit\year{yr}
\DeclareSIUnit\parsec{pc}
\DeclareSIUnit\Msun{M_\odot}
\DeclareSIUnit\angstrom{\text {\AA}}

\usepackage{lettrine}
\input Zallman.fd

\newcommand{\osuaffil}{Department of Astronomy, The Ohio State University, 140 W. 18th Ave, Columbus, OH 43210, USA}
\newcommand{\ccappaffil}{Center for Cosmology and AstroParticle Physics, The Ohio State University, 191 W. Woodruff Ave., Columbus, OH 43210, USA}

\newcommand{\figpath}[1]{figures/#1}

\newcommand{\bracket}[2][H]{{\rm [#2/#1]}}
\newcommand{\pc}{\,{\rm pc}}
\newcommand{\kpc}{\,{\rm kpc}}
\newcommand{\Myr}{\,{\rm Myr}}
\newcommand{\Gyr}{\,{\rm Gyr}}
\newcommand{\dex}{\,{\rm dex}}
\newcommand{\Msun}{\,{\rm M}_\odot}
\newcommand{\kms}{\,{\rm km}\,{\rm s}^{-1}}

\begin{document}

% Title
\title{A Broken Clock Is Right Twice a Day: [Ce/Mg] Is Not a Universal Chemical Clock}
\shorttitle{Right Twice a Day}

% Author list
\author[0000-0003-3781-0747]{Liam O.\ Dubay}
\affiliation{\osuaffil}
\affiliation{\ccappaffil}
\email{dubay.11@osu.edu}
\correspondingauthor{Liam Dubay}

\author[0000-0001-7258-1834]{Jennifer A.\ Johnson}
\affiliation{\osuaffil}
\affiliation{\ccappaffil}
\email{}

\author[0000-0002-6534-8783]{James W.\ Johnson}
\affiliation{Observatories of the Carnegie Institution for Science, 813 Santa Barbara St., Pasadena CA 91101, USA}
\email{}

\author[0000-0002-1423-2174]{Keith Hawkins}
\affiliation{Department of Astronomy, The University of Texas at Austin, 2515 Speedway Boulevard, Austin, TX 78712, USA}
\email{}

\author[0000-0002-7539-1638]{Sofia Feltzing}
\affiliation{Lund Observatory, Department of Earth and Environmental Sciences, S\"olvegatan 12, SE-223 62 Lund, Sweden}
\affiliation{Lund Observatory, Department of Geology, S\"olvegatan 12, SE-223 62 Lund, Sweden}
\email{}

\author[0000-0001-9738-4829]{Natalie R.\ Myers}
\affiliation{Department of Physics and Astronomy, Johns Hopkins University, 3400 N Charles St, Baltimore, MD 21218, USA}
\email{}

\author[0000-0003-0872-7098]{Adrian M.\ Price-Whelan}
\affiliation{Center for Computational Astrophysics, Flatiron Institute, 162~Fifth~Ave., New~York, NY~10010,~USA}
\email{}

\author[0000-0001-8208-9755]{Tawny Sit}
\affiliation{\osuaffil}
\affiliation{\ccappaffil}
\email{}

\author[0000-0003-0174-0564]{Andrew R.\ Casey}
\affiliation{Center for Computational Astrophysics, Flatiron Institute, 162~Fifth~Ave., New~York, NY~10010,~USA}
\affiliation{School of Physics and Astronomy, Monash University VIC 3800, Australia}
\email{}

\author[0000-0001-6476-0576]{Katia Cunha}
\affiliation{Steward Observatory, University of Arizona, 933 North Cherry Avenue, Tucson, AZ 85721-0065, USA}
\affiliation{Observat\'orio Nacional/MCTIC, R. Gen. Jos\'e Cristino, 77, 20921-400, Rio de Janeiro, Brazil}
\email{}

\author[0000-0002-0900-9760]{Jos\'e G.\ Fern\'andez-Trincado}
\affiliation{Centro de investigaci\'on en Astronom\'ia, Facultad de Ingenier\'ia, Ciencia y Tecnolog\'ia, Universidad Bernardo O’Higgins, Av.\ Viel 1497, Santiago, 8370993, Chile}
\email{}

\author[0000-0003-1856-2151]{Danny Horta}
\affiliation{Institute for Astronomy, University of Edinburgh, Royal Observatory, Blackford Hill, Edinburgh EH9 3HJ, UK}
\email{}

\author[0000-0002-6972-6411]{Jos\'e Eduardo M\'endez-Delgado}
\affiliation{Universidad Nacional Aut\'onoma de M\'exico, Instituto de Astronom\'ia, A.P.\ 70-264, 04510, Ciudad de M\'exico, M\'exico}
\email{}

\author[0000-0002-1379-4204]{Alexandre Roman Lopes}
\affiliation{Department of Astronomy, Universidad de La Serena, Av.\ Raul Bitran 1302, La Serena, Chile}
\email{}

\author[0009-0005-0182-7186]{Amaya Sinha}
\affiliation{Department of Physics and Astronomy, University of Utah, Salt Lake City, UT 84112, USA}
\email{}

\author[0009-0009-0081-4323]{Rodolfo de J.\ Zerme\~no}
\affiliation{Universidad Nacional Aut\'onoma de M\'exico, Instituto de Astronom\'ia, A.P.\ 70-264, 04510, Ciudad de M\'exico, M\'exico}
\email{}

\author[0000-0003-2868-8276]{Jingkun Zhao}
\affiliation{National Astronomical Observatories, Chinese Academy of Sciences, Beijing 100101, People's Republic of China}
\email{}

\shortauthors{Dubay et al.}
\shorttitle{A Broken Clock is Right Twice a Day}

\submitjournal{AAS Journals}
\received{26 August 2026}

\begin{abstract}
    The ratio of $s$-process to $\alpha$-element enrichment ($[s/\alpha]$) has been proposed as a ``chemical clock,'' or a means to estimate stellar ages. However, the age--$[s/\alpha]$ relation varies with metallicity and location in the Galaxy, and the observed trends are not well predicted by galactic chemical evolution models. We quantify the age--[Ce/Mg] correlation across the Galactic disk in roughly 100,000 red giant stars observed with APOGEE in the SDSS-V Milky Way Mapper survey. We find that the slope of the correlation varies significantly with metallicity and guiding radius in the chemical thin disk. The trend is steepest in the outer disk and below Solar metallicity, while the most metal-rich stars and those in the inner disk show no correlation with age. In contrast, [Ce/Mg] patterns in the chemical thick disk are consistent across the Galaxy. Halo stars have higher [Ce/Mg] than the chemical thick disk, suggesting that asymptotic giant branch (AGB) enrichment is important even at low metallicity. Overall, patterns in [Ce/Mg] trace both the local star formation history and AGB nucleosynthesis. The complex interplay between [Ce/Mg], age, metallicity, and Galactic position means that [Ce/Mg] (and by extension $[s/\alpha]$) is not a universal chemical clock. 
\end{abstract}

\section{Introduction}
\label{sec:introduction}

\lettrine{S}{tellar ages} are a vital component of Galactic archaeology studies. Stellar age data provide direct measurements of the Galaxy's evolution over time and allow for better comparisons with chemical evolution models. Recent years have seen an explosion of large stellar catalogs with age measurements for $10^5-10^6$ stars \citep[e.g.,][]{queiroz_starhorse_2020,xiang_time_2022,nataf_accurate_2024,stone-martinez_starflow_2025,wang_spectroscopic_2025,roberts_cn_2026}. However, age estimation techniques typically apply to a limited range of stellar parameters, and systematic offsets exist between different methods. ``Chemical clocks,'' or empirically-calibrated relations between stellar age and atmospheric abundance ratios, promise reliable spectroscopic age estimates over a wider parameter space \citep[e.g.,][]{hayden_galah_2022}. 

The $[s/\alpha]$ ratio, or the ratio of elements produced by the slow neutron-capture ($s$) process in asymptotic giant branch (AGB) stars to the $\alpha$-elements produced in core-collapse supernovae (CCSNe), was first proposed to be a chemical clock by \citet{nissen_high-precision_2015}. Typical $s$-process-dominated elements include Y, Ba, and Ce, while Mg is often the representative $\alpha$-element. A remarkably tight correlation between [Y/Mg] and age was found in solar twins \citep{nissen_high-precision_2015,tucci_maia_solar_2016,spina_temporal_2018} and red clump stars \citep{slumstrup_ymg_2017}, and subsequent studies found that this correlation extends to many $s$-process and $\alpha$-elements \citep{nissen_high-precision_2016,jofre_traits_2020,casali_tracing_2025}. \citet{casali_tracing_2025} found that [Ce/Mg] is especially promising due to its tight correlation with age and availability in the Apache Point Observatory Galactic Evolution Experiment (APOGEE) survey \citep{majewski_apogee_2017}. Recent studies have calibrated $[s/\alpha]$-based chemical clocks using main-sequence turn-off stars \citep{hayden_galah_2022} and asteroseismic ages for red giants \citep[e.g.,][]{casali_tracing_2025,pakstiene_calibration_2026,mikolaitis_non-lte_2026}, achieving a reported age precision of $\sim1-3\Gyr$. 

The age--$[s/\alpha]$ relation is observed to vary with metallicity and Galactic position, which complicates its usage as a genuine chemical clock. \citet{feltzing_metallicity_2017} found that the age--[Y/Mg] relation depends strongly on metallicity for solar neighborhood dwarfs. Other studies have confirmed this variation for both Y and Ce in open clusters \citep[e.g.,][]{casali_gaia-eso_2020,sales-silva_exploring_2022} and for [Ba/Fe] in red giants \citep{horta_neutron-capture_2022}, although \citet{titarenko_ambre_2019} detected no metallicity dependence in [Y/Mg] among main-sequence turn-off stars. Studies of open clusters \citep{casali_gaia-eso_2020,sales-silva_exploring_2022,viscasillas_vazquez_gaia-eso_2022} and red giants \citep{casali_time_2023,ratcliffe_chemical_2024,pakstiene_calibration_2026} have revealed variation in the age--$[s/\alpha]$ relation with Galactocentric radius, which may be connected to the metallicity dependence or variation in the local star formation history. Detailed studies have had samples that are typically small, limited to the chemical thin disk, and confined to the solar neighborhood (for dwarfs and subgiants) or to relatively young stars (for open clusters). In light of this evidence, a more complete picture of $[s/\alpha]$ across the Milky Way disk is needed to determine if it is a reliable chemical clock.

The basic mechanism of the $[s/\alpha]$ chemical clock comes from the difference in the enrichment timescales of the two elements. The progenitors of CCSNe are short-lived, so the production of $\alpha$-element abundances closely traces the star formation history. The $s$-process elements are primarily produced in the AGB phase of low- to intermediate-mass stars \citep[$\sim1-10\Msun$; see review by][]{karakas_dawes_2014} whose main-sequence lifetimes range from $\sim30\Myr-10\Gyr$. Therefore, in a simple picture of chemical evolution, $[s/\alpha]$ should generally increase over time. In detail, the evolution of $[s/\alpha]$ is affected by the star formation history and the mass- and metallicity-dependence of the AGB yields. Many $s$-process elements are also produced by the $r$-process in neutron star mergers and some CCSNe, which may dominate the chemical enrichment at low metallicity \citep[e.g.,][]{kobayashi_origin_2020}.

Galactic chemical evolution models currently do not capture the full picture of $s$-process evolution. While most surveys find that $s$-process ratios are highest in young stars, chemical evolution models with standard AGB yield prescriptions predict that $[s/\alpha]$ should plateau or even decrease over time in the thin disk \citep[e.g.,][]{grisoni_modelling_2020,ratcliffe_chemical_2024,molero_modelling_2025,casali_modelling_2026}. Modifications to the AGB stellar yields may improve the agreement with the data for some elements \citep{ratcliffe_chemical_2024,molero_constraining_2025}, but such {\it ad hoc} modifications cannot fully resolve the discrepancy \citep{casali_modelling_2026}. Therefore, theoretical models do not currently support the $[s/\alpha]$ chemical clock.

In this paper, we obtain the most comprehensive view to date of the [Ce/Mg] abundance patterns in the Milky Way, with the aim of determining its usefulness as a chemical clock. Utilizing the expanded sample size and coverage of the Milky Way Mapper (MWM), a program of the Sloan Digital Sky Survey \citep[SDSS-V;][]{kollmeier_sdss-v_2017,kollmeier_sloan_2026}, we quantify the age--[Ce/Mg] and [Mg/H]--[Ce/Mg] relations as a function of metallicity, $\alpha$-enhancement, and Galactic position. We select Ce as our representative $s$-process element because of its availability in the near-infrared APOGEE spectra and previous studies in the literature. In Section \ref{sec:data}, we describe our observational dataset. In Section \ref{sec:results}, we present the chemical abundance and age trends in our Milky Way disk sample. In Section \ref{sec:halo}, we present abundance trends in the halo. We summarize our conclusions in Section \ref{sec:conclusions}.

\section{Data}
\label{sec:data}

We use MWM data from Data Release 19 \citep[DR19;][]{sdss_collaboration_nineteenth_2025} of SDSS-V.
The catalog includes all data from the previous APOGEE DR17 \citep{abdurrouf_seventeenth_2022}. MWM DR19 data were obtained from the infrared APOGEE spectrographs \citep{wilson_apache_2019} mounted on the 2.5-meter Sloan Foundation Telescope \citep{gunn_25_2006} at Apache Point Observatory and the Ir{\'e}n{\'e}e du Pont Telescope \citep{bowen_optical_1973} at Las Campanas Observatory. 

Atmospheric parameters and chemical abundances were determined with the APOGEE Stellar Parameter and Chemical Abundance Pipeline \citep[ASPCAP;][]{garcia_perez_aspcap_2016}, which compares a grid of synthetic spectra \citep{hubeny_tlusty_2021} to the observed spectra using $\chi^2$ minimization. The atomic and molecular line lists are presented in \citet{hasselquist_identification_2016}, \citet{cunha_adding_2017}, and \citet{smith_apogee_2021}. 
Cerium abundances specifically are derived using three Ce II lines present in the APOGEE spectral windows: $\lambda\num{15784.75}$, $\lambda\num{16376.48}$, and $\lambda\SI{16595.18}{\angstrom}$ \citep{cunha_adding_2017}. All DR19 abundances assume local thermodynamic equilibrium (LTE). The spectral synthesis is further described by \citet{holtzman_apogee_2018} and \citet{jonsson_apogee_2020}. \citet{meszaros_sdss-v_2025} describe the precision, accuracy, and systematics of the DR19 stellar parameters and abundances in detail.

We limit our study to red giant branch (RGB) stars because the Ce abundances in MWM are unreliable on the main sequence \citep{meszaros_sdss-v_2025}. Following recommendations by \citet{meszaros_sdss-v_2025}, we select our main RGB sample using the following quality cuts:
\begin{itemize}
    \item $1<\log(g)<3$ (we observe that [Ce/H] measurements become unreliable for $\log(g)>3$);
    \item $\SI{4000}{\kelvin}<T_{\rm eff}<\SI{5500}{\kelvin}$;
    \item $\bracket{M}>-1.5$;
    \item $S/N>100$;
    \item No bad flag or spectrum flags;
    \item No Mg, Fe, or Ce abundance flags (avoiding stars close to theoretical measurement limits or grid edges);
    \item $\sigma_{\bracket{X}}<0.2\dex$ for Mg, Fe, and Ce;
    \item No SDSS-IV EXTRATARG flag (where available, limit to APOGEE DR17 Main Red Star Sample);
    \item Uncalibrated [Ce/H] above theoretical upper limit (see Section \ref{sec:ce-abundances}).
    \item Age training space density $>3\times10^9$ (``good ages''; see Section \ref{sec:ages}).
\end{itemize}
This produces a primary sample of \num{94134}stars.\footnote{Out of the $\sim\num{900000}$ total stars in DR19, the largest reductions in sample size are due to the quality and targeting cuts ($\sim60\%$), RGB selection ($\sim58\%$), and $S/N$ requirement ($\sim26\%$), with smaller reductions due to the abundance quality cuts.} In cases where stellar ages are not needed for our analysis, we omit the age quality cut for a slightly larger sample of \num{107688}stars. All of our stars have radial velocity scatter $\sigma(v_{\rm rad})<1\kms$ and 84\% have $\sigma(v_{\rm rad})<0.1\kms$, indicating minimal contamination from variable stars or binaries. We also perform sensitivity tests by running our analysis on more limited samples, including a stricter $S/N$ limit, higher age quality cut, and narrower $\log(g)$ range, and found no significant difference in our results. Appendix \ref{app:sensitivity} presents the results of these sensitivity tests.

\subsection{Positions, Kinematics, and Orbits}
\label{sec:orbits}

The 6-D kinematic properties for stars in our sample are calculated using position, proper motion, parallax, and radial velocity measurements from {\it Gaia} DR3 \citep{gaia_collaboration_gaia_2023}. We adopt distance estimates from \citet{zhang_parameters_2023}, the {\it Gaia} parallax, and \citet{bailer-jones_estimating_2021}, taking the first finite value from these sources in the order listed. These coordinates are transformed into the Galactocentric cylindrical frame $(R,\phi,z)_{\rm gal}$ with {\tt astropy} \citep{astropy_collaboration_astropy_2013,astropy_collaboration_astropy_2018,astropy_collaboration_astropy_2022} using updated solar position and motion parameters from \citet{hunt_multiple_2022}. In particular, we adopt the Galactic position of the Sun $(R,z)_\odot=(8275,20.8)\pc$ \citep{gravity_collaboration_improved_2021,bennett_vertical_2019} and solar velocity with respect to the Galactic center $(v_x,v_y,v_z)_\odot=(8.4,251.8,8.4)\kms$ \citep{reid_proper_2020,gravity_collaboration_improved_2021}. 

Orbital parameters such as the angular momentum $L_z$, orbital energy $E$, and actions $(J_R,J_\phi,J_z)$ are calculated using the {\tt gala} package \citep{price-whelan_gala_2017} with the {\tt MilkyWayPotential2022} potential model. This axisymmetric model lacks a bar or spiral perturbations. We use these orbital parameters primarily to bin the data on kpc scales, and we exclude the central $3\kpc$ of the Milky Way from our analysis, so our results should be minimally affected by limitations in the adopted potential model.

Throughout this study, we locate stars using their guiding center radius $R_{\rm guide}$ and maximum midplane distance $z_{\rm max}$ rather than their Galactocentric coordinates. A star's guiding radius is defined as the radius of a circular orbit with equal $L_z$ to the star, which we calculate assuming an \citet{eilers_circular_2019} circular velocity curve. This minimizes the effect of orbital ``blurring,'' where a star's eccentricity causes it to pass through a range of $R_{\rm gal}$. Similarly, binning by $z_{\rm max}$ reduces the contamination of stars on inclined orbits that happen to be passing near the midplane at the present day. We eliminate $\sim1200$ stars from our sample for which a good orbital solution could not be found.

\subsection{Low- and High-Ia Populations}
\label{sec:alpha-pops}

Milky Way disk stars can be divided into two populations based on their $[\alpha/{\rm Fe}]$ ratios, often termed low- and high-$\alpha$ or the chemical thin and thick disks. We refer to these populations as high- and low-Ia, respectively \citep[adopting the nomenclature of][]{griffith_abundance_2019}, out of recognition that the former population is enriched in Fe by Type Ia supernovae (SNe Ia) relative to the latter. These populations experienced different nucleosynthesis histories and show distinct abundance trends. Therefore, we define high- and low-Ia populations in our sample and analyze them separately.
We divide the sample using the following boundary condition:
\begin{equation}
    \bracket[Mg]{Fe} = 
    \begin{cases}
        -0.09, & \bracket{Fe}>0 \\
        -0.09+0.13\times\bracket{Fe}, & \bracket{Fe}<0.
    \end{cases}
    \label{eq:alpha-cut}
\end{equation}
High-Ia stars lie above this boundary, while low-Ia stars lie below. We exclude stars within $\pm0.02\dex$ of this boundary from either population. Figure \ref{fig:logg-calibrations} plots this boundary after a transformation into [Mg/H].

\subsection{Corrective Abundance Offsets}
\label{sec:logg-corrections}

\begin{figure}
    \centering
    \includegraphics[width=\linewidth]{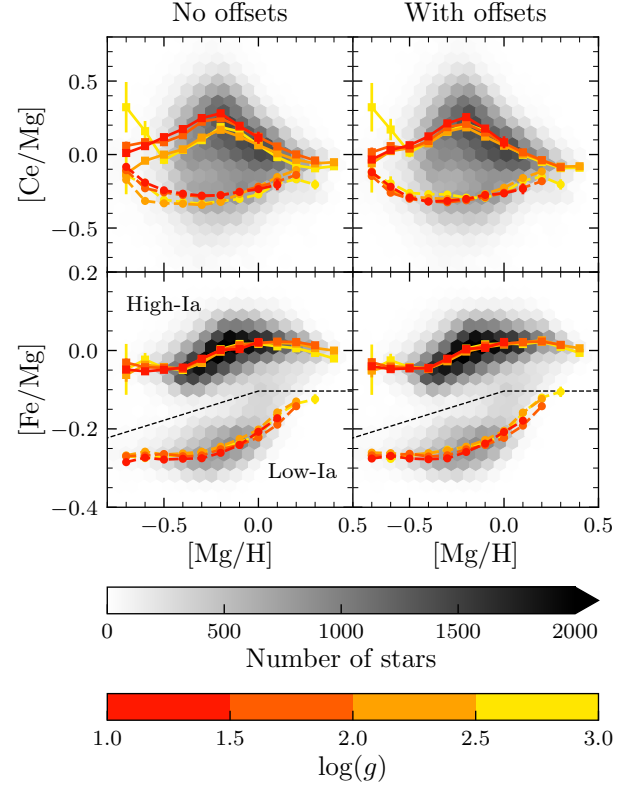}
    \caption{Demonstration of corrective $\log(g)$ offsets applied to Ce and Fe abundances in our full sample. Each panel plots the median trend in [X/Mg] as a function of [Mg/H] in multiple $\log(g)$ bins. Separate trends are shown for low-Ia (circles, dashed lines) and high-Ia (squares, solid lines) populations. The right-hand panels use the base DR19 abundances, while the left-hand panels incorporate the corrective offsets. The dashed black line in the lower-left panel indicates the division between low- and high-Ia populations.}
    \label{fig:logg-calibrations}
\end{figure}

Stellar abundance trends should be independent from evolutionary stage: stars on the lower RGB, red clump, and upper RGB should have similar abundance trends at the population level. However, because surface gravity has a significant effect on the strength of spectral lines (see Section \ref{sec:ce-abundances}), abundance measurements can show spurious correlations with $\log(g)$. Figure \ref{fig:logg-calibrations} shows that the median trends in un-corrected [Ce/Mg] as a function of [Mg/H] vary with $\log(g)$ by as much as $\sim0.3\dex$.

We follow the procedure outlined in \citet{sit_chemical_2024} to remove $\log(g)$ trends in Fe and Ce (see Appendix \ref{app:logg-calibration-procedure} for details). Figure \ref{fig:logg-calibrations} shows that the corrective offsets reduce variation in the Ce trends to the $\sim0.1\dex$ level, ensuring that the abundance trends are consistent across the full sample. We see similar levels of improvement when applying the analysis in Figure \ref{fig:logg-calibrations} to specific regions of the Galaxy (e.g., inner disk, solar neighborhood, and outer disk). We have tested running our analysis on the un-corrected abundances and found minimal differences in our results. We have also compared our approach to that of \citet{meszaros_sdss-v_2025}, who correct for trends in $T_{\rm eff}$ within open clusters. While both approaches reduce abundance correlations with $T_{\rm eff}$, our method better reduces the correlations with $\log(g)$, allowing us to more accurately compare abundance trends across the Galactic disk. Throughout the remainder of this paper, we will use these corrected abundances unless otherwise noted.

\subsection{Cerium Abundance Precision}
\label{sec:ce-abundances}

\begin{figure*}[h]
    \centering
    \includegraphics[width=\textwidth]{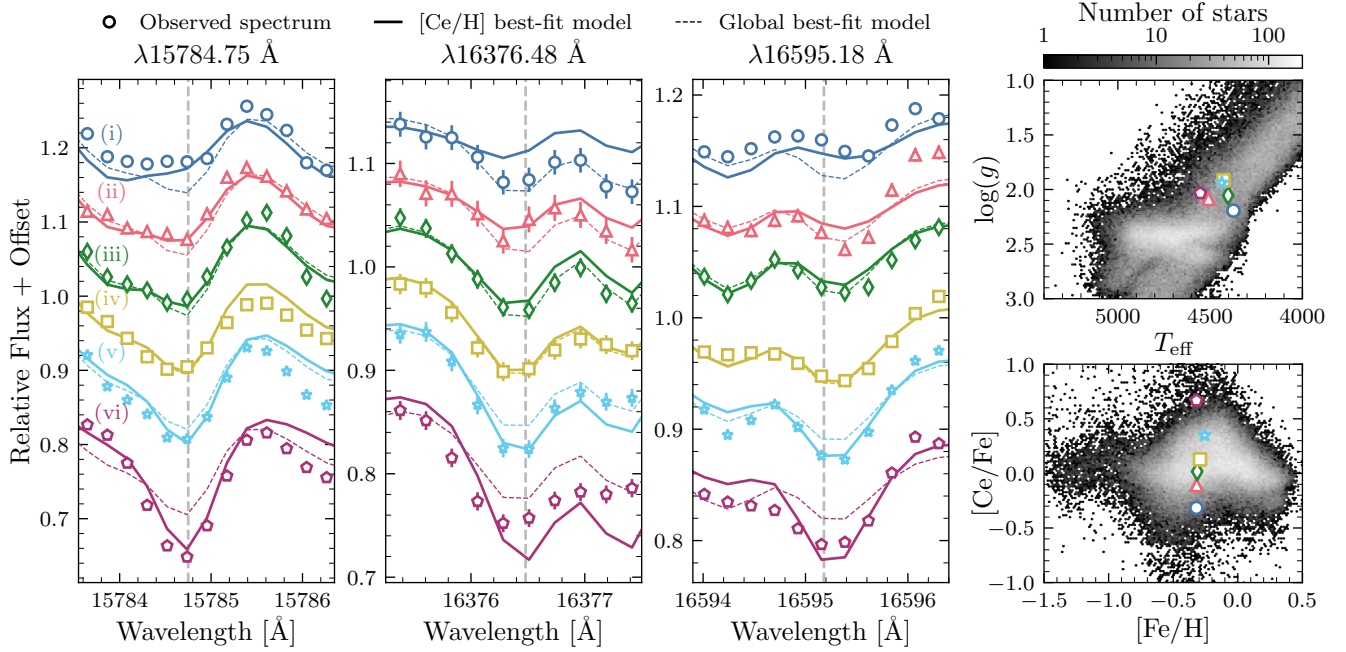}
    \caption{Ce II absorption lines in the APOGEE spectra for selected stars with similar $\log(g)$, $T_{\rm eff}$, [Fe/H], and $S/N\sim200$. From the top, the 2-MASS IDs for the selected stars are: (i) 2M04581987-0433056, (ii) 2M14415263+0953418, (iii) 2M20471037-0002171, (iv) 2M02183299+5434456, (v) 2M09413146+5939283, and (vi) 2M02390392+6132329. Open points plot the continuum-normalized observed flux, and solid (dashed) curves plot the [Ce/H] (global) best-fit model spectrum. Wavelengths are in the stellar rest frame and in air. 
    Stellar parameters are compared against the full sample in the right-hand panels.
    }
    \label{fig:ce-lines-1}
\end{figure*}

\begin{figure*}[h]
    \centering
    \includegraphics[width=\textwidth]{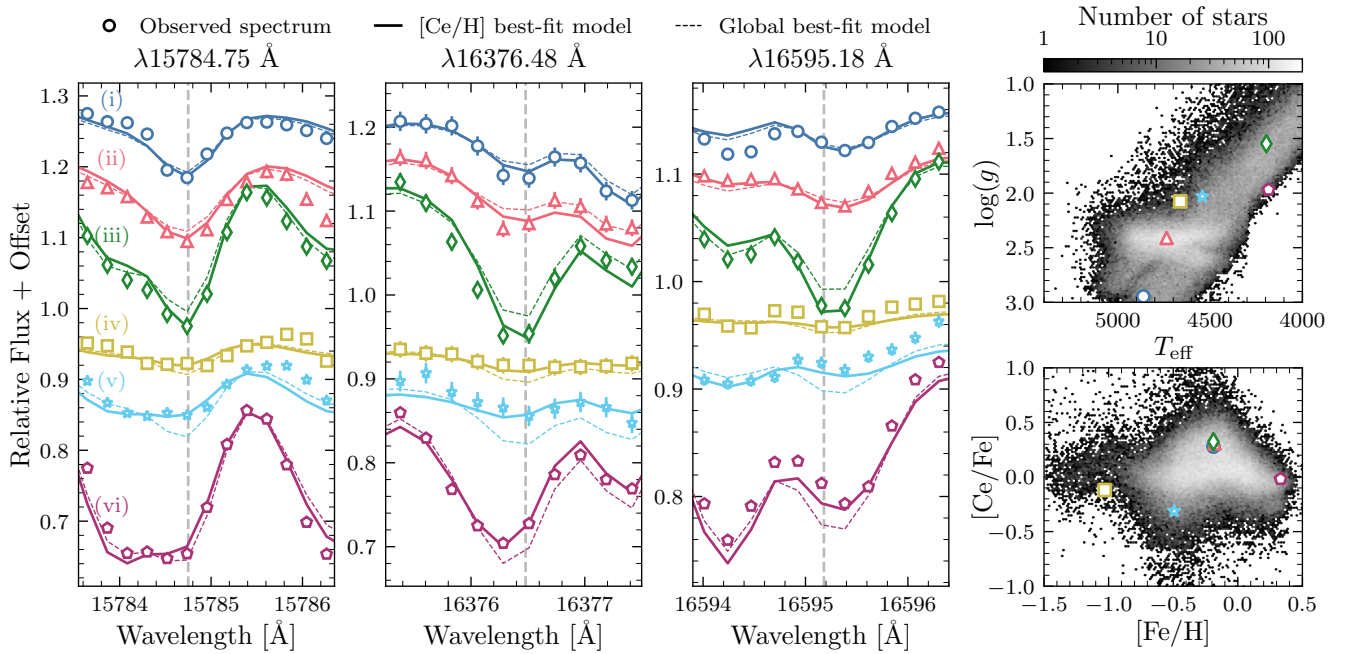}
    \caption{Similar to Figure \ref{fig:ce-lines-1} for six additional stars with $S/N\sim200$ that span a range of (i--iii) $\log(g)$ and (iv--vi) [Fe/H]. From the top, the 2-MASS IDs for the selected stars are: (i) 2M09093066+2420548, (ii) 2M04364872+5016286, (iii) 2M03343165+5329192, (iv) 2M21314472-0350388, (v) 2M15025976+7827316, and (vi) 2M09112399-4841166.}
    \label{fig:ce-lines-2}
\end{figure*}

The ASPCAP abundances for giant stars in DR19 have a precision of $0.02-0.04\dex$ for [M/H], [$\alpha$/M], and [Mg/H] \citep{meszaros_sdss-v_2025}. Due to its weaker lines, [Ce/H] is less precisely measured, with a median reported uncertainty  of $\sigma_{\bracket{Ce}}\approx0.047\dex$ for our sample. However, ASPCAP-derived abundance errors are known to be underestimated \citep[see][]{pinsonneault_apokasc-3_2025}. \citet{meszaros_sdss-v_2025} measured the scatter in [Ce/M] for open clusters, wide binaries, and the solar neighborhood, and estimated a precision of $\sim0.23\dex$ across all three methods.  The factor of $\sim5$ difference between the reported and externally-calibrated precision is comparable to other elements in DR19 \citep{meszaros_sdss-v_2025}.

Figure \ref{fig:ce-lines-1} presents the spectra for 6 ``sibling'' stars with similar $\log(g)$, $T_{\rm eff}$, and [Fe/H] that span a range of [Ce/Fe] (not $\log(g)$-corrected). All of the stars have $S/N\sim200$, close to the median $S/N$ of the full sample. Each of the three spectral windows used for Ce abundance determination are shown in the figure. The relative Ce enrichment is reflected in the difference between the observed flux and the global fit, which accounts for stellar parameters such as [M/H], $\log(g)$, and $T_{\rm eff}$, but not individual abundances. The [Ce/H] best-fit model is computed by re-scaling the model [M/H] to the individual element abundance, and therefore matches the data more closely near each of the the Ce II lines. The ASPCAP abundances are fit using $\chi^2$ minimization over all the lines simultaneously, so the quality of the fit varies between the individual windows for some stars. We also note that all three Ce II lines are blended with nearby lines: CO near $\lambda\SI{15784}{\angstrom}$, CN and Fe I near $\lambda\SI{16376}{\angstrom}$, and CN and Mg I near the $\lambda\SI{16595}{\angstrom}$ line \citep[][Figure 6]{cunha_adding_2017}.

Figure \ref{fig:ce-lines-2} similarly presents the spectra for 6 selected stars spanning a range of $\log(g)$, $T_{\rm eff}$, and [Fe/H]. The top three spectra span a range of $\log(g)$ at similar [Fe/H] and [Ce/H], illustrating that the Ce II lines get fainter with increasing surface gravity but are still detectable at $\log(g)\approx3$. The bottom three spectra span a range of [Fe/H] and $T_{\rm eff}$ at similar $\log(g)$. The lines are weak at $\bracket{Fe}\approx-1$ and strongly blended at $\bracket{Fe}\approx+0.3$, but are still fit well by the model at both extremes. Overall, Figures \ref{fig:ce-lines-1} and \ref{fig:ce-lines-2} demonstrate that the APOGEE spectra can accurately capture variation in [Ce/H] across the full parameter space in the sample, and that the ASPCAP Ce abundances for our high-quality sample are trustworthy.

ASPCAP does not include a treatment for upper limits, so the reported abundances may in some cases be theoretically unmeasurable. By comparing a grid of synthetic spectra with and without each element, \citet{shetrone_apache_2026} calibrated a function to determine an upper limit on the abundance for a given $T_{\rm eff}$ and $S/N$. For most of the stars in our sample, the measured [Ce/H] is above this limit due to our $S/N>100$ requirement. However, at lower metallicities, a larger fraction of our sample falls below these upper limits, which affects our analysis of the Milky Way halo in Section \ref{sec:halo}. Therefore, we calculate upper limits on [Ce/H] for each star using the method of \citet{shetrone_apache_2026} and eliminate \num{620} stars with uncalibrated abundances below the limit. We also compare the ASPCAP abundances against the BAWLAS catalog \citep{hayes_bacchus_2022} in Appendix \ref{app:bawlas-comparison}.

\subsection{Stellar Ages}
\label{sec:ages}

\subsubsection{The StarFlow Catalog}

\begin{figure}
    \centering
    \includegraphics[width=\linewidth]{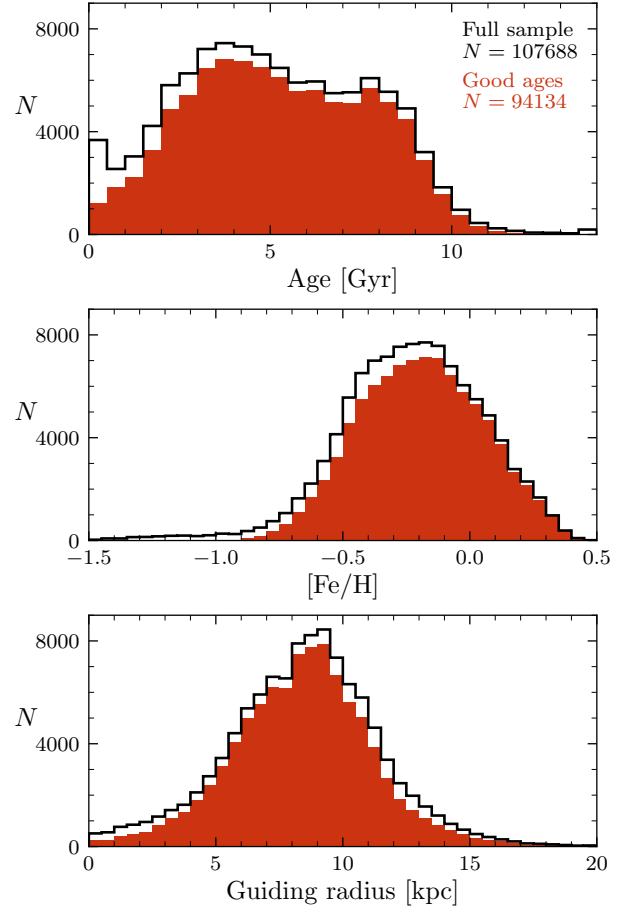}
    \caption{Distributions of stellar age, [Fe/H], and guiding radius of our sample. The open black histogram plots the full sample described in Section \ref{sec:data}, and the filled red histogram includes only stars that exceed the recommended age training density \citep[Section \ref{sec:ages};][]{stone-martinez_starflow_2025}.}
    \label{fig:sample-distributions}
\end{figure}

We use the StarFlow \citep{stone-martinez_starflow_2025} value-added catalog of stellar ages for MWM DR19. \citet{stone-martinez_starflow_2025} use a deep generative model trained on asteroseismic data to estimate ages for evolved stars. Specifically, they use stellar masses and ages from the APOGEE-Kepler Asteroseismic Catalog \citep[APOKASC-3;][]{pinsonneault_apokasc-3_2025} and APO-K2 catalog \citep{schonhut-stasik_apo-k2_2024,warfield_apo-k2_2024} and MWM DR19 stellar parameters ($T_{\rm eff}$ and $\log(g)$) and abundances ([Fe/H], [C/Fe], and [N/Fe]). This makes StarFlow essentially a [C/N]-derived age catalog. The \citet{stone-martinez_starflow_2025} model retains information about the parameter space distribution of the training data and accounts for asymmetric uncertainties. 

We adopt the recommended training space density threshold of $>3\times10^9$, resulting in good age estimates for  stars. Figure \ref{fig:sample-distributions} compares the distributions of age, [Fe/H], and guiding radius for the stars with good ages against our full sample. The age quality cut eliminates $\sim\num{16000}$ stars, including most $<1\Gyr$ old stars and nearly all stars with $\bracket{Fe}<-0.8$. Metal-poor stars are excluded from the StarFlow catalog entirely due to the sparse sampling of asteroseismic data. 

The ``good age'' sample serves as the primary sample for this work. However, the age quality cut introduces a bias against low-metallicity stars, as illustrated in Figure \ref{fig:sample-distributions}. This effect is minor for disk stars, but it eliminates nearly all halo stars. For cases where we calibrate abundance trends as a function of metallicity (e.g., Sections \ref{sec:logg-corrections} and \ref{sec:galactic-trends}) or investigate the halo (Section \ref{sec:halo}), we instead use the full sample defined in Section \ref{sec:data}.

\subsubsection{Age Trend Validation}

\begin{figure}
    \centering
    \includegraphics[width=\linewidth]{\figpath{dataset_comparison.pdf}}
    \caption{Comparison of the age--[Ce/Mg] trend between MWM+StarFlow \citep{stone-martinez_starflow_2025}, APOKASC-3 \citep{pinsonneault_apokasc-3_2025}, and OCCAM \citep{otto_open_2026}. Each survey uses different age estimation techniques, but the abundances on the y-axis all come from MWM DR19. For the sake of comparison, we do not apply $\log(g)$ corrections to the abundances in this plot. The solid curves plot the running median [Ce/Mg] as a function of age for each data set, and the dashed curves plot the StarFlow running median for comparison. The APOKASC-3 sample is restricted to ``gold sample'' (highest quality) RGB and red clump stars. The OCCAM sample includes only stars with $\log(g)<3.5$.}
    \label{fig:dataset-comparison}
\end{figure}

We validate the StarFlow ages against other age catalogs for DR19. Figure \ref{fig:dataset-comparison} compares the age--[Ce/Mg] relation from our sample against two other age catalogs: APOKASC-3 \citep{pinsonneault_apokasc-3_2025} and the Open Cluster Chemical Abundances and Mapping Survey \citep[OCCAM;][]{otto_open_2026}. Although APOKASC-3 uses DR17 stellar parameters and abundances, in Figure \ref{fig:dataset-comparison} we plot the DR19 abundances due to the difference in zero-point offsets between the data releases. For APOKASC-3 we use ``gold sample'' RGB and red clump stars, and for OCCAM we include all clusters with at least one [Ce/H] measurement at $\log(g)<3.5$.

The median trend in our sample matches the APOKASC-3 and OCCAM trends well. Agreement with APOKASC-3 is expected because StarFlow uses training data from APOKASC-3. \citet{stone-martinez_starflow_2025} directly compare their ages against APOKASC-3 on a star-by-star level and find good agreement for ages $< 8\Gyr$, but for ages $>8\Gyr$ the StarFlow ages diverge from the asteroseismic validation sample. 

The StarFlow age trend shows a bump in [Ce/Mg] at an age of $\sim7\Gyr$, and a qualitatively similar bump is seen in the APOKASC-3 trend at an age of $\sim8\Gyr$. This bump is likely due to the transition between the high-Ia and low-Ia dominated disk, as the two populations are separated by $\sim0.2\dex$ in [Ce/Mg] at this age (see Section \ref{sec:residual-abundances}). The quantitative differences between the StarFlow and APOKASC-3 trends are likely driven by sample selection effects, as the low-Ia population is more prominent and has a larger age dispersion in StarFlow compared to APOKASC-3.

The OCCAM cluster ages are more informative because they are fully independent. The OCCAM survey adopts cluster ages from \citet{cantat-gaudin_painting_2020}, which are calculated from {\it Gaia} DR2 \citep{gaia_dr2_2018} photometry and parallaxes using an artificial neural network. The OCCAM data included in Figure \ref{fig:dataset-comparison} are limited to clusters with at least one RGB star with a measured [Ce/H]. Most of the clusters are young ($\lesssim3\Gyr$) and these clusters also show the largest spread in [Ce/Mg], but the median trend agrees with StarFlow to $\sim0.1\dex$. The old clusters are sparse but agree with the StarFlow trend up to ages of $\sim7\Gyr$. Overall, this suggests that the StarFlow ages are suitable for studying broad age--abundance trends, though perhaps not yet for addressing finer details in the shape of the relation, such as possible breaks or subtle curvature.

\section{Disk Age--Abundance Patterns}
\label{sec:results}

\subsection{Global Metallicity Dependence}
\label{sec:metallicity-trends}

\begin{figure*}
    \centering
    \includegraphics[width=\linewidth]{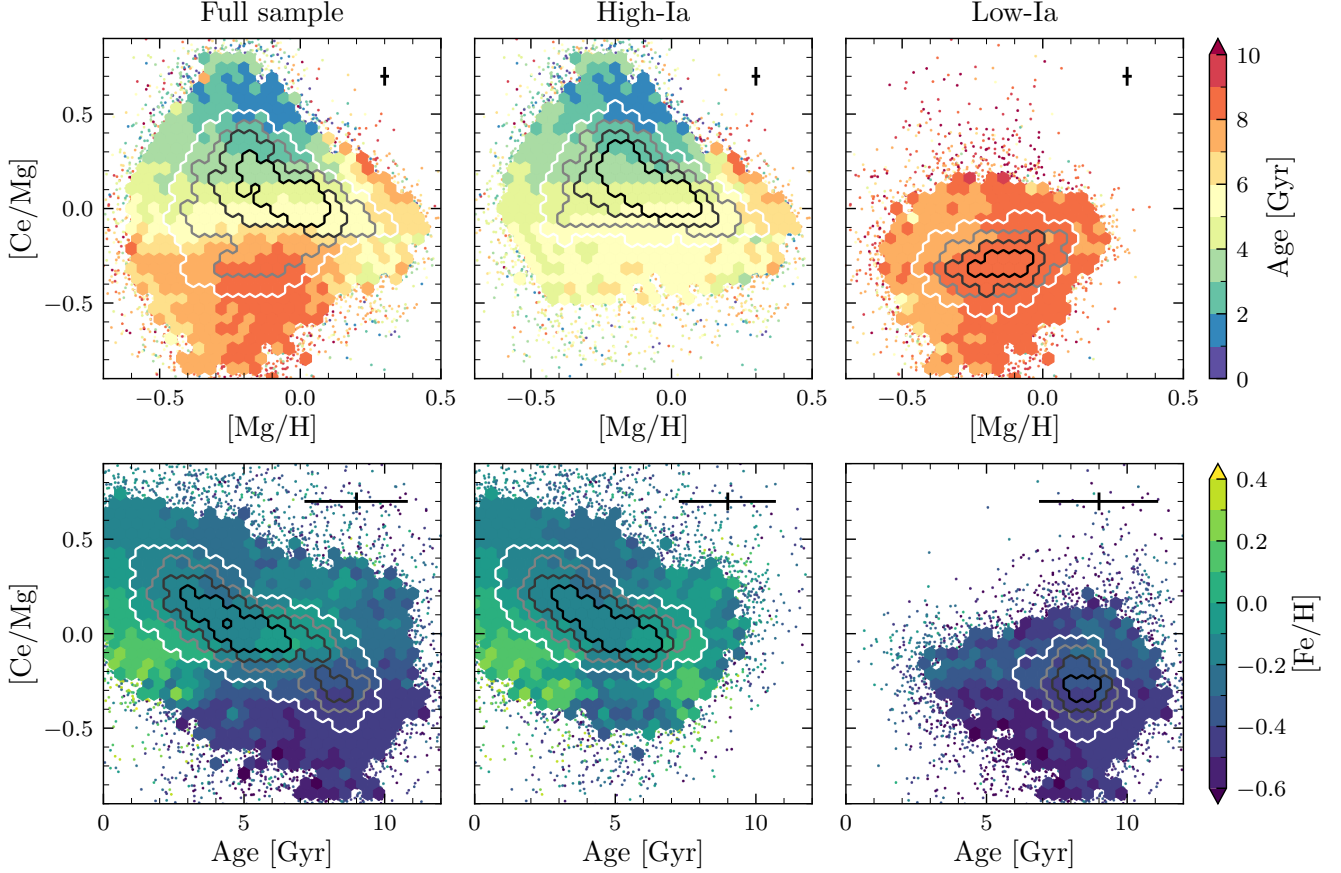}
    \caption{{\it Top row:} The distribution of abundances in the [Ce/Mg]--[Mg/H] plane color-coded by the median stellar age. The top-left panel plots the full sample, while the top-middle (top-right) panel plots the high-Ia (low-Ia) population only. Within each panel, each hexagonal bin contains a minimum of 10 stars; in lower-density regions, each star is plotted individually. The grayscale outlines represent linearly-spaced density contours, with the black contour encompassing the highest density of stars. The error bars in the upper right corner represent the median uncertainty in [Ce/Mg] and [Mg/H]. {\it Bottom row:} Similar, but showing the median stellar [Fe/H] as a function of [Ce/Mg] and age.}
    \label{fig:cemg-mgh-age}
\end{figure*}

We first explore the global age--metallicity--[Ce/Mg] relation. The top-left panel of Figure \ref{fig:cemg-mgh-age} shows the median stellar age as a function of [Mg/H] and [Ce/Mg]. A broad age trend is immediately apparent: stars with low [Ce/Mg] are on average the oldest in the sample, while stars with high [Ce/Mg] are the youngest. Stars with solar [Ce/Mg] have median ages of $\sim4-6\Gyr$, across a wide range of metallicity ($-0.5\lesssim\bracket{Mg}\lesssim+0.2$). The spread in [Ce/Mg] and age is broadest at $\bracket{Mg}\approx-0.2$, with most of the old, low-[Ce/Mg] stars belonging to the low-Ia population. The most Mg-rich stars occupy a narrower range of near-solar [Ce/Mg] with median ages of $\sim6-7\Gyr$, and some of the abundance bins with $\bracket[Mg]{Ce}>0$ are older on average than their low-[Ce/Mg] counterparts. 

The middle and right-hand panels of Figure \ref{fig:cemg-mgh-age} split the data into the high- and low-Ia populations according to Equation \ref{eq:alpha-cut}. The high-Ia population exhibits higher [Ce/Mg] ratios and a clear median age gradient, while the low-Ia population has lower [Ce/Mg] and a uniform median age. The two populations clearly occupy distinct parts of [Ce/Mg]--[Mg/H] space, although they overlap more than in [Fe/Mg], possibly due to the larger abundance errors.

The bottom-left panel of Figure \ref{fig:cemg-mgh-age} shows that the youngest stars span a broad range of [Ce/Mg] values, and that the median metallicity is inversely proportional to their [Ce/Mg]. For old stars, this trend is flipped, and stars with high [Ce/Mg] have higher metallicities on average than stars with lower [Ce/Mg]. The latter trend results from the transition from high-Ia to low-Ia-dominated space, and disappears when the two populations are plotted separately. The bottom middle panel shows that high-Ia, low-[Ce/Mg] stars tend to be metal-rich at the old and young ends of the distribution. Figure \ref{fig:cemg-mgh-age} provides the first hints in these data that the global age--[Ce/Mg] trend has an additional dependence on metallicity.

\subsection{Linear Fits to Age Trends}

\begin{figure*}
    \centering
    \includegraphics[width=\textwidth]{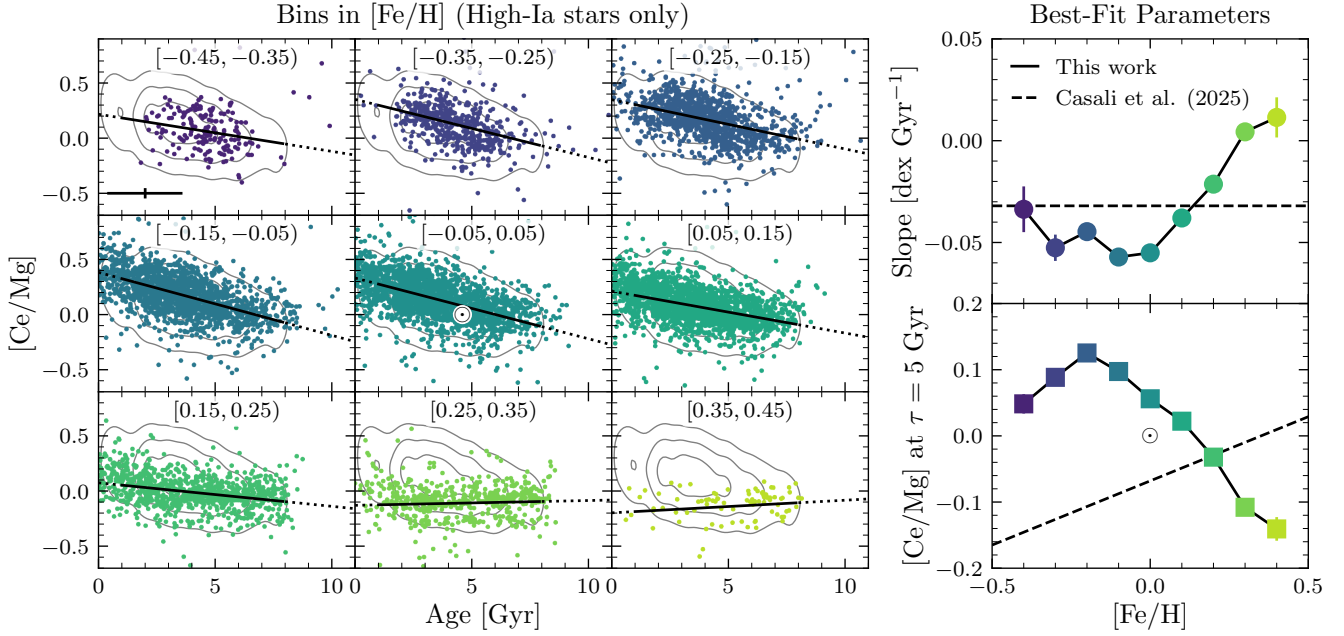}
    \caption{The variation in the age--[Ce/Mg] relation as a function of metallicity for high-Ia stars in the solar neighborhood ($7\leq R_{\rm guide}<9\kpc$, $z_{\rm max}<0.5\kpc$). {\it Left:} Each panel plots the high-Ia stars within the given metallicity bin (colored points) against a Gaussian kernel density estimate for the full solar neighborhood sample (gray contours). The solid line shows the linear regression to the data in that metallicity bin, and the dashed line shows the extrapolated relation outside the fitted age range of $1-8\Gyr$. The $\odot$ symbol marks the location of the Sun. {\it Right:} The best-fit parameters in the solar neighborhood as a function of metallicity. The dashed line represents the \citet{casali_tracing_2025} age--[Ce/Mg]--[Fe/H] relation.}
    \label{fig:local-metallicity-trends}
\end{figure*}

To quantify the metallicity dependence in the solar neighborhood, we fit a linear trend to [Ce/Mg] as a function of age in the high-Ia data in bins of metallicity. We bin stars by [Fe/H] with a width of $0.1\dex$ and fit a weighted least-squares regression in each bin. We weight each star by its error in [Ce/Mg] and age according to the weighting function
\begin{equation}
    w = 1 / (\sigma_{\bracket[Mg]{Ce}}^2 + b_0^2\sigma_{\rm age}^2),
\end{equation}
where $b$ is the slope computed without weighting by $\sigma_{\rm age}$. We iteratively compute the best-fit parameters, substituting $b_0$ for the slope computed in the last iteration, until the value converges to $<1\%$. We caution that the StarFlow age uncertainties are non-Gaussian, and the Ce abundance uncertainties in MWM are likely underestimated (see discussion in Section \ref{sec:ce-abundances}), both of which could lead to systematic errors in our fits.

We limit the sample in each bin to obtain the highest-quality fit. We only fit stars between $1-8\Gyr$ of age; outside this range, the StarFlow ages diverge from the asteroseismic test sample \citep[Figure 5,][]{stone-martinez_starflow_2025} due to intrinsic limitations in the accuracy of [C/N]-derived ages \citep{roberts_cn_2026}. We limit our fits to high-Ia stars because the low-Ia population has a smaller age spread and shows different trends with [Mg/H] and $R_{\rm guide}$ (see Section \ref{sec:galactic-trends}). To reduce the degeneracy between the regression slope $m$ and intercept $b$, we center the data at $\tau=5\Gyr$:
\begin{equation}
    \bracket[Mg]{Ce} = m\times(\tau-5\Gyr)+b.
\end{equation}

Both the slope and intercept of the fits vary with metallicity, as shown in Figure \ref{fig:local-metallicity-trends}. Metal-rich stars have shallow or flat slopes with $\bracket[Mg]{Ce}\leq0$ at $\tau=5\Gyr$. Stars with slightly sub-solar metallicity ($-0.3\leq\bracket{M}\leq0.0$) have the steepest trends and the highest [Ce/Mg] at $\tau=5\Gyr$. We note that even in the optimal range of [Fe/H], with a slope of $-0.05\dex\Gyr^{-1}$, the reported median [Ce/H] uncertainty of $0.05\dex$ places a floor of $1\Gyr$ on the precision of [Ce/Mg]-derived ages using MWM abundances. The situation is worse for metal-rich stars, where the flat age--[Ce/Mg] is essentially useless as a chemical clock.

We compare our fits against the metallicity-dependent age--[Ce/Mg] relation of \citet{casali_tracing_2025}, who calibrated their chemical clock using a small sample of {\it Kepler} field stars with asteroseismic ages and high-precision abundances. \citet{casali_tracing_2025} allowed their intercept to vary with [Fe/H] but not the slope, as illustrated in Figure \ref{fig:local-metallicity-trends}. This assumption does not account for the flattening of the relation at high metallicity and could lead to systematic effects in the estimated ages.

\begin{figure}
    \centering
    \includegraphics[width=\linewidth]{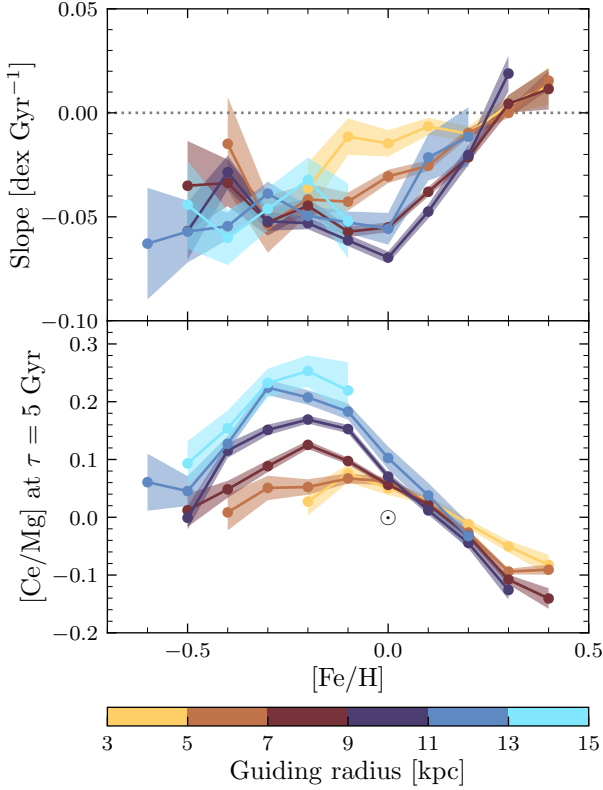}
    \caption{Best-fit parameters for the linear regression to the age--[Ce/Mg] relation as a function of metallicity and guiding radius. All fits are restricted to high-Ia stars with $z_{\rm max}<0.5\kpc$. The shaded bands indicate the uncertainty in the best-fit parameters at each metallicity bin. The dark brown curve plots the relation in the solar neighborhood and is the same as Figure \ref{fig:local-metallicity-trends}. {\it Top:} The slope as a function of metallicity. The dotted gray line indicates a flat slope. {\it Bottom:} The intercept at $\tau=5\Gyr$ as a function of metallicity. The $\odot$ symbol marks the Sun's location.}
    \label{fig:global-metallicity-fits}
\end{figure}

Figure \ref{fig:global-metallicity-fits} duplicates our linear fit procedure across multiple bins in guiding radius between $3-15\kpc$. If the metallicity variation was universal, Figure \ref{fig:global-metallicity-fits} would show similar behavior in each bin. Instead, we observe similar patterns in some regimes, but not others. Metal-rich stars ($\bracket{Fe}\ge+0.2$) have shallow or flat trends across the disk, with similar slopes and intercepts. Near solar metallicity, however, outer-disk stars show significantly steeper slopes and larger intercepts than their inner-disk counterparts. At $\bracket{Fe}=-0.2$, the best-fit [Ce/Mg] at $\tau=5\Gyr$ differs by $\sim0.25\dex$ between the inner- and outer-most bins. As a consequence, a [Ce/Mg] chemical clock calibrated in the solar neighborhood might systematically under-predict the age of an outer-disk star by $\sim2\Gyr$.

\subsection{Median Trends across the Galaxy}
\label{sec:galactic-trends}

\begin{figure*}
    \centering
    \includegraphics[width=\linewidth]{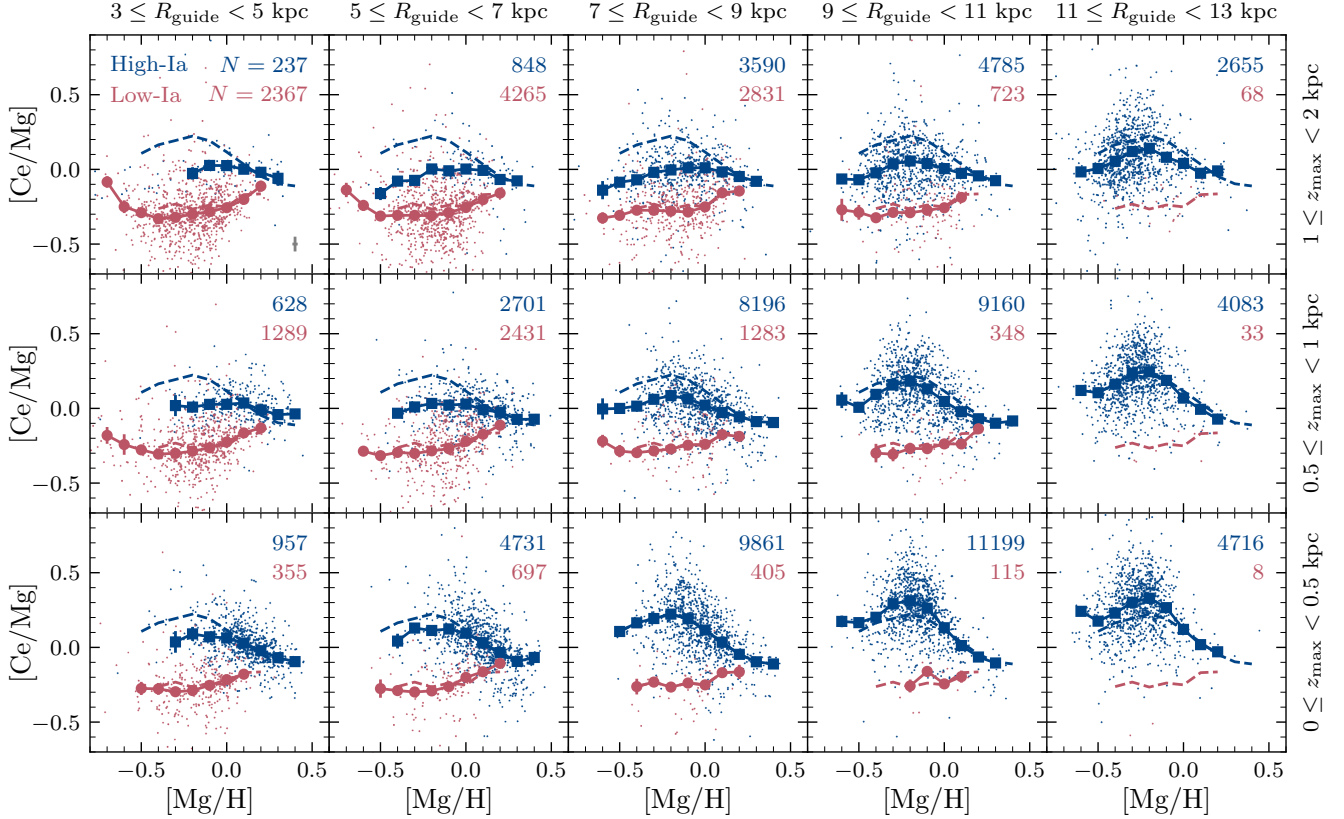}
    \caption{The median [Ce/Mg] as a function of [Mg/H] for high- and low-Ia populations (defined in Section \ref{sec:alpha-pops}) across the Galaxy. Each panel corresponds to a bin in $R_{\rm guide}$ (increasing left to right) and $z_{\rm max}$ (increasing bottom to top). The small blue (red) points plot individual abundances of high-Ia (low-Ia) stars. For visual clarity, only a random sample of 1,000 stars are plotted in each panel. The number of high- and low-Ia stars in each region is indicated in the top-right corner of each panel. The solid lines with large points plot the median [Ce/Mg] in bins of [Mg/H] for each region, with error bars indicating the bootstrapped standard error of the median. The dashed lines indicate the median trends in the Solar neighborhood ($7\leq R_{\rm guide}<9\kpc$, $z_{\rm max}<0.5\kpc$) for reference. The gray error bars in the top left panel (lower right corner) indicate the median abundance errors for the full DR19 sample.}
    \label{fig:median-trends-grid}
\end{figure*}

We next investigate positional variation in the [Ce/Mg]--[Mg/H] relation directly.
Figure \ref{fig:median-trends-grid} shows how the [Ce/Mg] median trend with [Mg/H] changes with position in the Galaxy for the high- and low-Ia populations (see Section \ref{sec:alpha-pops}). The figure layout is inspired by Figure 4 of \citet{hayden_chemical_2015}, but we use orbital parameters $R_{\rm guide}$ and $z_{\rm max}$ instead of present-day coordinates to better separate the chemical thin and thick disks. The standard errors of the median trend are small thanks to large sample sizes in each region. We observe similar patterns in the [Ce/H]--[Mg/H] plane, suggesting that it is Ce and not Mg that is driving these trends.

In the solar neighborhood, the median [Ce/Mg] of high-Ia stars peaks at $\bracket{Mg}=-0.2$ and is lowest at the metal-rich end. The location of the maximum [Ce/Mg] is consistent across the Galaxy, which suggests that this feature is driven by AGB star physics rather than the local star formation history. The maximum [Ce/Mg] is lowest at small $R_{\rm guide}$ and highest at large $R_{\rm guide}$, mirroring the variation with the linear fit intercept in Figure \ref{fig:global-metallicity-fits}. This behavior may be consistent with [Ce/Mg] as a chemical clock: in the inside-out formation scenario of the Milky Way, the inner Galaxy formed first and is therefore dominated by older stars, while the outer Galaxy formed more recently \citep[e.g.,][]{larson_models_1976,bird_inside_2021}. However, differences in the star formation history between regions may also contribute to the trend variation. 

Notably, the most metal-rich stars in each region, including the solar neighborhood, have slightly sub-solar [Ce/Mg] ratios. These stars are most likely migrators from inner regions, populating the metal-rich tail of the distribution in each region, so their similar [Ce/Mg] ratios could be because the stars have similar birth radii. This also lines up with the observation in Figure \ref{fig:global-metallicity-fits} that the most metal-rich stars have flat age--[Ce/Mg] trends.

At greater distances from the midplane, the high-Ia trends are flatter and have lower [Ce/Mg] on average, leading to less separation between the high- and low-Ia sequences. One explanation is that stars with higher $z_{\rm max}$ have experienced more vertical heating and are therefore older, so they have lower [Ce/Mg] on average. Alternatively, these stars could have formed out of a vertically heated gas disk, in which case their trends indicate an additional environmental dependence for the chemical clock. Both mechanisms could be at play in the high-$z_{\rm max}$ population \citep[e.g.,][]{bird_inside_2021}.

The low-Ia trend, by contrast, is remarkably consistent in each region: roughly constant at $\bracket[Mg]{Ce}\approx-0.3$ below $\bracket{Mg}<-0.2$, and increasing up to $\bracket[Mg]{Ce}\approx-0.1$ at higher metallicity. While the high-Ia trend in most regions is concave-down, the low-Ia trend is concave-up. The metal-rich tail of the high-Ia and low-Ia populations show very similar [Ce/Mg] ratios, and we note that the true agreement may be closer than shown in Figure \ref{fig:median-trends-grid} because by definition the populations are separated by $\ge0.04\dex$ in [Fe/Mg] (Equation \ref{eq:alpha-cut}). The low-Ia abundance patterns are consistent with the picture that this population formed rapidly and in a well-mixed environment in the inner Galaxy, with radial migration contributing low-Ia stars to the solar neighborhood and outer regions \citep[e.g.,][]{schonrich_chemical_2009,kubryk_evolution_2015,sharma_chemical_2021,chen_chemical_2023,prantzos_on_2023}.

The median sequences of [X/Mg] as a function of [Mg/H] can be used to decompose the element abundances into a combination of nucleosynthetic processes, such as SNe Ia and CCSNe via the ``two-process model'' \citep{weinberg_chemical_2019}. The residual abundances obtained after subtracting out the two-process predictions contain information about un-modeled physics, potentially including additional nucleosynthetic sources such as AGB stars \citep[e.g.,][]{griffith_residual_2022,weinberg_chemical_2022}. \citet{griffith_many_2025} observed strong correlations between the Ce residual abundance and position in the high-Ia disk that mirror the patterns seen in Figure \ref{fig:median-trends-grid}: a deficit of Ce at small $R_{\rm guide}$ and large $z_{\rm max}$, and an excess at large $R_{\rm guide}$ and small $z_{\rm max}$. This pattern likely reflects the lack of an AGB nucleosynthesis channel in the two-process model, but it also reflects variation in the median [Ce/Mg] trend across the galaxy, which may be driven by both AGB physics and chemical evolution. More work is needed to understand these trends from a theoretical perspective.

Overall, Figures \ref{fig:global-metallicity-fits} and \ref{fig:median-trends-grid} complicate the picture of [Ce/Mg] as a chemical clock. Broadly, old populations have low [Ce/Mg] and young populations have high [Ce/Mg], but there are also clear trends with metallicity and location in the Galaxy. These effects are also hard to disentangle: the highest [Ce/Mg] ratios are found in the outer Galaxy, where stellar populations are young {\it and} low-metallicity. These results imply that a chemical clock calibrated in the solar neighborhood, even accounting for the metallicity dependence, will fail when applied in other regions of the Galaxy.

\subsection{Residual Abundances}
\label{sec:residual-abundances}

\begin{figure}
    \centering
    \includegraphics[width=\linewidth]{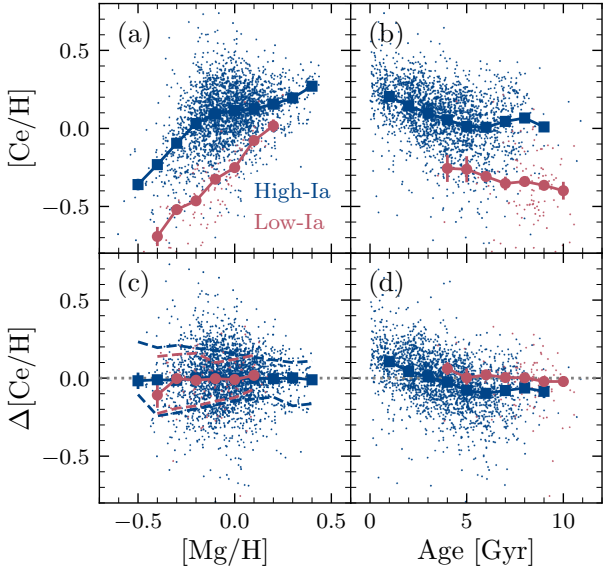}
    \caption{Demonstration of the residual abundance calculation for the solar neighborhood. Top row: median trend in [Ce/H] as a function of (a) [Mg/H] and (b) age for the high- and low-Ia populations. The format of each panel is similar to Figure \ref{fig:median-trends-grid}. Bottom row: median trend in $\Delta\bracket{Ce}$ as a function of (c) [Mg/H] and (d) age. In panel (c), the dashed lines plot the 16th and 84th percentiles. Both populations show similar trends with age in [Ce/H] and $\Delta\bracket{Ce}$.}
    \label{fig:residual-explainer}
\end{figure}

To disentangle the evolution of Ce from that of Mg, we perform a simplified version of the residual abundance analysis of \citet{weinberg_chemical_2022} and \citet{griffith_many_2025}. For each star, we calculate the residual abundance from the median trend with [Mg/H],
\begin{equation}
    \Delta\bracket{Ce} \equiv \bracket{Ce} - {\rm med}(\bracket{Ce} | \bracket{Mg}),
\end{equation}
with the median trend computed separately for the low- and high-Ia populations. Figure \ref{fig:residual-explainer} illustrates the residual abundance calculation for the solar neighborhood. The residual abundances show a flat trend with [Mg/H] but the age trend persists, demonstrating the utility of $\Delta\bracket{Ce}$ to separate age and metallicity trends.

We found in Section \ref{sec:galactic-trends} that abundance trends vary with position in the Galaxy, so we compute the median high-Ia trend separately for each region presented in Figure \ref{fig:median-trends-grid}. For low-Ia stars, we use the global median trend because variations between regions are small and some regions have few low-Ia stars (see Figure \ref{fig:median-trends-grid}). 

\begin{figure}
    \centering
    \includegraphics[width=\linewidth]{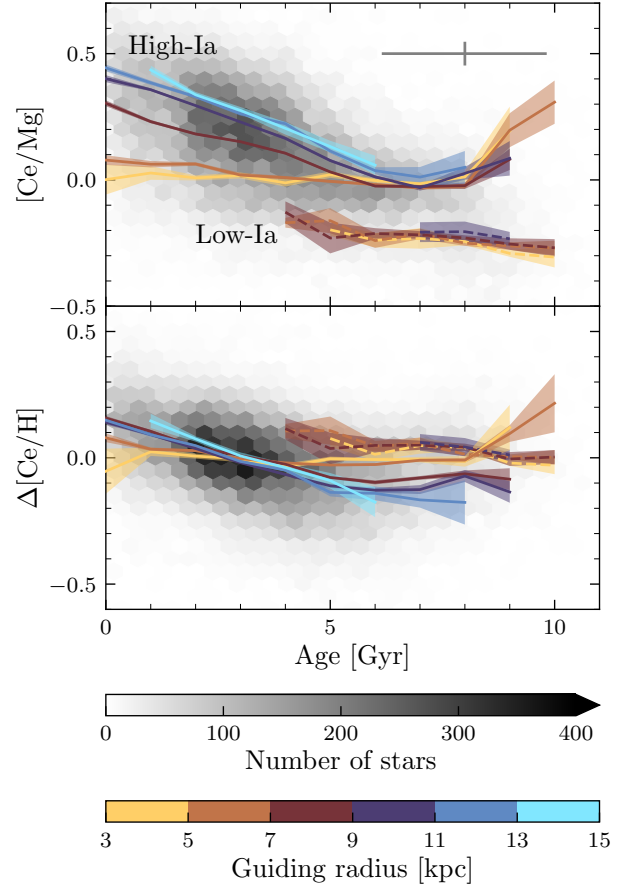}
    \caption{Median age--abundance trends in bins of guiding radius. Solid (dashed) lines plot the median trends in age bins of 1 Gyr for the high-Ia (low-Ia) sub-samples. Semi-transparent bands represent the standard error on the median abundance in each age bin. The underlying 2-D histogram plots the density of the full sample in age--abundance space. Stars are restricted to $z_{\rm max}<0.5$ kpc in both panels. The gray error bars in the top panel indicate the median age and abundance uncertainties.}
    \label{fig:median-age-trends}
\end{figure}

Figure \ref{fig:median-age-trends} shows the variation in the [Ce/Mg]--age and $\Delta\bracket{Ce}$--age trends with $R_{\rm guide}$. In the high-Ia population, the [Ce/Mg]--age trends in the outer galaxy ($R_{\rm guide}>9\kpc$) are both significantly steeper and normalized to higher [Ce/Mg] values than in the solar neighborhood, while in the inner galaxy ($R_{\rm guide}<7\kpc$) the trends are shallow or flat. This matches the findings of \citet{casali_time_2023} and \citet{ratcliffe_chemical_2024}, who both observe a weaker trend in the inner Galaxy. By contrast, the low-Ia population shows a uniformly flat [Ce/Mg]--age trend that is consistent across $R_{\rm guide}=3-11\kpc$. The low-Ia trend extends to ages as low as $4\Gyr$, but this mostly reflects scatter due to the larger age uncertainties for these stars.

Figure \ref{fig:median-age-trends} also shows evidence for a possible break in the age--[Ce/Mg] relation at $\tau\approx6\Gyr$. This feature has also been observed by \citet{casali_time_2023} at similar radii in APOGEE DR17, and \citet{ratcliffe_chemical_2024} also observed non-linear trends in the [Ce/Mg]--age relation. However, this feature may be influenced by systematic effects, such as the decreasing density of the asteroseismic training sample at old ages and low metallicities. The age range of $6-8\Gyr$ is also where the sample transitions from high-Ia to low-Ia dominated, which could affect the trends in each of the sub-samples.

The bottom panel of Figure \ref{fig:median-age-trends} shows the residual abundance trends with age across the Galaxy. The residual abundance $\Delta\bracket{Ce}$ normalizes the trends to 0 at the median of the distribution, but variation in slope with $R_{\rm guide}$ is still present. The outer galaxy trends overlap with the solar neighborhood for ages younger than $\tau=6\Gyr$, while the inner galaxy trends are still flat. Even with metallicity trends subtracted out independently in each region, the age trend still generally steepens with increasing $R_{\rm guide}$. Figure \ref{fig:median-age-trends} illustrates that even after factoring out the metallicity dependence, the Ce age--abundance trends still show variation across the Galaxy. Any age inference from [Ce/Mg] is therefore tied to the local star formation history, and at best is limited to outer-disk high-Ia stars with ages $<6\Gyr$.

\subsection{Radial Abundance Gradients}

\begin{figure}
    \centering
    \includegraphics[width=\linewidth]{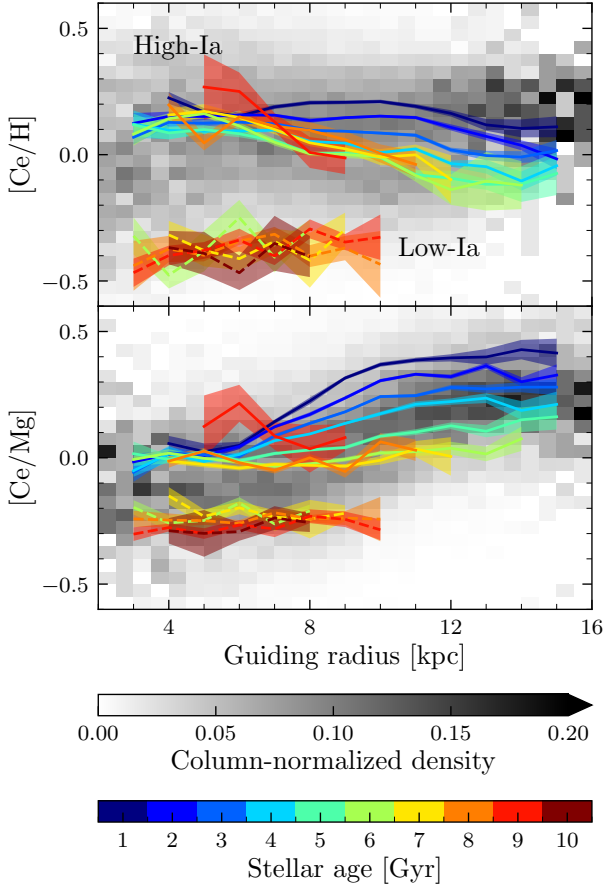}
    \caption{Radial gradients of [Ce/H] and [Ce/Mg] as a function of stellar age. Solid (dashed) lines plot the median abundance in bins of guiding radius for the high-Ia (low-Ia) sub-sample, color-coded by stellar age. The shaded areas indicate the standard error of the median at each bin. Stars are restricted to $z_{\rm max}<0.5$ kpc for both sub-samples. The underlying 2-D histogram plots the density of stars in each bin of abundance and radius, normalized to 1 in each radial bin.}
    \label{fig:gradients}
\end{figure}

Figure \ref{fig:gradients} shows the radial [Ce/H] and [Ce/Mg] gradients in bins of stellar age. Within the high-Ia population, the radial [Ce/H] gradient is negative with a remarkably consistent slope for $\ge5\Gyr$ old stars, whereas it is nearly flat for the youngest stars. The [Ce/Mg] gradient evolves in a similar direction, except the oldest stars show a mostly flat gradient (from $6-8\Gyr$, excluding the $9\Gyr$ bin because it contains few stars), while the youngest stars exhibit a steep {\it positive} gradient. Between $R_{\rm guide}=3-5\kpc$, the median [Ce/H] and [Ce/Mg] are consistent to within $\pm0.1\dex$ in all age bins. The low-Ia population exhibits a flat or slightly positive gradient in both [Ce/H] and [Ce/Mg] that is consistent across multiple age bins. Notably, the Ce abundances of low-Ia stars are significantly lower (by $0.3\dex$ in [Ce/Mg] or up to $0.5\dex$ in [Ce/H]) than high-Ia stars with similar age and $R_{\rm guide}$.

We stress that the gradient evolution in Figure \ref{fig:gradients} is driven by Ce and not Mg. The stellar [Mg/H] gradient only exhibits variation at the $\sim0.1\dex$ level in our sample, which is in line with other stellar age catalogs \citep[e.g.,][]{roberts_cn_2026}, and we measure a gradient of ${\rm d}\bracket{Mg}/{\rm d}R_{\rm gal}=-0.055\pm0.002\dex\kpc^{-1}$ for the youngest age bin. Compared to recent measurements by \citet{martinez-hernandez_desired_2026}, our stellar [Mg/H] gradient is slightly steeper than the [O/H] gradient traced by nebular gas ($-0.045\pm0.003\dex\kpc^{-1}$) and Cepheids ($-0.045\pm0.002$), and consistent with the gradient measured by OB stars ($-0.06\pm0.01\dex\kpc^{-1}$). These independent zero-age anchors strengthen the case that the positive [Ce/Mg] gradient is driven by Ce rather than by residual systematics in Mg.

The age-decomposition of the radial gradients indicates little evolution in the inner Galaxy across the full age range, and also minimal evolution in the outer galaxy for $\gtrsim6\Gyr$ old stars. The evolution to higher [Ce/H] and [Ce/Mg] ratios is significant for young stars in the outer Galaxy, but this highlights the localized nature of true ``chemical clock'' behavior. Using data from the OCCAM survey, \citet{sales-silva_exploring_2022} similarly found that young open clusters exhibit a flatter radial [Ce/H] gradient and higher overall [Ce/H] than older clusters, and recently \citet{otto_open_2026} found that the gradient in [Ce/Fe] is positive and steepest for the youngest clusters. In the bar and inner bulge, \citet{sales-silva_chemical_2026} measured a slightly positive [Ce/H] gradient using APOGEE DR17 for low-Ia stars with typical values of $\bracket{Ce}\approx-0.5$, consistent with our measurements in the low-Ia disk.

\subsection{Effects of Radial Migration}

The results in this section could suggest that, even if a universal age--[Ce/Mg] relation does not exist, a chemical clock could nonetheless be calibrated on [Fe/H] and guiding radius within a specified parameter window (say, $\bracket{Fe}\leq0$ and $R_{\rm guide}\geq7\kpc$). However, a star's abundance pattern reflects its birth environment, which may not be the same as its present-day environment. Gas abundance measurements show no evidence for azimuthal variations in [O/H] at the $>0.1\dex$ level \citep{mendez-delgado_gradients_2022,martinez-hernandez_desired_2026}, so some amount of radial migration is necessary to explain the abundance scatter in mono-$R_{\rm guide}$ stellar populations. Radial migration and vertical heating processes cause stars to move far from their birthplace over relatively short timescales, mixing up local abundance trends \citep{sellwood_radial_2002,schonrich_chemical_2009}. The strength of radial migration in the Milky Way is debated \citep[e.g.,][]{frankel_measuring_2018,chen_open_2020,lian_quantifying_2022,lehmann_probing_2024}, but it is clear that the solar neighborhood contains both inward and outward migrators \citep{schonrich_chemical_2009,johnson_stellar_2021}. Therefore, controlling for the environmental dependence of the age--[Ce/Mg] relation requires knowledge of the star's birth environment.

Stellar birth radius estimates attempt to undo the effects of radial migration. \citet{ratcliffe_chemical_2024} found that clear trends in [Ce/Mg] with birth radius are blurred when stars are instead grouped by present-day guiding radius. However, current methods for estimating birth radii require stellar age measurements \citep{ratcliffe_unveiling_2023,lu_there_2024}. The radial metallicity profile only varies at the $\sim0.1\dex$ level between mono-age populations, even up to ages of $\sim10\Gyr$ \citep{johnson_milky_2025}, which raises the possibility of age-independent birth radius measurements, but no such catalog currently exists. As the utility of a chemical clock lies in estimating stellar ages that are otherwise difficult to measure, requiring accurate birth radii to estimate stellar ages is a non-starter. The non-universality of the age--[Ce/Mg] relation is therefore an effect that cannot be calibrated away.

\section{Halo Abundance Patterns}
\label{sec:halo}

In this section, we investigate [Ce/Mg] trends in the Milky Way halo. Halo stars fall outside of the optimal parameter space of the StarFlow age catalog, so we do not investigate trends with age in this section. Furthermore, we do not apply the corrective $\log(g)$ abundance offsets to the halo sample because many of the stars fall outside the metallicity range of the calibration sample (see Section \ref{sec:logg-corrections}). We additionally exclude a small number of stars with ASPCAP abundance flags for [Al/H] or [Mn/H].

\subsection{Halo Selection}
\label{sec:halo-selection}

\begin{deluxetable*}{lCcl}
    \tablecaption{Selection criteria for the halo sub-samples plotted in Figure \ref{fig:halo}.\label{tab:halo-selection}}
    \tablehead{
        \colhead{Sub-sample} & \colhead{$N$} & \colhead{Selection Criteria} & \colhead{References}
    }
    \startdata
    \multirow{3}{*}{GSE} & \multirow{3}{*}{173} & $30\leq\sqrt{J_R}\leq\SI{50}{\kilo\parsec^{1/2}\kilo\meter^{1/2}\second^{-1/2}}$   & \multirow{2}{*}{\citet{feuillet_gse_2020}} \\
    & & $-500\leq L_z\leq\SI{500}{\kilo\parsec\kilo\meter\per\second}$ & \\
    & & $\bracket[Mg]{Mn} \leq -0.65-2\times\bracket[Fe]{Al}$ & \citet{hawkins_using_2015} \\
    \hline
    \multirow{3}{*}{Accreted halo\tablenotemark{a}}  & \multirow{3}{*}{284} & $|L_z/J_{\rm tot}|\leq0.3$ & \citet{feltzing_metal-weak_2023} \\
    & & $\bracket[Mg]{Mn} \leq -0.65-2\times\bracket[Fe]{Al}$ & \citet{hawkins_using_2015} \\
    & & $R_{\rm gal} > 3\kpc$ & \\
    \hline
    \multirow{3}{*}{{\it In situ} halo}  & \multirow{3}{*}{549} & $|L_z/J_{\rm tot}|\leq0.3$ & \citet{feltzing_metal-weak_2023} \\
    & & $-0.55-2\times\bracket[Fe]{Al} \leq \bracket[Mg]{Mn} \leq -0.2$ & \citet{hawkins_using_2015} \\
    & & $R_{\rm gal} > 3\kpc$ & \\
    \hline
    \multirow{2}{*}{Disk} & \multirow{2}{*}{103263} & $L_z/J_{\rm tot}\geq0.6$  & \citet{feltzing_metal-weak_2023} \\
    & & $R_{\rm gal} > 3\kpc$ & \\
    \enddata
    \tablenotetext{a}{Excluding GSE stars.}
\end{deluxetable*}

\begin{figure*}
    \centering
    \includegraphics[width=0.8\linewidth]{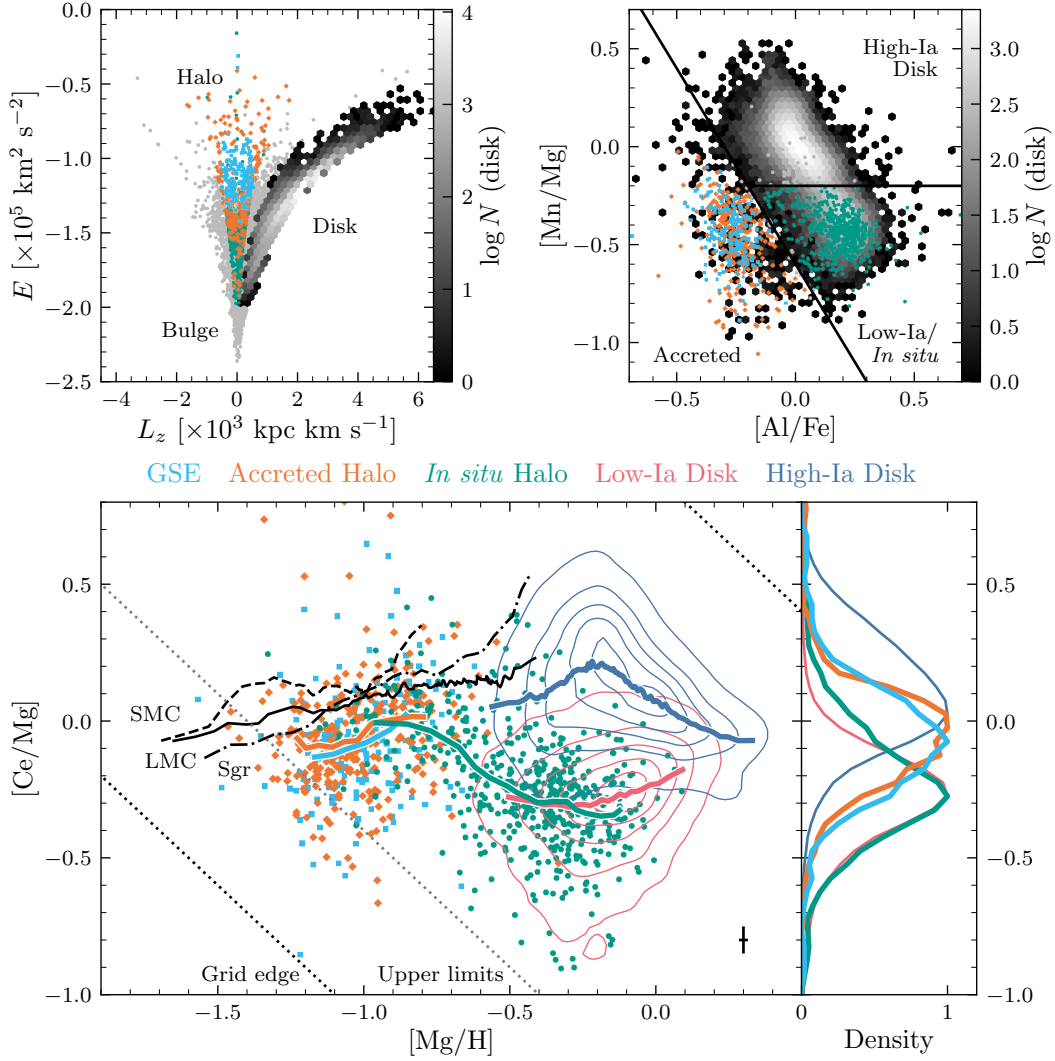}
    \caption{{\it Top left:} Our halo sub-samples plotted in the orbital $E-L_z$ plane (see selection criteria in Table \ref{tab:halo-selection}). The 2-D histogram plots the density of stars on disk-like orbits, and the colored points plot stars with halo-like orbits. The gray points do not fall within either the halo or disk selections and are not plotted in other panels. {\it Top right:} Chemical selection for accreted stars. The gray points fall outside of the chemical cuts or too close to the boundaries and are excluded from analysis. {\it Bottom left:} Abundance ratios (non-$\log(g)$-corrected) for {\it in situ} halo stars (orange diamonds), accreted halo stars (turquoise circles), and GSE stars (cyan squares). The blue and pink contours indicate the density of high-Ia and low-Ia disk stars, respectively. The solid colored curves plot a rolling median for each of the five sub-samples. The thin black curves plot the rolling median trends for three Milky Way satellites from \citet{hasselquist_apogee_2021} with updated DR19 zero-point offsets. The black dotted lines indicate the boundary where abundances are flagged for being too close to a grid edge, and the gray dotted line indicates the approximate region where the \citet{shetrone_apache_2026} upper limits dominate the sample. {\it Bottom right:} Normalized distributions of [Ce/Mg] for the five sub-samples. Distributions are boxcar-smoothed by 0.2 dex for clarity.}
    \label{fig:halo}
\end{figure*}

We define an initial halo sample using the orbital ``action diamond,'' which compresses the three orbital actions into a scaled, 2-D space \citep[see][]{lane_kinematic_2022}. Following \citet{feltzing_metal-weak_2023}, we select halo stars with $|L_z/J_{\rm tot}|<0.3$, where $J_{\rm tot}\equiv|J_R|+|L_z|+|J_z|$. We impose an additional cut on present-day Galactocentric radius of $R_{\rm gal}>3\kpc$ to avoid contamination from stars in the bulge or bar. We also select a comparison sample of disk stars with $|L_z/J_{\rm tot}|>0.6$ and $R_{\rm gal}>3\kpc$.

We further sub-divide the halo sample in abundance space.
Following \citet{hawkins_using_2015}, we use [Mn/Mg]\footnote{Because we use Mg as our reference element, we note that the sign convention in the top-right panel of Figure \ref{fig:halo} is reversed from the typical convention in the literature, $\bracket[Mn]{Mg}$.} and [Al/Fe] to distinguish stars accreted from dwarf galaxies from stars that were born {\it in situ} in the disk or proto-Milky Way. We also select a subset of accreted stars belonging to the disrupted dwarf galaxy {\it Gaia} Sausage/Enceladus (GSE) \citep[e.g.,][]{belokurov_co-formation_2018,helmi_merger_2018} using the orbital selection criteria of \citet{feuillet_gse_2020}. Stars in GSE are removed from the accreted halo sub-sample so the two populations can be examined independently. Table \ref{tab:halo-selection} presents the selection criteria for each of our halo sub-samples, and the top two panels of Figure \ref{fig:halo} compare the halo-subsamples against the disk in orbital and chemical space.

Our sample was not optimized for a representative census of the halo, so a number of selection effects could influence our results. The Ce II lines are hard to detect at low metallicities (see Figure \ref{fig:ce-lines-2}), so the abundance upper limits preferentially exclude stars with $\bracket{Mg}<-1$. Our hard cut at $\bracket{M}=-1.5$ prohibits us from extending the analysis to the very metal-poor regime. Finally, because we do not apply $\log(g)$ corrections, ASPCAP systematics could be affecting our observed trends. We expect these $\log(g)$ systematics should still be minimal because only 13\% of the halo sample has $\log(g)>2.5$, the regime where the corrections are most severe, on account of the distance to a typical halo star. For these reasons our findings in this section are more tentative than those in Section \ref{sec:results}.

\subsection{[Ce/Mg] Trends in the Halo}

The lower left panel in Figure \ref{fig:halo} plots [Ce/Mg] as a function of [Mg/H] for each of the halo sub-samples as compared to the low- and high-Ia disk. Most of the GSE and other chemically accreted stars have $\bracket{Mg}<-0.7$, with similar respective means of $\langle\bracket[Mg]{Ce}\rangle=-0.07$ and $\langle\bracket[Mg]{Ce}\rangle=-0.06$, and their [Ce/Mg] distributions lie in between the low- and high-Ia disks, as shown in the lower right panel. Accreted stars, including GSE, have a positive correlation between [Ce/Mg] and [Mg/H], with [Ce/Mg] increasing by $\sim0.2\dex$ over $\sim0.4\dex$ in [Mg/H]. 
The roughly solar [Ce/Mg] ratios in the accreted halo population suggest that AGB enrichment is important even below $\bracket{Mg}\leq-1$. While the $r$-process channel does produce some Ce, it accounts for only $\sim20\%$ of Ce in the Sun \citep{arlandini_neutron_1999}, so it is unlikely to be responsible for the high [Ce/Mg] in the accreted halo. 

The {\it in situ} halo shows distinct trends from the accreted stars. These stars have $\bracket{Mg}>-1$, a lower mean $\langle\bracket[Mg]{Ce}\rangle=-0.25$, and a [Ce/Mg] distribution that is similar to the low-Ia disk. In contrast to the accreted stars, the {\it in situ} population has a negative correlation between [Ce/Mg] and [Mg/H]. The median trend overlaps with the accreted and GSE trends between $-1\lesssim\bracket{Mg}\lesssim-0.8$, and between $-0.8\lesssim\bracket{Mg}\lesssim-0.5$ it bridges the space between the accreted halo and low-Ia disk. Due to the $\alpha$-enhancement of $\bracket[Fe]{Mg}\approx+0.3$, the metal-rich end of the {\it in situ} halo at $\bracket{Mg}=-0.2$ corresponds to $\bracket{Fe}=-0.5$, approximately the metal-poor end of the high-Ia disk.

The decrease in [Ce/Mg] within the {\it in situ} halo is interesting and could point to the Milky Way's unique evolutionary history. This could be caused by an increase in the star formation efficiency at the onset of the low-Ia epoch, which would temporarily increase the production of elements with short enrichment timescales, such as Mg, relative to the delayed enrichment from AGB stars. \citet{conroy_birth_2022} showed that a significant increase in the star formation efficiency can explain the $\bracket[Fe]{\alpha}$ ratios at the transition from the halo to the thick disk. Future work should explore the consequences of this model for the evolution of $[s/\alpha]$.

\subsection{Literature Comparison}

In Figure \ref{fig:halo}, we compare to trends measured by \citet{hasselquist_apogee_2021} for the Large and Small Magellanic Clouds (L/SMC) and the Sagittarius dwarf (Sgr), applying their selection criteria to the APOGEE catalog with updated abundances from DR19. The LMC and SMC show a similar positive trend over the same metallicity range as the accreted halo population, albeit with a $\sim+0.1-0.2\dex$ offset, while Sgr has a steeper trend extending to higher [Ce/Mg]. \citet{hasselquist_apogee_2021} found that GSE is deficient in [Ce/Mg] compared to other dwarf galaxies, in agreement with our own findings. Above $\bracket{Mg}\ge-0.8$, the LMC and Sgr trends continue to increase, similar to the median trend of the high-Ia disk rather than the {\it in situ} halo. The divergence between the satellite trends and the Milky Way halo and low-Ia disk might indicate a unique event in our Galaxy's evolution, possibly associated with the birth of the low-Ia disk.

Our halo trends are generally consistent with the literature. Several studies have found a declining trend in [Ce/Fe] with increasing [Fe/H] in the Galactic bulge, matching the trend in our {\it in situ} population \citep{razera_abundance_2022,sales-silva_perspective_2024,ernandes_abundances_2026}. In the GALAH survey, \citet{myeong_milky_2022} found solar or super-solar [Y/Fe] abundances in accreted halo populations and a declining [Y/Fe]--[Fe/H] trend for {\it in situ} halo stars. \citet{carrillo_detailed_2022} report an average $\bracket[Fe]{Ce}=0.31$ for accreted stars, much higher than our measurement, although they also note a systematic $0.35\dex$ offset between their optical abundance measurements and APOGEE. Accounting for this offset, their findings are consistent with our accreted halo trends.

In contrast to the disk, mean [$s$/Fe] trends in the halo and local dwarfs differ depending on the choice of $s$-process element \citep{van_der_swaelmen_chemical_2013,skuladottir_neutron-capture_2020,carrillo_detailed_2022}, and abundance patterns also vary between dwarf galaxies \citep[e.g.,][]{shetrone_apache_2026}. Further study is needed to connect $s$-process abundance trends with the star formation history in the halo and local dwarf galaxies.

\section{Conclusions}
\label{sec:conclusions}

We have explored the relationship between [Ce/Mg], age, metallicity, and Galactic position for $\sim100,000$ stars in MWM DR19. We focused our analysis on the age--[Ce/Mg] relation in the high-Ia disk and the [Mg/H]--[Ce/Mg] relation in the low-Ia disk and halo. Our conclusions are summarized as follows:

\begin{itemize}
    \item The slope of the local age--[Ce/Mg] relation for high-Ia stars varies substantially with metallicity. Stars with $-0.3\leq\bracket{Fe}\leq0.0$ have the steepest trend, whereas stars with $\bracket{Fe}\ge+0.3$ have a flat trend (Figure \ref{fig:local-metallicity-trends}). The metallicity dependence also varies with guiding radius (Figure \ref{fig:global-metallicity-fits}).
    \item The median trend in [Ce/Mg] with [Mg/H] varies across the Galaxy. High-Ia stars have generally solar or super-solar [Ce/Mg] ratios, and the abundance trends vary with guiding radius and midplane distance. Low-Ia stars have sub-solar [Ce/Mg] ratios, and show consistent abundance patterns across the disk (Figure \ref{fig:median-trends-grid}).
    \item The age--[Ce/Mg] trend varies substantially with guiding radius. The trend is steepest in the outer disk, at $R_{\rm guide}\ge13\kpc$, and flat in the inner disk, at $R_{\rm guide}\le5\kpc$. After calculating a residual Ce abundance to account for the effect of metallicity, the radial trends persist, suggesting that the age--[Ce/Mg] relation is influenced by the local star formation history (Figure \ref{fig:median-age-trends}). This finding is reinforced by the age-decomposition of the radial [Ce/Mg] gradient (Figure \ref{fig:gradients}).
    \item Low-metallicity halo stars have $\bracket[Mg]{Ce}\approx0$, and this is consistent between accreted and {\it in situ} populations. At higher metallicities, chemically-tagged {\it in situ} halo stars have [Ce/Mg] consistent with the low-Ia population, which diverges from trends in Milky Way satellites. The relatively high [Ce/Mg] of halo stars suggests that AGB enrichment is important even at $\bracket{Mg}\leq-1$ (Figure \ref{fig:halo}).
\end{itemize}

Overall, the data show that the [Ce/Mg] ratio is influenced by both the local star formation history and the physics of AGB star nucleosynthesis, complicating its use as a chemical clock. Given our current understanding of Galactic chemical evolution, we expect these conclusions to extend to any given $s$-process element. We therefore caution against using $[s/\alpha]$ as a universal tool to estimate stellar ages.

The main limitation of this study is the precision of the age and abundance measurements. Previous high-precision studies have focused on $s$-process abundances near solar metallicity \citep[e.g.,][]{nissen_high-precision_2015,tucci_maia_solar_2016,slumstrup_ymg_2017,spina_temporal_2018,nissen_high-precision_2020}. Our results highlight the need for more high-resolution studies of neutron-capture elements at super-solar metallicity and in the metal-poor halo, where the Ce abundance patterns differ from solar-metallicity stars. A more precise understanding of the metallicity dependence will help constrain AGB star nucleosynthesis and galactic chemical evolution models.

Our study focused on Ce as the representative neutron-capture element in the MWM survey. While the age--$[s/\alpha]$ trends appear generally consistent across many elements \citep{casali_tracing_2025}, elements in the first and second $s$-process peaks should show differences in their age and metallicity dependence due to internal AGB star physics \citep{karakas_dawes_2014}. A complementary large-scale study done with optical spectroscopy \citep[e.g., the GALAH survey;][]{buder_galah_2021} would allow for comparisons with first-peak elements such as Y or Zr, providing a window to test our understanding of the structure and evolution of AGB stars.

The chemical evolution of $s$-process elements is not well understood. Galactic chemical evolution models that adopt the best available AGB star yield prescriptions are unsuccessful at reproducing the observed trends in the Milky Way \citep[e.g.,][]{grisoni_modelling_2020,ratcliffe_chemical_2024,molero_constraining_2025}. In a future work, we will use these results to constrain a chemical evolution model of the disk and explore modifications to standard AGB star yield assumptions. 

\section*{Acknowledgements}

The authors thank Dr.\ Alexander Stone-Martinez for assistance with the StarFlow age catalog. L.D.\ thanks Dr.\ Giulia Cinquegrana, Dr.\ Madeleine McKenzie, Dr.\ Lucy Lu, Prof.\ David Weinberg, and the attendees of OSU's Galaxy Hour for many useful conversations about this work. L.D.\ is also grateful to Phil for moral support.

L.O.D. and J.A.J. acknowledge support from National Science Foundation (NSF) grant AST-2307621. 
J.W.J. acknowledges support from a Carnegie Theoretical Astrophysics Center postdoctoral fellowship.
S.F. was supported by a project grant from the Knut and Alice Wallenberg Foundation (KAW 2020.0061 Galactic Time Machine  KAW 2013 The New Milky Way, PI: Feltzing) and a project grant from the Swedish Research Council (2024-04619\_VR, Galaxy formation and evolution: solving the thick disk conundrum).
J.G.F-T gratefully acknowledges the support provided by ANID Fondecyt Regular No. 1260371.
J.E.M.-D. gratefully acknowledges support from the Secretaria de Ciencia, Humanidades, Tecnologia e Innovacion (SECIHTI) project CBF-2025-I-2048, ``Resolviendo la Fisica Interna de las Galaxias: De las Escalas Locales a la Estructura Global con el SDSS-V Local Volume Mapper'', and from the UNAM/DGAPA/PAPIIT project IA103326, ``DESIRED (DEep Spectra of ionised Regions Database): de las emisiones mas sutiles a la fisica fundamental del universo.''

Funding for the Sloan Digital Sky Survey V has been provided by the Alfred P. Sloan Foundation, the Heising-Simons Foundation, the National Science Foundation, and the Participating Institutions. SDSS acknowledges support and resources from the Center for High-Performance Computing at the University of Utah. SDSS telescopes are located at Apache Point Observatory, funded by the Astrophysical Research Consortium and operated by New Mexico State University, and at Las Campanas Observatory, operated by the Carnegie Institution for Science. The SDSS web site is \url{www.sdss.org}.

SDSS is managed by the Astrophysical Research Consortium for the Participating Institutions of the SDSS Collaboration, including the Carnegie Institution for Science, Chilean National Time Allocation Committee (CNTAC) ratified researchers, Caltech, the Gotham Participation Group, Harvard University, Heidelberg University, The Flatiron Institute, The Johns Hopkins University, L'Ecole polytechnique f\'{e}d\'{e}rale de Lausanne (EPFL), Leibniz-Institut f\"{u}r Astrophysik Potsdam (AIP), Max-Planck-Institut f\"{u}r Astronomie (MPIA Heidelberg), Max-Planck-Institut f\"{u}r Extraterrestrische Physik (MPE), Nanjing University, National Astronomical Observatories of China (NAOC), New Mexico State University, The Ohio State University, Pennsylvania State University, Smithsonian Astrophysical Observatory, Space Telescope Science Institute (STScI), the Stellar Astrophysics Participation Group, Universidad Nacional Aut\'{o}noma de M\'{e}xico, University of Arizona, University of Colorado Boulder, University of Illinois at Urbana-Champaign, University of Toronto, University of Utah, University of Virginia, Yale University, and Yunnan University.

% From the Center for Belonging and Social Change, https://cbsc.osu.edu/about-us/land-acknowledgement
We acknowledge the land that The Ohio State University occupies is the ancestral and contemporary territory of the Shawnee, Potawatomi, Delaware, Miami, Peoria, Seneca, Wyandotte, Ojibwe, and many other Indigenous peoples. Specifically, the university resides on land ceded in the 1795 Treaty of Greeneville and the forced removal of tribes through the Indian Removal Act of 1830. As a land grant institution, we honor the resiliency of these tribal nations and recognize the historical contexts that have and continue to affect the Indigenous peoples of this land.

\software{Astropy \citep{astropy_collaboration_astropy_2013,astropy_collaboration_astropy_2018,astropy_collaboration_astropy_2022}, gala \citep{price-whelan_gala_2017}, Matplotlib \citep{hunter_matplotlib_2007}, NumPy \citep{harris_array_2020}, pandas \citep{reback_pandas-devpandas_2021,the_pandas_development_team_pandas-devpandas_2025}, scikit-learn \citep{pedregosa_scikit-learn_2011}, SciPy \citep{virtanen_scipy_2020}, sdss-access (\url{https://github.com/sdss/sdss_access}), statsmodels \citep{seabold2010statsmodels}.}

\appendix

\section{Reproducibility}
\label{app:reproducibility}

The DR19 stellar parameters and abundances can be found in the ASPCAP summary file at \url{https://data.sdss.org/sas/dr19/spectro/astra/0.6.0/summary/astraAllStarASPCAP-0.6.0.fits.gz}. The stellar age data from the StarFlow value-added catalog can be found at \url{https://data.sdss.org/sas/dr19/vac/mwm/starflow/}. The complete merged catalog of stellar abundances, ages, and orbits can be furnished upon request. Scripts used to generate the figures are stored in a GitHub repository at \url{https://github.com/lodubay/chemical-clocks}, and release \texttt{v1.0.0} can be used to re-build this manuscript.

\section{Abundance Calibration Procedure}
\label{app:logg-calibration-procedure}

Following \citet{sit_chemical_2024}, we define a calibration sample of stars in DR19 with $-0.75 < \bracket{Mg} < 0.45$, $3<R_{\rm gal}<13\kpc$, and $|z| < 2\kpc$, in addition to the cuts listed in Section \ref{sec:data}. Corrective offsets are applied to each individual [Fe/H] and [Ce/H] such that the global median [X/Mg] trend with [Mg/H] is the same over all $\log(g)$. [X/Mg] medians are calculated in $0.1\dex$-wide bins of [Mg/H] and $0.5\dex$-wide bins of $\log(g)$. Differences in the median values of each bin from a fiducial value (selected to be $\bracket{Mg} = 0$ and $\log(g) = 1.75$) are calculated separately for the low-Ia and high-Ia sequences. Then a number-weighted average of median differences over the two alpha sequences is calculated at each ([Mg/H], $\log(g)$) bin pair. The final offset value for a given star is calculated by interpolating the binned median differences to that star's [Mg/H] and $\log(g)$. We finally subtract a zero-point offset equal to the median difference between the corrected and un-corrected abundances for the entire sample, to avoid shifting the overall distribution of [Ce/H] and [Fe/H]. Tables \ref{tab:ce-offset-table} and \ref{tab:fe-offset-table} present the abundance offsets for Ce and Fe, respectively, before applying the final zero-point offset.

\begin{table*}[]
    \centering
    \begin{tabular}{r|cccccccccccc}
\hline\hline
[Mg/H] & -0.7 & -0.6 & -0.5 & -0.4 & -0.3 & -0.2 & -0.1 & 0.0 & 0.1 & 0.2 & 0.3 & 0.4 \\
$\log(g)$ &  &  &  &  &  &  &  &  &  &  &  &  \\
\hline
1.25 & -0.00094 & 0.032 & -0.012 & -0.01 & 0.017 & 0.039 & 0.047 & 0.039 & 0.022 & 0 & 0 & 0 \\
1.75 & 0 & 0 & 0 & 0 & 0 & 0 & 0 & 0 & 0 & 0 & 0 & 0 \\
2.25 & 0.13 & 0.14 & 0.1 & 0.096 & 0.099 & 0.073 & 0.053 & 0.044 & 0.039 & 0.044 & 0.015 & 0.015 \\
2.75 & 0.038 & -0.0078 & 0.1 & 0.094 & 0.11 & 0.059 & 0.039 & 0.056 & 0.065 & 0.083 & 0.048 & 0.048 \\
\hline
\end{tabular}

    \caption{Corrective offsets to [Ce/H] as a function of $\log(g)$ and [Mg/H].}
    \label{tab:ce-offset-table}
\end{table*}

\begin{table*}[]
    \centering
    \begin{tabular}{r|cccccccccccc}
\hline\hline
[Mg/H] & -0.7 & -0.6 & -0.5 & -0.4 & -0.3 & -0.2 & -0.1 & 0.0 & 0.1 & 0.2 & 0.3 & 0.4 \\
$\log(g)$ &  &  &  &  &  &  &  &  &  &  &  &  \\
\hline
1.25 & 0.012 & 0.0079 & 0.0071 & 0.00015 & 0.0016 & 0.0058 & 0.0076 & 0.0017 & -0.012 & 0 & 0 & 0 \\
1.75 & 0 & 0 & 0 & 0 & 0 & 0 & 0 & 0 & 0 & 0 & 0 & 0 \\
2.25 & 0.012 & -0.0063 & 0.00037 & 0.00055 & 0.0062 & 0.0022 & -0.0016 & -0.0044 & -0.0003 & 0.0085 & 0.005 & 0.005 \\
2.75 & 0.00019 & -0.00022 & 0.0042 & 0.0044 & 0.0074 & -0.0013 & -0.0028 & 0.00094 & 0.0063 & 0.021 & 0.018 & 0.018 \\
\hline
\end{tabular}

    \caption{Corrective offsets to [Fe/H] as a function of $\log(g)$ and [Mg/H].}
    \label{tab:fe-offset-table}
\end{table*}

\section{Comparison with BAWLAS}
\label{app:bawlas-comparison}

\begin{figure}
    \centering
    \includegraphics[width=\linewidth]{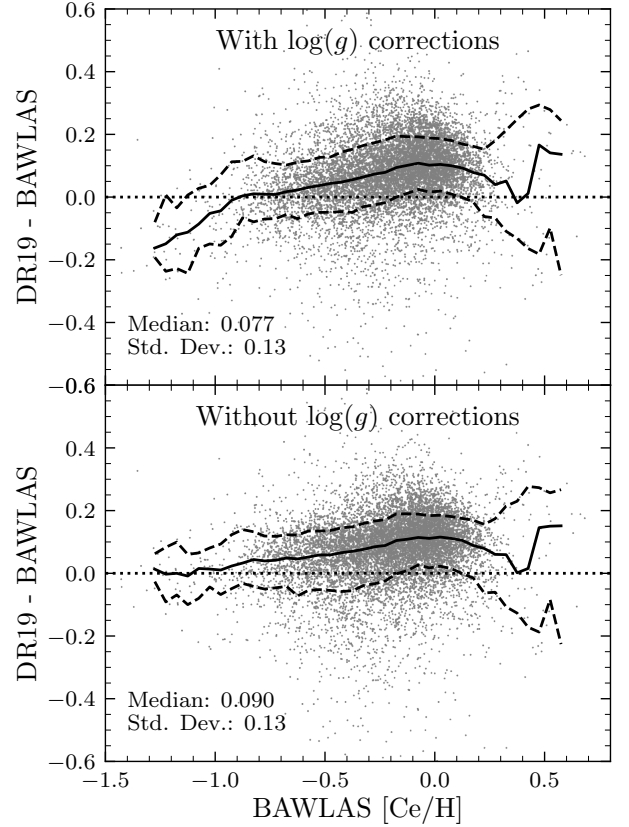}
    \caption{Difference between DR19 ASPCAP and DR17 BAWLAS \citep{hayes_bacchus_2022} abundances. The solid (dashed) black line plots the binned median (16th and 84th percentiles) of the abundance differences. The dotted line indicates zero for reference. The gray points indicate individual stellar abundances, and we plot a random sample of \num{10000} stars for visual clarity. In the top (bottom) panel, DR19 abundances are presented with (without) the corrective $\log(g)$ abundance offsets (Section \ref{sec:logg-corrections}).}
    \label{fig:bawlas-comparison}
\end{figure}

We compare the DR19 ASPCAP [Ce/H] abundances with and without the corrective $\log(g)$ offsets against measurements from the BACCHUS Analysis of Weak Lines in APOGEE Spectra \citep[BAWLAS;][]{hayes_bacchus_2022} catalog. BAWLAS is a value-added catalog for APOGEE DR17 that contains line-by-line abundance measurements for elements with weak and blended lines, including Ce. The BAWLAS methodology offers greater precision, quality flagging, and a prescription for upper limits, but requires higher $S/N$ spectra and therefore results in a smaller sample. Figure \ref{fig:bawlas-comparison} plots the difference between BAWLAS and DR19 [Ce/H] measurements. The median offset is $+0.077$ ($+0.090$) dex with (without) the $\log(g)$ corrections, and we observe a systematic trend of larger differences at higher [Ce/H]. Overall, the zero line falls within $1\sigma$ of the median across most of the abundance space. We note that extrapolation of the $\log(g)$ corrections causes the median to diverge significantly at low metallicity, which is why we use un-corrected abundances for our analysis of the halo (Section \ref{sec:halo}).

\section{Sensitivity Tests}
\label{app:sensitivity}

\begin{figure*}
    \centering
    \includegraphics[width=\linewidth]{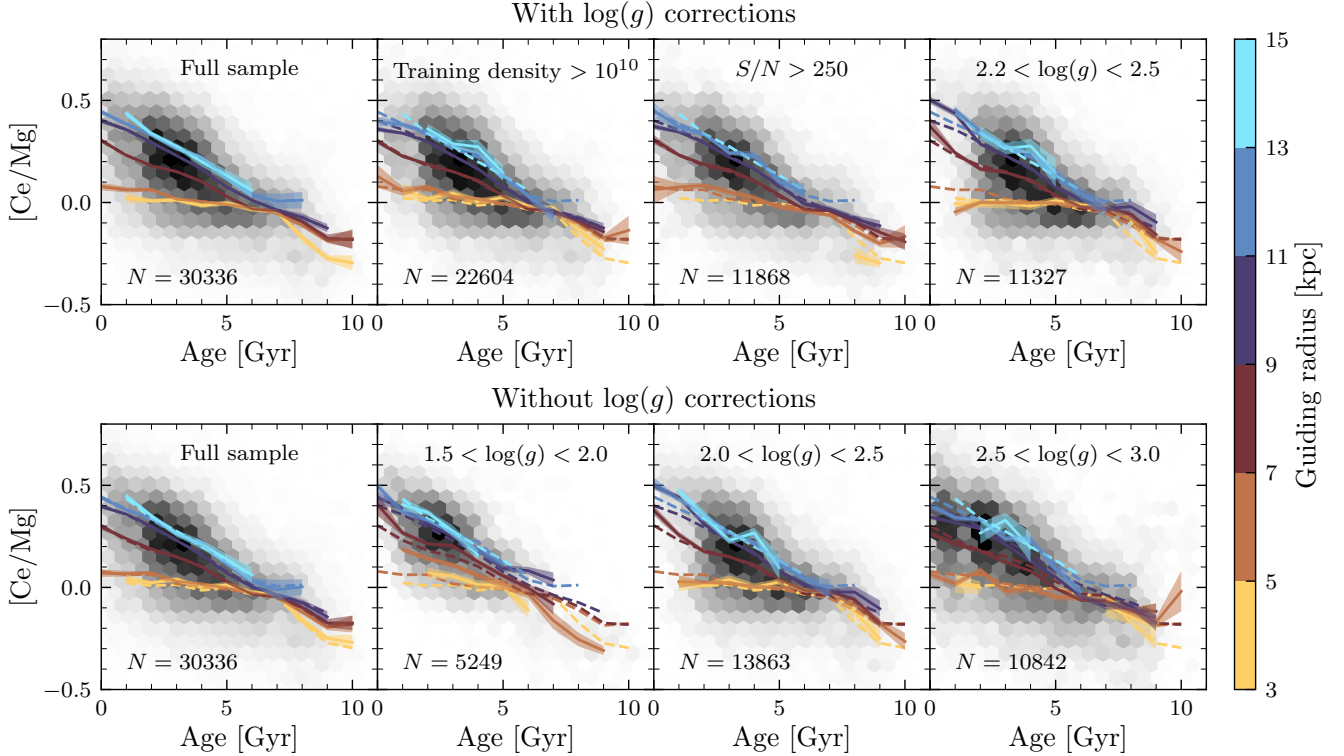}
    \caption{The age--[Ce/Mg] relation in bins of guiding radius for different sample selection criteria. Each panel is similar to the top panel of Figure \ref{fig:median-age-trends}, except the high- and low-Ia populations are not plotted separately here. {\it Top:} adopting the corrective $\log(g)$ abundance offsets (see Section \ref{sec:logg-corrections}), from left to right, the full sample with $z_{\rm max}<0.5\kpc$; a stricter age quality cut; a higher signal-to-noise cut; and a narrow $\log(g)$ range selecting the red clump. {\it Bottom:} adopting the original ASPCAP abundances, the left-most panel plots the full sample with $z_{\rm max}<0.5\kpc$, and the remaining three panels plot narrower bins in $\log(g)$. The dashed lines indicate the median trends in the upper-left panel.}
    \label{fig:age-trend-sensitivity}
\end{figure*}

\begin{figure}
    \centering
    \includegraphics[width=\linewidth]{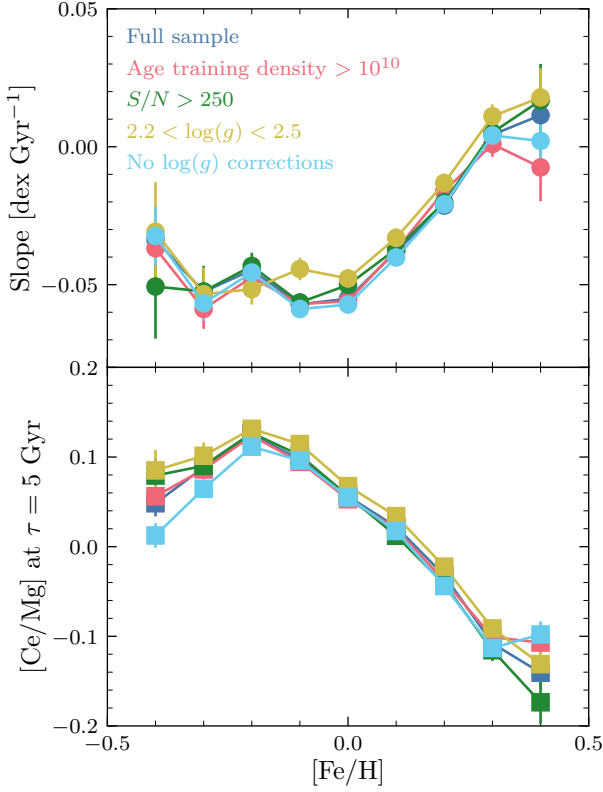}
    \caption{Linear fit parameters to the age--[Ce/Mg] relation in the solar neighborhood for different sample selection choices (see Figure \ref{fig:local-metallicity-trends}).}
    \label{fig:linear-fit-sensitivity}
\end{figure}

\begin{figure}
    \centering
    \includegraphics[width=\linewidth]{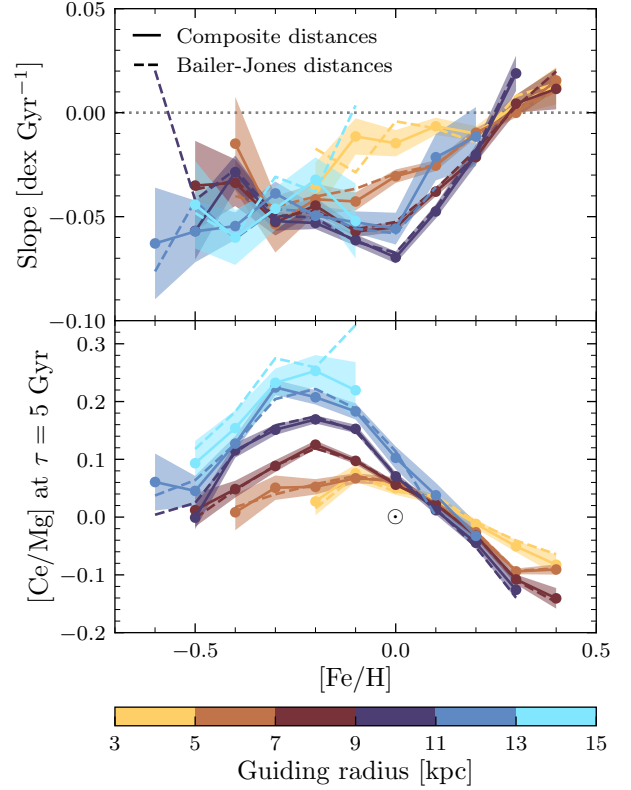}
    \caption{Linear fit parameters to the age--[Ce/Mg] relation in bins of $R_{\rm guide}$ for two different orbital catalogs (see Figure \ref{fig:global-metallicity-fits}). The solid lines with points adopt a composite catalog of distances as described in Section \ref{sec:orbits}. The dashed lines plot the best-fit parameters for an alternate orbital catalog based on just the \citet{bailer-jones_estimating_2021} photo-geometric distances.}
    \label{fig:distance-source-sensitivity}
\end{figure}

We test the sensitivity of our results to the sample selection criteria. Figure \ref{fig:age-trend-sensitivity} plots the median age--[Ce/Mg] trends in bins of $R_{\rm guide}$ (similar to Figure \ref{fig:median-age-trends}) for samples with a stricter age quality cut, a higher signal-to-noise threshold, a narrow range of $\log(g)$ around the red clump, and limited ranges of $\log(g)$ without the corrective $\log(g)$ offsets. The principal qualitative result that the trends are steepest in the outer galaxy and flat in the inner galaxy holds across all panels, and the quantitative trends are also generally consistent. Similarly, Figure \ref{fig:linear-fit-sensitivity} duplicates the right-hand panels of Figure \ref{fig:local-metallicity-trends}. The slopes are generally consistent at the level of $\lesssim0.01\dex\Gyr^{-1}$, and the intercepts are consistent to within $\lesssim0.05\dex$, suggesting that our findings are not affected by sample selection choices.

In our main sample, we combine three separate distance catalogs to obtain orbits for the greatest number of stars possible (see Section \ref{sec:orbits}). However, distance systematics could change discontinuously across subsets of stars. The \citet{zhang_parameters_2023} distances also have a metallicity dependence, which could in principle masquerade as a metallicity-dependent radial trend. We test for these systematic effects by repeating our analysis on a sample restricted to just the \citet{bailer-jones_estimating_2021} distances. Figure \ref{fig:distance-source-sensitivity} duplicates the results in Figure \ref{fig:global-metallicity-fits} on this restricted distance catalog. For most metallicity and guiding radius bins, the fit parameters for the two catalogs agree within uncertainties, demonstrating that distance systematics do not have a large effect on our results.

\bibliography{references}

@ARTICLE{casali_modelling_2026,
       author = {{Casali}, G. and {Molero}, M.},
        title = "{Modelling s-process chemical clocks: insights from high-precision Kepler data}",
      journal = {arXiv e-prints},
         year = 2026,
        month = aug,
          eid = {arXiv:2608.17480},
        pages = {arXiv:2608.17480},
archivePrefix = {arXiv},
       eprint = {2608.17480},
 primaryClass = {astro-ph.SR},
       adsurl = {https://ui.adsabs.harvard.edu/abs/2026arXiv260817480C}
}

@ARTICLE{prantzos_on_2023,
       author = {{Prantzos}, Nikos and {Abia}, Carlos and {Chen}, Tianxiang and {de Laverny}, Patrick and {Recio-Blanco}, Alejandra and {Athanassoula}, E. and {Roberti}, Lorenzo and {Vescovi}, Diego and {Limongi}, Marco and {Chieffi}, Alessandro and {Cristallo}, Sergio},
        title = "{On the origin of the Galactic thin and thick discs, their abundance gradients and the diagnostic potential of their abundance ratios}",
      journal = {\mnras},
         year = 2023,
        month = aug,
       volume = {523},
       number = {2},
        pages = {2126-2145},
          doi = {10.1093/mnras/stad1551},
archivePrefix = {arXiv},
       eprint = {2305.13431},
 primaryClass = {astro-ph.GA},
       adsurl = {https://ui.adsabs.harvard.edu/abs/2023MNRAS.523.2126P}
}

@ARTICLE{chen_chemical_2023,
       author = {{Chen}, Boquan and {Hayden}, Michael R. and {Sharma}, Sanjib and {Bland-Hawthorn}, Joss and {Kobayashi}, Chiaki and {Karakas}, Amanda I.},
        title = "{Chemical evolution with radial mixing redux: a detailed model for formation and evolution of the Milky Way}",
      journal = {\mnras},
         year = 2023,
        month = aug,
       volume = {523},
       number = {3},
        pages = {3791-3811},
          doi = {10.1093/mnras/stad1568},
archivePrefix = {arXiv},
       eprint = {2204.11413},
 primaryClass = {astro-ph.GA},
       adsurl = {https://ui.adsabs.harvard.edu/abs/2023MNRAS.523.3791C}
}

@ARTICLE{sharma_chemical_2021,
       author = {{Sharma}, Sanjib and {Hayden}, Michael R. and {Bland-Hawthorn}, Joss},
        title = "{Chemical enrichment and radial migration in the Galactic disc - the origin of the [{\ensuremath{\alpha}}Fe] double sequence}",
      journal = {\mnras},
         year = 2021,
        month = nov,
       volume = {507},
       number = {4},
        pages = {5882-5901},
          doi = {10.1093/mnras/stab2015},
archivePrefix = {arXiv},
       eprint = {2005.03646},
 primaryClass = {astro-ph.GA},
       adsurl = {https://ui.adsabs.harvard.edu/abs/2021MNRAS.507.5882S}
}

@ARTICLE{kubryk_evolution_2015,
       author = {{Kubryk}, M. and {Prantzos}, N. and {Athanassoula}, E.},
        title = "{Evolution of the Milky Way with radial motions of stars and gas. I. The solar neighbourhood and the thin and thick disks}",
      journal = {\aap},
         year = 2015,
        month = aug,
       volume = {580},
          eid = {A126},
        pages = {A126},
          doi = {10.1051/0004-6361/201424171},
archivePrefix = {arXiv},
       eprint = {1412.0585},
 primaryClass = {astro-ph.GA},
       adsurl = {https://ui.adsabs.harvard.edu/abs/2015A&A...580A.126K}
}

@ARTICLE{majewski_apogee_2017,
       author = {{Majewski}, Steven R. and {Schiavon}, Ricardo P. and {Frinchaboy}, Peter M. and {Allende Prieto}, Carlos and {Barkhouser}, Robert and {Bizyaev}, Dmitry and {Blank}, Basil and {Brunner}, Sophia and {Burton}, Adam and {Carrera}, Ricardo and {Chojnowski}, S. Drew and {Cunha}, K{\'a}tia and {Epstein}, Courtney and {Fitzgerald}, Greg and {Garc{\'\i}a P{\'e}rez}, Ana E. and {Hearty}, Fred R. and {Henderson}, Chuck and {Holtzman}, Jon A. and {Johnson}, Jennifer A. and {Lam}, Charles R. and {Lawler}, James E. and {Maseman}, Paul and {M{\'e}sz{\'a}ros}, Szabolcs and {Nelson}, Matthew and {Nguyen}, Duy Coung and {Nidever}, David L. and {Pinsonneault}, Marc and {Shetrone}, Matthew and {Smee}, Stephen and {Smith}, Verne V. and {Stolberg}, Todd and {Skrutskie}, Michael F. and {Walker}, Eric and {Wilson}, John C. and {Zasowski}, Gail and {Anders}, Friedrich and {Basu}, Sarbani and {Beland}, Stephane and {Blanton}, Michael R. and {Bovy}, Jo and {Brownstein}, Joel R. and {Carlberg}, Joleen and {Chaplin}, William and {Chiappini}, Cristina and {Eisenstein}, Daniel J. and {Elsworth}, Yvonne and {Feuillet}, Diane and {Fleming}, Scott W. and {Galbraith-Frew}, Jessica and {Garc{\'\i}a}, Rafael A. and {Garc{\'\i}a-Hern{\'a}ndez}, D. An{\'\i}bal and {Gillespie}, Bruce A. and {Girardi}, L{\'e}o and {Gunn}, James E. and {Hasselquist}, Sten and {Hayden}, Michael R. and {Hekker}, Saskia and {Ivans}, Inese and {Kinemuchi}, Karen and {Klaene}, Mark and {Mahadevan}, Suvrath and {Mathur}, Savita and {Mosser}, Beno{\^\i}t and {Muna}, Demitri and {Munn}, Jeffrey A. and {Nichol}, Robert C. and {O'Connell}, Robert W. and {Parejko}, John K. and {Robin}, A.~C. and {Rocha-Pinto}, Helio and {Schultheis}, Matthias and {Serenelli}, Aldo M. and {Shane}, Neville and {Silva Aguirre}, Victor and {Sobeck}, Jennifer S. and {Thompson}, Benjamin and {Troup}, Nicholas W. and {Weinberg}, David H. and {Zamora}, Olga},
        title = "{The Apache Point Observatory Galactic Evolution Experiment (APOGEE)}",
      journal = {\aj},
         year = 2017,
        month = sep,
       volume = {154},
       number = {3},
          eid = {94},
        pages = {94},
          doi = {10.3847/1538-3881/aa784d},
archivePrefix = {arXiv},
       eprint = {1509.05420},
 primaryClass = {astro-ph.IM},
       adsurl = {https://ui.adsabs.harvard.edu/abs/2017AJ....154...94M}
}

@ARTICLE{kobayashi_origin_2020,
       author = {{Kobayashi}, Chiaki and {Karakas}, Amanda I. and {Lugaro}, Maria},
        title = "{The Origin of Elements from Carbon to Uranium}",
      journal = {\apj},
         year = 2020,
        month = sep,
       volume = {900},
       number = {2},
          eid = {179},
        pages = {179},
          doi = {10.3847/1538-4357/abae65},
archivePrefix = {arXiv},
       eprint = {2008.04660},
 primaryClass = {astro-ph.GA},
       adsurl = {https://ui.adsabs.harvard.edu/abs/2020ApJ...900..179K}
}

@inproceedings{seabold2010statsmodels,
  title={statsmodels: Econometric and statistical modeling with python},
  author={Seabold, Skipper and Perktold, Josef},
  booktitle={9th Python in Science Conference},
  year={2010},
}

@ARTICLE{jofre_traits_2020,
       author = {{Jofr{\'e}}, Paula and {Jackson}, Holly and {Tucci Maia}, Marcelo},
        title = "{Traits for chemical evolution in solar twins. Trends of neutron-capture elements with stellar age}",
      journal = {\aap},
         year = 2020,
        month = jan,
       volume = {633},
          eid = {L9},
        pages = {L9},
          doi = {10.1051/0004-6361/201937140},
archivePrefix = {arXiv},
       eprint = {1912.02800},
 primaryClass = {astro-ph.SR},
       adsurl = {https://ui.adsabs.harvard.edu/abs/2020A&A...633L...9J}
}

@ARTICLE{martinez-hernandez_desired_2026,
       author = {{Mart{\'\i}nez-Hern{\'a}ndez}, I. Rafael and {M{\'e}ndez-Delgado}, J. Eduardo and {Esteban}, C{\'e}sar and {Garc{\'\i}a-Rojas}, Jorge and {Carigi}, Leticia and {Rodr{\'\i}guez}, Luis F. and {Zapata}, Luis A. and {Rosales-Ortega}, F. Fabi{\'a}n and {Orte-Garc{\'\i}a}, Maialen and {Reyes-Rodr{\'\i}guez}, Elena and {Arellano-C{\'o}rdova}, Karla Z. and {Kreckel}, Kathryn and {Sattler}, Natascha and {Morisset}, Christophe and {Peimbert}, Manuel and {Torres-Peimbert}, Silvia and {Pe{\~n}a}, Miriam and {Chrob{\'a}kov{\'a}}, {\v{Z}}ofia and {Zari}, Eleonora and {Espinoza-Galeas}, David A.},
        title = "{The DESIRED temperature─metallicity relations in star-forming regions: probing the Galactic radial and azimuthal metallicity distributions}",
      journal = {\mnras},
         year = 2026,
        month = mar,
       volume = {546},
       number = {4},
          eid = {stag149},
        pages = {stag149},
          doi = {10.1093/mnras/stag149},
archivePrefix = {arXiv},
       eprint = {2601.13337},
 primaryClass = {astro-ph.GA},
       adsurl = {https://ui.adsabs.harvard.edu/abs/2026MNRAS.546ag149M}
}

@ARTICLE{sales-silva_chemical_2026,
       author = {{Sales-Silva}, J.~V. and {Cunha}, K. and {Smith}, V.~V. and {Daflon}, S. and {Souto}, D. and {Guer{\c{c}}o}, R. and {Loaiza-Tacuri}, V. and {Queiroz}, A. and {Chiappini}, C. and {Minchev}, I. and {Majewski}, S.~R. and {Barbuy}, B. and {Bizyaev}, D. and {Fern{\'a}ndez-Trincado}, Jos{\'e} G. and {Frinchaboy}, Peter M. and {Hasselquist}, S. and {Horta}, D. and {J{\"o}nsson}, Henrik and {Masseron}, T. and {Prantzos}, N. and {Schiavon}, R.~P. and {Schultheis}, M. and {Zoccali}, M.},
        title = "{Chemical Radial Gradients for the Bulge Bar Stellar Populations from the APOGEE Survey}",
      journal = {\apj},
         year = 2026,
        month = apr,
       volume = {1001},
       number = {1},
          eid = {79},
        pages = {79},
          doi = {10.3847/1538-4357/ae505e},
archivePrefix = {arXiv},
       eprint = {2603.12070},
 primaryClass = {astro-ph.GA},
       adsurl = {https://ui.adsabs.harvard.edu/abs/2026ApJ..1001...79S}
}

@ARTICLE{hayden_chemical_2015,
       author = {{Hayden}, Michael R. and {Bovy}, Jo and {Holtzman}, Jon A. and {Nidever}, David L. and {Bird}, Jonathan C. and {Weinberg}, David H. and {Andrews}, Brett H. and {Majewski}, Steven R. and {Allende Prieto}, Carlos and {Anders}, Friedrich and {Beers}, Timothy C. and {Bizyaev}, Dmitry and {Chiappini}, Cristina and {Cunha}, Katia and {Frinchaboy}, Peter and {Garc{\'\i}a-Her{\'n}andez}, D.~A. and {Garc{\'\i}a P{\'e}rez}, Ana E. and {Girardi}, L{\'e}o and {Harding}, Paul and {Hearty}, Fred R. and {Johnson}, Jennifer A. and {M{\'e}sz{\'a}ros}, Szabolcs and {Minchev}, Ivan and {O'Connell}, Robert and {Pan}, Kaike and {Robin}, Annie C. and {Schiavon}, Ricardo P. and {Schneider}, Donald P. and {Schultheis}, Mathias and {Shetrone}, Matthew and {Skrutskie}, Michael and {Steinmetz}, Matthias and {Smith}, Verne and {Wilson}, John C. and {Zamora}, Olga and {Zasowski}, Gail},
        title = "{Chemical Cartography with APOGEE: Metallicity Distribution Functions and the Chemical Structure of the Milky Way Disk}",
      journal = {\apj},
         year = 2015,
        month = aug,
       volume = {808},
       number = {2},
          eid = {132},
        pages = {132},
          doi = {10.1088/0004-637X/808/2/132},
archivePrefix = {arXiv},
       eprint = {1503.02110},
 primaryClass = {astro-ph.GA},
       adsurl = {https://ui.adsabs.harvard.edu/abs/2015ApJ...808..132H}
}

@ARTICLE{griffith_residual_2022,
       author = {{Griffith}, Emily J. and {Weinberg}, David H. and {Buder}, Sven and {Johnson}, Jennifer A. and {Johnson}, James W. and {Vincenzo}, Fiorenzo},
        title = "{Residual Abundances in GALAH DR3: Implications for Nucleosynthesis and Identification of Unique Stellar Populations}",
      journal = {\apj},
         year = 2022,
        month = may,
       volume = {931},
       number = {1},
          eid = {23},
        pages = {23},
          doi = {10.3847/1538-4357/ac5826},
archivePrefix = {arXiv},
       eprint = {2110.06240},
 primaryClass = {astro-ph.GA},
       adsurl = {https://ui.adsabs.harvard.edu/abs/2022ApJ...931...23G}
}

@ARTICLE{mikolaitis_non-lte_2026,
       author = {{Mikolaitis}, {\v{S}}. and {Tautvai{\v{s}}ien{\.{e}}}, G. and {Pak{\v{s}}tien{\.{e}}}, E. and {Drazdauskas}, A. and {Bagdonas}, V. and {Viscasillas V{\'a}zquez}, C. and {Ambrosch}, M. and {Chorniy}, Y. and {Minkevi{\v{c}}i{\={u}}t{\.{e}}}, R. and {Stonkut{\.{e}}}, E. and {Bale}, B. and {{\'C}urjuri{\'c}}, B. and {Sharma}, A. and {Diktanait{\.{e}}}, K.},
        title = "{The non-LTE abundances of magnesium and yttrium and asteroseismic ages for the chemical clock calibration}",
      journal = {arXiv e-prints},
         year = 2026,
        month = jul,
          eid = {arXiv:2607.15017},
        pages = {arXiv:2607.15017},
          doi = {10.48550/arXiv.2607.15017},
archivePrefix = {arXiv},
       eprint = {2607.15017},
 primaryClass = {astro-ph.SR},
       adsurl = {https://ui.adsabs.harvard.edu/abs/2026arXiv260715017M}
}

@ARTICLE{queiroz_starhorse_2020,
       author = {{Queiroz}, A.~B.~A. and {Anders}, F. and {Chiappini}, C. and {Khalatyan}, A. and {Santiago}, B.~X. and {Steinmetz}, M. and {Valentini}, M. and {Miglio}, A. and {Bossini}, D. and {Barbuy}, B. and {Minchev}, I. and {Minniti}, D. and {Garc{\'\i}a Hern{\'a}ndez}, D.~A. and {Schultheis}, M. and {Beaton}, R.~L. and {Beers}, T.~C. and {Bizyaev}, D. and {Brownstein}, J.~R. and {Cunha}, K. and {Fern{\'a}ndez-Trincado}, J.~G. and {Frinchaboy}, P.~M. and {Lane}, R.~R. and {Majewski}, S.~R. and {Nataf}, D. and {Nitschelm}, C. and {Pan}, K. and {Roman-Lopes}, A. and {Sobeck}, J.~S. and {Stringfellow}, G. and {Zamora}, O.},
        title = "{From the bulge to the outer disc: StarHorse stellar parameters, distances, and extinctions for stars in APOGEE DR16 and other spectroscopic surveys}",
      journal = {\aap},
         year = 2020,
        month = jun,
       volume = {638},
          eid = {A76},
        pages = {A76},
          doi = {10.1051/0004-6361/201937364},
archivePrefix = {arXiv},
       eprint = {1912.09778},
 primaryClass = {astro-ph.GA},
       adsurl = {https://ui.adsabs.harvard.edu/abs/2020A&A...638A..76Q}
}

@ARTICLE{helmi_merger_2018,
       author = {{Helmi}, Amina and {Babusiaux}, Carine and {Koppelman}, Helmer H. and {Massari}, Davide and {Veljanoski}, Jovan and {Brown}, Anthony G.~A.},
        title = "{The merger that led to the formation of the Milky Way's inner stellar halo and thick disk}",
      journal = {\nat},
         year = 2018,
        month = oct,
       volume = {563},
       number = {7729},
        pages = {85-88},
          doi = {10.1038/s41586-018-0625-x},
archivePrefix = {arXiv},
       eprint = {1806.06038},
 primaryClass = {astro-ph.GA},
       adsurl = {https://ui.adsabs.harvard.edu/abs/2018Natur.563...85H}
}

@ARTICLE{belokurov_co-formation_2018,
       author = {{Belokurov}, V. and {Erkal}, D. and {Evans}, N.~W. and {Koposov}, S.~E. and {Deason}, A.~J.},
        title = "{Co-formation of the disc and the stellar halo}",
      journal = {\mnras},
         year = 2018,
        month = jul,
       volume = {478},
       number = {1},
        pages = {611-619},
          doi = {10.1093/mnras/sty982},
archivePrefix = {arXiv},
       eprint = {1802.03414},
 primaryClass = {astro-ph.GA},
       adsurl = {https://ui.adsabs.harvard.edu/abs/2018MNRAS.478..611B}
}

@ARTICLE{feuillet_gse_2020,
       author = {{Feuillet}, Diane K. and {Feltzing}, Sofia and {Sahlholdt}, Christian L. and {Casagrande}, Luca},
        title = "{The SkyMapper-Gaia RVS view of the Gaia-Enceladus-Sausage - an investigation of the metallicity and mass of the Milky Way's last major merger}",
      journal = {\mnras},
         year = 2020,
        month = sep,
       volume = {497},
       number = {1},
        pages = {109-124},
          doi = {10.1093/mnras/staa1888},
archivePrefix = {arXiv},
       eprint = {2003.11039},
 primaryClass = {astro-ph.GA},
       adsurl = {https://ui.adsabs.harvard.edu/abs/2020MNRAS.497..109F}
}

@ARTICLE{lane_kinematic_2022,
       author = {{Lane}, James M.~M. and {Bovy}, Jo and {Mackereth}, J. Ted},
        title = "{The kinematic properties of Milky Way stellar halo populations}",
      journal = {\mnras},
         year = 2022,
        month = mar,
       volume = {510},
       number = {4},
        pages = {5119-5141},
          doi = {10.1093/mnras/stab3755},
archivePrefix = {arXiv},
       eprint = {2106.09699},
 primaryClass = {astro-ph.GA},
       adsurl = {https://ui.adsabs.harvard.edu/abs/2022MNRAS.510.5119L}
}

@ARTICLE{feltzing_metal-weak_2023,
       author = {{Feltzing}, Sofia and {Feuillet}, Diane},
        title = "{The Metal-weak Milky Way Stellar Disk Hidden in the Gaia-Sausage-Enceladus Debris: The APOGEE DR17 View}",
      journal = {\apj},
         year = 2023,
        month = aug,
       volume = {953},
       number = {2},
          eid = {143},
        pages = {143},
          doi = {10.3847/1538-4357/ace185},
archivePrefix = {arXiv},
       eprint = {2303.00016},
 primaryClass = {astro-ph.GA},
       adsurl = {https://ui.adsabs.harvard.edu/abs/2023ApJ...953..143F}
}

@ARTICLE{larson_models_1976,
       author = {{Larson}, R.~B.},
        title = "{Models for the formation of disc galaxies.}",
      journal = {\mnras},
         year = 1976,
        month = jul,
       volume = {176},
        pages = {31-52},
          doi = {10.1093/mnras/176.1.31},
       adsurl = {https://ui.adsabs.harvard.edu/abs/1976MNRAS.176...31L}
}

@ARTICLE{warfield_apo-k2_2024,
       author = {{Warfield}, Jack T. and {Zinn}, Joel C. and {Schonhut-Stasik}, Jessica and {Johnson}, James W. and {Pinsonneault}, Marc H. and {Johnson}, Jennifer A. and {Stello}, Dennis and {Beaton}, Rachael L. and {Elsworth}, Yvonne and {Garc{\'\i}a}, Rafael A. and {Mathur}, Savita and {Mosser}, Beno{\^\i}t and {Serenelli}, Aldo and {Tayar}, Jamie},
        title = "{The APO-K2 Catalog. II. Accurate Stellar Ages for Red Giant Branch Stars across the Milky Way}",
      journal = {\aj},
         year = 2024,
        month = may,
       volume = {167},
       number = {5},
          eid = {208},
        pages = {208},
          doi = {10.3847/1538-3881/ad33bb},
archivePrefix = {arXiv},
       eprint = {2403.16250},
 primaryClass = {astro-ph.GA},
       adsurl = {https://ui.adsabs.harvard.edu/abs/2024AJ....167..208W}
}

@ARTICLE{schonhut-stasik_apo-k2_2024,
       author = {{Schonhut-Stasik}, Jessica and {Zinn}, Joel C. and {Stassun}, Keivan G. and {Pinsonneault}, Marc and {Johnson}, Jennifer A. and {Warfield}, Jack T. and {Stello}, Dennis and {Elsworth}, Yvonne and {Garc{\'\i}a}, Rafael A. and {Mathur}, Savita and {Mosser}, Benoit and {Hon}, Marc and {Tayar}, Jamie and {Stringfellow}, Guy S. and {Beaton}, Rachael L. and {J{\"o}nsson}, Henrik and {Minniti}, Dante},
        title = "{The APO-K2 Catalog. I.  7500 Red Giants with Fundamental Stellar Parameters from APOGEE DR17 Spectroscopy and K2-GAP Asteroseismology}",
      journal = {\aj},
         year = 2024,
        month = feb,
       volume = {167},
       number = {2},
          eid = {50},
        pages = {50},
          doi = {10.3847/1538-3881/ad0b13},
archivePrefix = {arXiv},
       eprint = {2304.10654},
 primaryClass = {astro-ph.SR},
       adsurl = {https://ui.adsabs.harvard.edu/abs/2024AJ....167...50S}
}

@ARTICLE{eilers_circular_2019,
       author = {{Eilers}, Anna-Christina and {Hogg}, David W. and {Rix}, Hans-Walter and {Ness}, Melissa K.},
        title = "{The Circular Velocity Curve of the Milky Way from 5 to 25 kpc}",
      journal = {\apj},
         year = 2019,
        month = jan,
       volume = {871},
       number = {1},
          eid = {120},
        pages = {120},
          doi = {10.3847/1538-4357/aaf648},
archivePrefix = {arXiv},
       eprint = {1810.09466},
 primaryClass = {astro-ph.GA},
       adsurl = {https://ui.adsabs.harvard.edu/abs/2019ApJ...871..120E}
}

@ARTICLE{hayden_galah_2022,
       author = {{Hayden}, Michael R. and {Sharma}, Sanjib and {Bland-Hawthorn}, Joss and {Spina}, Lorenzo and {Buder}, Sven and {Ciuc{\u{a}}}, Ioana and {Asplund}, Martin and {Casey}, Andrew R. and {De Silva}, Gayandhi M. and {D'Orazi}, Valentina and {Freeman}, Ken C. and {Kos}, Janez and {Lewis}, Geraint F. and {Lin}, Jane and {Lind}, Karin and {Martell}, Sarah L. and {Schlesinger}, Katharine J. and {Simpson}, Jeffrey D. and {Zucker}, Daniel B. and {Zwitter}, Toma{\v{z}} and {Chen}, Boquan and {{\v{C}}otar}, Klemen and {Feuillet}, Diane and {Horner}, Jonti and {Joyce}, Meridith and {Nordlander}, Thomas and {Stello}, Dennis and {Tepper-Garcia}, Thor and {Ting}, Yuan-sen and {Wang}, Purmortal and {Wittenmyer}, Rob and {Wyse}, Rosemary},
        title = "{The GALAH survey: chemical clocks}",
      journal = {\mnras},
         year = 2022,
        month = dec,
       volume = {517},
       number = {4},
        pages = {5325-5339},
          doi = {10.1093/mnras/stac2787},
archivePrefix = {arXiv},
       eprint = {2011.13745},
 primaryClass = {astro-ph.GA},
       adsurl = {https://ui.adsabs.harvard.edu/abs/2022MNRAS.517.5325H}
}

@ARTICLE{horta_neutron-capture_2022,
       author = {{Horta}, Danny and {Ness}, Melissa K. and {Rybizki}, Jan and {Schiavon}, Ricardo P. and {Buder}, Sven},
        title = "{Neutron-capture elements record the ordered chemical evolution of the disc over time}",
      journal = {\mnras},
         year = 2022,
        month = jul,
       volume = {513},
       number = {4},
        pages = {5477-5504},
          doi = {10.1093/mnras/stac953},
archivePrefix = {arXiv},
       eprint = {2111.01809},
 primaryClass = {astro-ph.GA},
       adsurl = {https://ui.adsabs.harvard.edu/abs/2022MNRAS.513.5477H}
}

@ARTICLE{griffith_many_2025,
       author = {{Griffith}, Emily J. and {Hogg}, David W. and {Hasselquist}, Sten and {Johnson}, James W. and {Price-Whelan}, Adrian and {Sit}, Tawny and {Stone-Martinez}, Alexander and {Weinberg}, David H.},
        title = "{Many Elements Matter: Detailed Abundance Patterns Reveal Star Formation and Enrichment Differences among Milky Way Structural Components}",
      journal = {\aj},
         year = 2025,
        month = may,
       volume = {169},
       number = {5},
          eid = {280},
        pages = {280},
          doi = {10.3847/1538-3881/adc07f},
archivePrefix = {arXiv},
       eprint = {2410.22121},
 primaryClass = {astro-ph.GA},
       adsurl = {https://ui.adsabs.harvard.edu/abs/2025AJ....169..280G}
}

@ARTICLE{buder_galah_2021,
       author = {{Buder}, Sven and {Sharma}, Sanjib and {Kos}, Janez and {Amarsi}, Anish M. and {Nordlander}, Thomas and {Lind}, Karin and {Martell}, Sarah L. and {Asplund}, Martin and {Bland-Hawthorn}, Joss and {Casey}, Andrew R. and {de Silva}, Gayandhi M. and {D'Orazi}, Valentina and {Freeman}, Ken C. and {Hayden}, Michael R. and {Lewis}, Geraint F. and {Lin}, Jane and {Schlesinger}, Katharine J. and {Simpson}, Jeffrey D. and {Stello}, Dennis and {Zucker}, Daniel B. and {Zwitter}, Toma{\v{z}} and {Beeson}, Kevin L. and {Buck}, Tobias and {Casagrande}, Luca and {Clark}, Jake T. and {{\v{C}}otar}, Klemen and {da Costa}, Gary S. and {de Grijs}, Richard and {Feuillet}, Diane and {Horner}, Jonathan and {Kafle}, Prajwal R. and {Khanna}, Shourya and {Kobayashi}, Chiaki and {Liu}, Fan and {Montet}, Benjamin T. and {Nandakumar}, Govind and {Nataf}, David M. and {Ness}, Melissa K. and {Spina}, Lorenzo and {Tepper-Garc{\'\i}a}, Thor and {Ting}, Yuan-Sen and {Traven}, Gregor and {Vogrin{\v{c}}i{\v{c}}}, Rok and {Wittenmyer}, Robert A. and {Wyse}, Rosemary F.~G. and {{\v{Z}}erjal}, Maru{\v{s}}a and {Galah Collaboration}},
        title = "{The GALAH+ survey: Third data release}",
      journal = {\mnras},
         year = 2021,
        month = sep,
       volume = {506},
       number = {1},
        pages = {150-201},
          doi = {10.1093/mnras/stab1242},
archivePrefix = {arXiv},
       eprint = {2011.02505},
 primaryClass = {astro-ph.GA},
       adsurl = {https://ui.adsabs.harvard.edu/abs/2021MNRAS.506..150B}
}

@ARTICLE{myeong_milky_2022,
       author = {{Myeong}, G.~C. and {Belokurov}, Vasily and {Aguado}, David S. and {Evans}, N. Wyn and {Caldwell}, Nelson and {Bradley}, James},
        title = "{Milky Way's Eccentric Constituents with Gaia, APOGEE, and GALAH}",
      journal = {\apj},
         year = 2022,
        month = oct,
       volume = {938},
       number = {1},
          eid = {21},
        pages = {21},
          doi = {10.3847/1538-4357/ac8d68},
archivePrefix = {arXiv},
       eprint = {2206.07744},
 primaryClass = {astro-ph.GA},
       adsurl = {https://ui.adsabs.harvard.edu/abs/2022ApJ...938...21M}
}

@ARTICLE{gaia_dr2_2018,
       author = {{Gaia Collaboration} and {Brown}, A.~G.~A. and {Vallenari}, A. and {Prusti}, T. and {de Bruijne}, J.~H.~J. and {Babusiaux}, C. and {Bailer-Jones}, C.~A.~L. and {Biermann}, M. and {Evans}, D.~W. and {Eyer}, L. and {Jansen}, F. and {Jordi}, C. and {Klioner}, S.~A. and {Lammers}, U. and {Lindegren}, L. and {Luri}, X. and {Mignard}, F. and {Panem}, C. and {Pourbaix}, D. and {Randich}, S. and {Sartoretti}, P. and {Siddiqui}, H.~I. and {Soubiran}, C. and {van Leeuwen}, F. and {Walton}, N.~A. and {Arenou}, F. and {Bastian}, U. and {Cropper}, M. and {Drimmel}, R. and {Katz}, D. and {Lattanzi}, M.~G. and {Bakker}, J. and {Cacciari}, C. and {Casta{\~n}eda}, J. and {Chaoul}, L. and {Cheek}, N. and {De Angeli}, F. and {Fabricius}, C. and {Guerra}, R. and {Holl}, B. and {Masana}, E. and {Messineo}, R. and {Mowlavi}, N. and {Nienartowicz}, K. and {Panuzzo}, P. and {Portell}, J. and {Riello}, M. and {Seabroke}, G.~M. and {Tanga}, P. and {Th{\'e}venin}, F. and {Gracia-Abril}, G. and {Comoretto}, G. and {Garcia-Reinaldos}, M. and {Teyssier}, D. and {Altmann}, M. and {Andrae}, R. and {Audard}, M. and {Bellas-Velidis}, I. and {Benson}, K. and {Berthier}, J. and {Blomme}, R. and {Burgess}, P. and {Busso}, G. and {Carry}, B. and {Cellino}, A. and {Clementini}, G. and {Clotet}, M. and {Creevey}, O. and {Davidson}, M. and {De Ridder}, J. and {Delchambre}, L. and {Dell'Oro}, A. and {Ducourant}, C. and {Fern{\'a}ndez-Hern{\'a}ndez}, J. and {Fouesneau}, M. and {Fr{\'e}mat}, Y. and {Galluccio}, L. and {Garc{\'\i}a-Torres}, M. and {Gonz{\'a}lez-N{\'u}{\~n}ez}, J. and {Gonz{\'a}lez-Vidal}, J.~J. and {Gosset}, E. and {Guy}, L.~P. and {Halbwachs}, J.-L. and {Hambly}, N.~C. and {Harrison}, D.~L. and {Hern{\'a}ndez}, J. and {Hestroffer}, D. and {Hodgkin}, S.~T. and {Hutton}, A. and {Jasniewicz}, G. and {Jean-Antoine-Piccolo}, A. and {Jordan}, S. and {Korn}, A.~J. and {Krone-Martins}, A. and {Lanzafame}, A.~C. and {Lebzelter}, T. and {L{\"o}ffler}, W. and {Manteiga}, M. and {Marrese}, P.~M. and {Mart{\'\i}n-Fleitas}, J.~M. and {Moitinho}, A. and {Mora}, A. and {Muinonen}, K. and {Osinde}, J. and {Pancino}, E. and {Pauwels}, T. and {Petit}, J.-M. and {Recio-Blanco}, A. and {Richards}, P.~J. and {Rimoldini}, L. and {Robin}, A.~C. and {Sarro}, L.~M. and {Siopis}, C. and {Smith}, M. and {Sozzetti}, A. and {S{\"u}veges}, M. and {Torra}, J. and {van Reeven}, W. and {Abbas}, U. and {Abreu Aramburu}, A. and {Accart}, S. and {Aerts}, C. and {Altavilla}, G. and {{\'A}lvarez}, M.~A. and {Alvarez}, R. and {Alves}, J. and {Anderson}, R.~I. and {Andrei}, A.~H. and {Anglada Varela}, E. and {Antiche}, E. and {Antoja}, T. and {Arcay}, B. and {Astraatmadja}, T.~L. and {Bach}, N. and {Baker}, S.~G. and {Balaguer-N{\'u}{\~n}ez}, L. and {Balm}, P. and {Barache}, C. and {Barata}, C. and {Barbato}, D. and {Barblan}, F. and {Barklem}, P.~S. and {Barrado}, D. and {Barros}, M. and {Barstow}, M.~A. and {Bartholom{\'e} Mu{\~n}oz}, S. and {Bassilana}, J.-L. and {Becciani}, U. and {Bellazzini}, M. and {Berihuete}, A. and {Bertone}, S. and {Bianchi}, L. and {Bienaym{\'e}}, O. and {Blanco-Cuaresma}, S. and {Boch}, T. and {Boeche}, C. and {Bombrun}, A. and {Borrachero}, R. and {Bossini}, D. and {Bouquillon}, S. and {Bourda}, G. and {Bragaglia}, A. and {Bramante}, L. and {Breddels}, M.~A. and {Bressan}, A. and {Brouillet}, N. and {Br{\"u}semeister}, T. and {Brugaletta}, E. and {Bucciarelli}, B. and {Burlacu}, A. and {Busonero}, D. and {Butkevich}, A.~G. and {Buzzi}, R. and {Caffau}, E. and {Cancelliere}, R. and {Cannizzaro}, G. and {Cantat-Gaudin}, T. and {Carballo}, R. and {Carlucci}, T. and {Carrasco}, J.~M. and {Casamiquela}, L. and {Castellani}, M. and {Castro-Ginard}, A. and {Charlot}, P. and {Chemin}, L. and {Chiavassa}, A. and {Cocozza}, G. and {Costigan}, G. and {Cowell}, S. and {Crifo}, F. and {Crosta}, M. and {Crowley}, C. and {Cuypers}, J. and {Dafonte}, C. and {Damerdji}, Y. and {Dapergolas}, A. and {David}, P. and {David}, M. and {de Laverny}, P. and {De Luise}, F.},
        title = "{Gaia Data Release 2. Summary of the contents and survey properties}",
      journal = {\aap},
         year = 2018,
        month = aug,
       volume = {616},
          eid = {A1},
        pages = {A1},
          doi = {10.1051/0004-6361/201833051},
archivePrefix = {arXiv},
       eprint = {1804.09365},
 primaryClass = {astro-ph.GA},
       adsurl = {https://ui.adsabs.harvard.edu/abs/2018A&A...616A...1G}
}

@ARTICLE{cantat-gaudin_painting_2020,
       author = {{Cantat-Gaudin}, T. and {Anders}, F. and {Castro-Ginard}, A. and {Jordi}, C. and {Romero-G{\'o}mez}, M. and {Soubiran}, C. and {Casamiquela}, L. and {Tarricq}, Y. and {Moitinho}, A. and {Vallenari}, A. and {Bragaglia}, A. and {Krone-Martins}, A. and {Kounkel}, M.},
        title = "{Painting a portrait of the Galactic disc with its stellar clusters}",
      journal = {\aap},
         year = 2020,
        month = aug,
       volume = {640},
          eid = {A1},
        pages = {A1},
          doi = {10.1051/0004-6361/202038192},
archivePrefix = {arXiv},
       eprint = {2004.07274},
 primaryClass = {astro-ph.GA},
       adsurl = {https://ui.adsabs.harvard.edu/abs/2020A&A...640A...1C}
}

@ARTICLE{xiang_time_2022,
       author = {{Xiang}, Maosheng and {Rix}, Hans-Walter},
        title = "{A time-resolved picture of our Milky Way's early formation history}",
      journal = {\nat},
         year = 2022,
        month = mar,
       volume = {603},
       number = {7902},
        pages = {599-603},
          doi = {10.1038/s41586-022-04496-5},
archivePrefix = {arXiv},
       eprint = {2203.12110},
 primaryClass = {astro-ph.GA},
       adsurl = {https://ui.adsabs.harvard.edu/abs/2022Natur.603..599X}
}

@article{kollmeier_sloan_2026,
	title = {Sloan {Digital} {Sky} {Survey}. {V}. {Pioneering} {Panoptic} {Spectroscopy}},
	volume = {171},
	issn = {0004-6256},
	url = {https://ui.adsabs.harvard.edu/abs/2026AJ....171...52K},
	doi = {10.3847/1538-3881/ae0576},
	urldate = {2026-06-24},
	journal = {\aj},
	publisher = {IOP},
	author = {Kollmeier, Juna A. and Rix, Hans-Walter and Aerts, Conny and Aird, James and Vera Alfaro, Pablo and Almeida, Andrés and Anderson, Scott F. and Arseneau, Stefan M. and Assef, Roberto J. and Aviram, Shir and Aydar, Catarina and Badenes, Carles and Bandyopadhyay, Avrajit and Barger, Kat and Barkhouser, Robert H. and Bauer, Franz E. and Behmard, Aida and Bender, Chad and Besser, Felipe and Bhattarai, Binod and Bilgi, Pavaman and Bird, Jonathan and Bizyaev, Dmitry and Blanc, Guillermo A. and Blanton, Michael R. and Bochanski, John and Bovy, Jo and Brandon, Christopher and Brandt, William Nielsen and Brownstein, Joel R. and Buchner, Johannes and Burchett, Joseph N. and Carlberg, Joleen and Casey, Andrew R. and Castaneda-Carlos, Lesly and Chakraborty, Priyanka and Chanamé, Julio and Chandra, Vedant and Cherinka, Brian and Chilingarian, Igor and Comparat, Johan and Cosens, Maren and Covey, Kevin and Crane, Jeffrey D. and Crumpler, Nicole R. and Cruz-Gonzalez, Irene and Cunha, Katia and Cunningham, Tim and Dai, Xinyu and Darling, Jeremy and Davidson, Jr., James W. and Davis, Megan C. and De Lee, Nathan and Deacon, Niall and Méndez Delgado, José Eduardo and Demasi, Sebastian and Demianenko, Mariia and Derwent, Mark and D'Onghia, Elena and Di Mille, Francesco and Dias, Bruno and Donor, John and Dow, Peter N. and Drory, Niv and Dwelly, Tom and Egorov, Oleg and Egorova, Evgeniya and El-Badry, Kareem and Engelman, Mike and Eracleous, Mike and Fan, Xiaohui and Farr, Emily and Fries, Logan and Frinchaboy, Peter and Froning, Cynthia S. and Gänsicke, Boris T. and García, Pablo and Gelfand, Joseph and Gentile Fusillo, Nicola Pietro and Glover, Simon and Grabowski, Katie and Grebel, Eva K. and Green, Paul J. and Grier, Catherine and Gupta, Pramod and Gray, Aidan C. and Häberle, Maximilian and Hall, Patrick B. and Hammond, Randolph P. and Hawkins, Keith and Harding, Albert C. and Hegedűs, Viola and Herbst, Tom and Hermes, J. J. and Rodríguez Hidalgo, Paola and Hilder, Thomas and Hogg, David W. and Holtzman, Jon A. and Horta, Danny and Huang, Yang and Hwang, Hsiang-Chih and Ibarra-Medel, Hector Javier and Imig, Julie and Inight, Keith and Jana, Arghajit and Ji, Alexander P. and Jiménez-Arranz, Óscar and Jofre, Paula and Johns, Matt and Johnson, Jennifer and Johnson, James W. and Johnston, Evelyn J. and Jones, Amy M. and Katkov, Ivan and Knapp, Gillian R. and Koekemoer, Anton M. and Kounkel, Marina and Kreckel, Kathryn and Krishnarao, Dhanesh and Krumpe, Mirko and Kumari, Nimisha and Kupfer, Thomas and Lacerna, Ivan and Laporte, Chervin and Lepine, Sebastien and Li, Jing and Liu, Xin and Loebman, Sarah and Long, Knox and Roman-Lopes, Alexandre and Lu, Yuxi and Majewski, Steven Raymond and Maoz, Dan and McKinnon, Kevin A. and Medan, Ilija and Merloni, Andrea and Minniti, Dante and Morrison, Sean and Myers, Natalie and Mészáros, Szabolcs and Nandra, Kirpal and Nayak, Prasanta K. and Ness, Melissa K. and Nidever, David L. and O'Brien, Thomas and Oeur, Micah and Oravetz, Audrey and Oravetz, Daniel and Otto, Jonah and Pallathadka, Gautham Adamane and Palunas, Povilas and Pan, Kaike and Pappalardo, Daniel and Pandey, Rakesh and Negrete Peñaloza, Castalia Alenka and Pinsonneault, Marc H. and Pogge, Richard W. and Taghizadeh Popp, Manuchehr and Price-Whelan, Adrian M. and Pulatova, Nadiia and Qiu, Dan and Ramirez, Solange and Rankine, Amy and Ricci, Claudio and Runnoe, Jessie C. and Sanchez, Sebastian and Salvato, Mara and Sarbadhicary, Sumit K. and Sattler, Natascha and Saydjari, Andrew K. and Sayres, Conor and Schinnerer, Eva and Schlaufman, Kevin C. and Schneider, Donald P. and Schreiber, Matthias R. and Schwope, Axel and Serna, Javier and Shen, Yue and Sifón, Cristóbal and Singh, Amrita and Sinha, Amaya and Smee, Stephen and Song, Ying-Yi and Souto, Diogo and Stassun, Keivan G. and Steinmetz, Matthias and Stone-Martinez, Alexander and Stringfellow, Guy and Stutz, Amelia and Sánchez-Gallego, José and Tan, Jonathan C. and Tayar, Jamie and Thai, Riley and Thakar, Ani and Ting, Yuan-Sen and Tkachenko, Andrew and Tovmassian, Gagik and Trakhtenbrot, Benny and Fernández-Trincado, José G. and Troup, Nicholas},
	month = jan,
	year = {2026},
	note = {ADS Bibcode: 2026AJ....171...52K},
	pages = {52},
}

@article{viscasillas_vazquez_gaia-eso_2022,
	title = {The {Gaia}-{ESO} survey: {Age}-chemical-clock relations spatially resolved in the {Galactic} disc},
	volume = {660},
	issn = {0004-6361},
	shorttitle = {The {Gaia}-{ESO} survey},
	url = {https://ui.adsabs.harvard.edu/abs/2022A&A...660A.135V},
	doi = {10.1051/0004-6361/202142937},
	urldate = {2026-06-23},
	journal = {\aap},
	publisher = {EDP},
	author = {Viscasillas Vázquez, C. and Magrini, L. and Casali, G. and Tautvaišienė, G. and Spina, L. and Van der Swaelmen, M. and Randich, S. and Bensby, T. and Bragaglia, A. and Friel, E. and Feltzing, S. and Sacco, G. G. and Turchi, A. and Jiménez-Esteban, F. and D'Orazi, V. and Delgado-Mena, E. and Mikolaitis, Š. and Drazdauskas, A. and Minkevičiūtė, R. and Stonkutė, E. and Bagdonas, V. and Montes, D. and Guiglion, G. and Baratella, M. and Tabernero, H. M. and Gilmore, G. and Alfaro, E. and Francois, P. and Korn, A. and Smiljanic, R. and Bergemann, M. and Franciosini, E. and Gonneau, A. and Hourihane, A. and Worley, C. C. and Zaggia, S.},
	month = apr,
	year = {2022},
	note = {ADS Bibcode: 2022A\&A...660A.135V},
	pages = {A135},
}

@article{frankel_measuring_2018,
	title = {Measuring {Radial} {Orbit} {Migration} in the {Galactic} {Disk}},
	volume = {865},
	issn = {0004-637X},
	url = {https://ui.adsabs.harvard.edu/abs/2018ApJ...865...96F},
	doi = {10.3847/1538-4357/aadba5},
	urldate = {2026-06-23},
	journal = {\apj},
	publisher = {IOP},
	author = {Frankel, Neige and Rix, Hans-Walter and Ting, Yuan-Sen and Ness, Melissa and Hogg, David W.},
	month = oct,
	year = {2018},
	note = {ADS Bibcode: 2018ApJ...865...96F},
	pages = {96},
}

@article{lian_quantifying_2022,
	title = {Quantifying radial migration in the {Milky} {Way}: inefficient over short time-scales but essential to the very outer disc beyond  15 kpc},
	volume = {511},
	issn = {0035-8711},
	shorttitle = {Quantifying radial migration in the {Milky} {Way}},
	url = {https://ui.adsabs.harvard.edu/abs/2022MNRAS.511.5639L},
	doi = {10.1093/mnras/stac479},
	urldate = {2026-06-23},
	journal = {\mnras},
	publisher = {OUP},
	author = {Lian, Jianhui and Zasowski, Gail and Hasselquist, Sten and Holtzman, Jon A. and Boardman, Nicholas and Cunha, Katia and Fernández-Trincado, José G. and Frinchaboy, Peter M. and Garcia-Hernandez, D. A. and Nitschelm, Christian and Lane, Richard R. and Thomas, Daniel and Zhang, Kai},
	month = apr,
	year = {2022},
	note = {ADS Bibcode: 2022MNRAS.511.5639L},
	pages = {5639--5655},
}

@article{lehmann_probing_2024,
	title = {Probing the strength of radial migration via churning by using metal-rich red giant stars from {APOGEE}},
	volume = {533},
	issn = {0035-8711},
	url = {https://ui.adsabs.harvard.edu/abs/2024MNRAS.533..538L},
	doi = {10.1093/mnras/stae1736},
	urldate = {2026-06-23},
	journal = {\mnras},
	publisher = {OUP},
	author = {Lehmann, Christian and Feltzing, Sofia and Feuillet, Diane and Kordopatis, Georges},
	month = sep,
	year = {2024},
	note = {ADS Bibcode: 2024MNRAS.533..538L},
	pages = {538--550},
}

@article{chen_open_2020,
	title = {Open clusters as tracers on radial migration of the galactic disc},
	volume = {495},
	issn = {0035-8711},
	url = {https://doi.org/10.1093/mnras/staa1079},
	doi = {10.1093/mnras/staa1079},
	number = {3},
	urldate = {2026-06-23},
	journal = {\mnras},
	author = {Chen, Y Q and Zhao, G},
	month = jul,
	year = {2020},
	pages = {2673--2681},
}

@article{schonrich_chemical_2009,
	title = {Chemical evolution with radial mixing},
	volume = {396},
	issn = {0035-8711},
	url = {https://ui.adsabs.harvard.edu/abs/2009MNRAS.396..203S},
	doi = {10.1111/j.1365-2966.2009.14750.x},
	urldate = {2022-02-07},
	journal = {\mnras},
	author = {Schönrich, Ralph and Binney, James},
	month = jun,
	year = {2009},
	note = {ADS Bibcode: 2009MNRAS.396..203S},
	pages = {203--222},
}

@article{sellwood_radial_2002,
	title = {Radial mixing in galactic discs},
	volume = {336},
	issn = {0035-8711},
	url = {https://ui.adsabs.harvard.edu/abs/2002MNRAS.336..785S},
	doi = {10.1046/j.1365-8711.2002.05806.x},
	urldate = {2024-01-10},
	journal = {\mnras},
	author = {Sellwood, J. A. and Binney, J. J.},
	month = nov,
	year = {2002},
	note = {ADS Bibcode: 2002MNRAS.336..785S},
	pages = {785--796},
}

@article{lu_there_2024,
	title = {There is no place like home – finding birth radii of stars in the {Milky} {Way}},
	volume = {535},
	issn = {0035-8711},
	url = {https://doi.org/10.1093/mnras/stae2364},
	doi = {10.1093/mnras/stae2364},
	number = {1},
	urldate = {2025-09-05},
	journal = {\mnras},
	author = {Lu, Yuxi (Lucy) and Minchev, Ivan and Buck, Tobias and Khoperskov, Sergey and Steinmetz, Matthias and Libeskind, Noam and Cescutti, Gabriele and Freeman, Ken C and Ratcliffe, Bridget},
	month = nov,
	year = {2024},
	pages = {392--405},
}

@article{ratcliffe_unveiling_2023,
	title = {Unveiling the time evolution of chemical abundances across the {Milky} {Way} disc with {APOGEE}},
	volume = {525},
	issn = {0035-8711},
	url = {https://ui.adsabs.harvard.edu/abs/2023MNRAS.525.2208R},
	doi = {10.1093/mnras/stad1573},
	urldate = {2026-06-23},
	journal = {\mnras},
	publisher = {OUP},
	author = {Ratcliffe, Bridget and Minchev, Ivan and Anders, Friedrich and Khoperskov, Sergey and Guiglion, Guillaume and Buck, Tobias and Cunha, Katia and Queiroz, Anna and Nitschelm, Christian and Meszaros, Szabolcs and Steinmetz, Matthias and de Jong, Roelof S. and Nepal, Samir and Lane, Richard R. and Sobeck, Jennifer},
	month = oct,
	year = {2023},
	note = {ADS Bibcode: 2023MNRAS.525.2208R},
	pages = {2208--2228},
}

@article{titarenko_ambre_2019,
	title = {The {AMBRE} {Project}: [{Y}/{Mg}] stellar dating calibration with {Gaia}},
	volume = {622},
	issn = {0004-6361},
	shorttitle = {The {AMBRE} {Project}},
	url = {https://ui.adsabs.harvard.edu/abs/2019A&A...622A..59T},
	doi = {10.1051/0004-6361/201833721},
	urldate = {2026-06-23},
	journal = {\aap},
	publisher = {EDP},
	author = {Titarenko, A. and Recio-Blanco, A. and de Laverny, P. and Hayden, M. and Guiglion, G.},
	month = feb,
	year = {2019},
	note = {ADS Bibcode: 2019A\&A...622A..59T},
	pages = {A59},
}

@article{nissen_high-precision_2020,
	title = {High-precision abundances of elements in solar-type stars. {Evidence} of two distinct sequences in abundance-age relations},
	volume = {640},
	issn = {0004-6361},
	url = {https://ui.adsabs.harvard.edu/abs/2020A&A...640A..81N},
	doi = {10.1051/0004-6361/202038300},
	urldate = {2026-06-23},
	journal = {\aap},
	publisher = {EDP},
	author = {Nissen, P. E. and Christensen-Dalsgaard, J. and Mosumgaard, J. R. and Silva Aguirre, V. and Spitoni, E. and Verma, K.},
	month = aug,
	year = {2020},
	note = {ADS Bibcode: 2020A\&A...640A..81N},
	pages = {A81},
}

@article{tucci_maia_solar_2016,
	title = {The {Solar} {Twin} {Planet} {Search}. {III}. {The} [{Y}/{Mg}] clock: estimating stellar ages of solar-type stars},
	volume = {590},
	issn = {0004-6361},
	shorttitle = {The {Solar} {Twin} {Planet} {Search}. {III}. {The} [{Y}/{Mg}] clock},
	url = {https://ui.adsabs.harvard.edu/abs/2016A&A...590A..32T},
	doi = {10.1051/0004-6361/201527848},
	urldate = {2026-06-23},
	journal = {\aap},
	publisher = {EDP},
	author = {Tucci Maia, M. and Ramírez, I. and Meléndez, J. and Bedell, M. and Bean, J. L. and Asplund, M.},
	month = may,
	year = {2016},
	note = {ADS Bibcode: 2016A\&A...590A..32T},
	pages = {A32},
}

@misc{conroy_birth_2022,
	title = {Birth of the {Galactic} {Disk} {Revealed} by the {H3} {Survey}},
	url = {https://ui.adsabs.harvard.edu/abs/2022arXiv220402989C},
	doi = {10.48550/arXiv.2204.02989},
	urldate = {2025-01-17},
	author = {Conroy, Charlie and Weinberg, David H. and Naidu, Rohan P. and Buck, Tobias and Johnson, James W. and Cargile, Phillip and Bonaca, Ana and Caldwell, Nelson and Chandra, Vedant and Han, Jiwon Jesse and Johnson, Benjamin D. and Speagle, Joshua S. and Ting, Yuan-Sen and Woody, Turner and Zaritsky, Dennis},
	month = apr,
	year = {2022},
	note = {Publication Title: arXiv e-prints
ADS Bibcode: 2022arXiv220402989C},
}

@article{van_der_swaelmen_chemical_2013,
	title = {Chemical abundances in {LMC} stellar populations. {II}. {The} bar sample},
	volume = {560},
	issn = {0004-6361},
	url = {https://ui.adsabs.harvard.edu/abs/2013A&A...560A..44V},
	doi = {10.1051/0004-6361/201321109},
	urldate = {2026-06-09},
	journal = {\aap},
	publisher = {EDP},
	author = {Van der Swaelmen, M. and Hill, V. and Primas, F. and Cole, A. A.},
	month = dec,
	year = {2013},
	note = {ADS Bibcode: 2013A\&A...560A..44V},
	pages = {A44},
}

@article{johnson_stellar_2021,
	title = {Stellar migration and chemical enrichment in the milky way disc: a hybrid model},
	volume = {508},
	issn = {0035-8711},
	shorttitle = {Stellar migration and chemical enrichment in the milky way disc},
	url = {https://ui.adsabs.harvard.edu/abs/2021MNRAS.508.4484J},
	doi = {10.1093/mnras/stab2718},
	urldate = {2026-06-02},
	journal = {\mnras},
	publisher = {OUP},
	author = {Johnson, James W. and Weinberg, David H. and Vincenzo, Fiorenzo and Bird, Jonathan C. and Loebman, Sarah R. and Brooks, Alyson M. and Quinn, Thomas R. and Christensen, Charlotte R. and Griffith, Emily J.},
	month = dec,
	year = {2021},
	note = {ADS Bibcode: 2021MNRAS.508.4484J},
	pages = {4484--4511},
}

@article{spina_temporal_2018,
	title = {The temporal evolution of neutron-capture elements in the {Galactic} discs},
	volume = {474},
	issn = {0035-8711},
	url = {https://ui.adsabs.harvard.edu/abs/2018MNRAS.474.2580S},
	doi = {10.1093/mnras/stx2938},
	urldate = {2026-06-01},
	journal = {\mnras},
	publisher = {OUP},
	author = {Spina, Lorenzo and Meléndez, Jorge and Karakas, Amanda I. and dos Santos, Leonardo and Bedell, Megan and Asplund, Martin and Ramírez, Ivan and Yong, David and Alves-Brito, Alan and Bean, Jacob L. and Dreizler, Stefan},
	month = feb,
	year = {2018},
	note = {ADS Bibcode: 2018MNRAS.474.2580S},
	pages = {2580--2593},
}

@article{wang_spectroscopic_2025,
	title = {Spectroscopic {Ages} for 4 {Million} {Main}-sequence {Dwarf} {Stars} from {LAMOST} {DR10} {Estimated} with a {Data}-driven {Approach}},
	volume = {280},
	issn = {0067-0049},
	url = {https://ui.adsabs.harvard.edu/abs/2025ApJS..280...13W},
	doi = {10.3847/1538-4365/aded16},
	urldate = {2026-05-29},
	journal = {\apjs},
	publisher = {IOP},
	author = {Wang, Jia-Hui and Xiang, Maosheng and Zhang, Meng and Xie, Ji-Wei and Ge, Jian and Zhang, Jinghua and Mou, Lanya and Liu, Ji-Feng},
	month = sep,
	year = {2025},
	note = {ADS Bibcode: 2025ApJS..280...13W},
	pages = {13},
}

@article{nataf_accurate_2024,
	title = {Accurate, {Precise}, and {Physically} {Self}-consistent {Ages} and {Metallicities} for 400,000 {Solar} {Neighborhood} {Subgiant} {Branch} {Stars}},
	volume = {976},
	issn = {0004-637X},
	url = {https://ui.adsabs.harvard.edu/abs/2024ApJ...976...87N},
	doi = {10.3847/1538-4357/ad7c4e},
	urldate = {2026-05-29},
	journal = {\apj},
	publisher = {IOP},
	author = {Nataf, David M. and Schlaufman, Kevin C. and Reggiani, Henrique and Hahn, Isabel},
	month = nov,
	year = {2024},
	note = {ADS Bibcode: 2024ApJ...976...87N},
	pages = {87},
}

@article{roberts_cn_2026,
	title = {[{C}/{N}] {Ages} for {Red} {Giants} and {Their} {Implications} for {Galactic} {Archaeology}},
	volume = {1002},
	issn = {0004-637X},
	url = {https://ui.adsabs.harvard.edu/abs/2026ApJ..1002..191R},
	doi = {10.3847/1538-4357/ae5bae},
	urldate = {2026-05-29},
	journal = {\apj},
	publisher = {IOP},
	author = {Roberts, John D. and Pinsonneault, Marc H. and Johnson, Jennifer A. and Dubay, Liam O. and Johnson, James W.},
	month = may,
	year = {2026},
	note = {ADS Bibcode: 2026ApJ..1002..191R},
	pages = {191},
}

@article{carrillo_detailed_2022,
	title = {The detailed chemical abundance patterns of accreted halo stars from the optical to infrared},
	volume = {513},
	issn = {0035-8711},
	url = {https://doi.org/10.1093/mnras/stac518},
	doi = {10.1093/mnras/stac518},
	number = {2},
	urldate = {2026-05-13},
	journal = {\mnras},
	author = {Carrillo, Andreia and Hawkins, Keith and Jofré, Paula and de Brito Silva, Danielle and Das, Payel and Lucey, Madeline},
	month = jun,
	year = {2022},
	pages = {1557--1580},
}

@article{reid_proper_2020,
	title = {The {Proper} {Motion} of {Sagittarius} {A}*. {III}. {The} {Case} for a {Supermassive} {Black} {Hole}},
	volume = {892},
	issn = {0004-637X},
	url = {https://ui.adsabs.harvard.edu/abs/2020ApJ...892...39R},
	doi = {10.3847/1538-4357/ab76cd},
	urldate = {2026-05-13},
	journal = {\apj},
	publisher = {IOP},
	author = {Reid, M. J. and Brunthaler, A.},
	month = mar,
	year = {2020},
	note = {ADS Bibcode: 2020ApJ...892...39R},
	pages = {39},
}

@article{bennett_vertical_2019,
	title = {Vertical waves in the solar neighbourhood in {Gaia} {DR2}},
	volume = {482},
	issn = {0035-8711},
	url = {https://ui.adsabs.harvard.edu/abs/2019MNRAS.482.1417B},
	doi = {10.1093/mnras/sty2813},
	urldate = {2026-05-13},
	journal = {\mnras},
	publisher = {OUP},
	author = {Bennett, Morgan and Bovy, Jo},
	month = jan,
	year = {2019},
	note = {ADS Bibcode: 2019MNRAS.482.1417B},
	pages = {1417--1425},
}

@article{gravity_collaboration_improved_2021,
	title = {Improved {GRAVITY} astrometric accuracy from modeling optical aberrations},
	volume = {647},
	issn = {0004-6361},
	url = {https://ui.adsabs.harvard.edu/abs/2021A&A...647A..59G},
	doi = {10.1051/0004-6361/202040208},
	urldate = {2026-05-13},
	journal = {\aap},
	publisher = {EDP},
	author = {{GRAVITY Collaboration} and Abuter, R. and Amorim, A. and Bauböck, M. and Berger, J. P. and Bonnet, H. and Brandner, W. and Clénet, Y. and Davies, R. and de Zeeuw, P. T. and Dexter, J. and Dallilar, Y. and Drescher, A. and Eckart, A. and Eisenhauer, F. and Förster Schreiber, N. M. and Garcia, P. and Gao, F. and Gendron, E. and Genzel, R. and Gillessen, S. and Habibi, M. and Haubois, X. and Heißel, G. and Henning, T. and Hippler, S. and Horrobin, M. and Jiménez-Rosales, A. and Jochum, L. and Jocou, L. and Kaufer, A. and Kervella, P. and Lacour, S. and Lapeyrère, V. and Le Bouquin, J.-B. and Léna, P. and Lutz, D. and Nowak, M. and Ott, T. and Paumard, T. and Perraut, K. and Perrin, G. and Pfuhl, O. and Rabien, S. and Rodríguez-Coira, G. and Shangguan, J. and Shimizu, T. and Scheithauer, S. and Stadler, J. and Straub, O. and Straubmeier, C. and Sturm, E. and Tacconi, L. J. and Vincent, F. and von Fellenberg, S. and Waisberg, I. and Widmann, F. and Wieprecht, E. and Wiezorrek, E. and Woillez, J. and Yazici, S. and Young, A. and Zins, G.},
	month = mar,
	year = {2021},
	note = {ADS Bibcode: 2021A\&A...647A..59G},
	pages = {A59},
}

@article{casali_gaia-eso_2020,
	title = {The {Gaia}-{ESO} survey: the non-universality of the age-chemical-clocks-metallicity relations in the {Galactic} disc},
	volume = {639},
	issn = {0004-6361},
	shorttitle = {The {Gaia}-{ESO} survey},
	url = {https://ui.adsabs.harvard.edu/abs/2020A&A...639A.127C},
	doi = {10.1051/0004-6361/202038055},
	urldate = {2026-04-25},
	journal = {\aap},
	publisher = {EDP},
	author = {Casali, G. and Spina, L. and Magrini, L. and Karakas, A. I. and Kobayashi, C. and Casey, A. R. and Feltzing, S. and Van der Swaelmen, M. and Tsantaki, M. and Jofré, P. and Bragaglia, A. and Feuillet, D. and Bensby, T. and Biazzo, K. and Gonneau, A. and Tautvaišienė, G. and Baratella, M. and Roccatagliata, V. and Pancino, E. and Sousa, S. and Adibekyan, V. and Martell, S. and Bayo, A. and Jackson, R. J. and Jeffries, R. D. and Gilmore, G. and Randich, S. and Alfaro, E. and Koposov, S. E. and Korn, A. J. and Recio-Blanco, A. and Smiljanic, R. and Franciosini, E. and Hourihane, A. and Monaco, L. and Morbidelli, L. and Sacco, G. and Worley, C. and Zaggia, S.},
	month = jul,
	year = {2020},
	note = {ADS Bibcode: 2020A\&A...639A.127C},
	pages = {A127},
}

@article{hayes_bacchus_2022,
	title = {{BACCHUS} {Analysis} of {Weak} {Lines} in {APOGEE} {Spectra} ({BAWLAS})},
	volume = {262},
	issn = {0067-0049},
	url = {https://ui.adsabs.harvard.edu/abs/2022ApJS..262...34H},
	doi = {10.3847/1538-4365/ac839f},
	urldate = {2026-04-07},
	journal = {\apjs},
	publisher = {IOP},
	author = {Hayes, Christian R. and Masseron, Thomas and Sobeck, Jennifer and García-Hernández, D. A. and Allende Prieto, Carlos and Beaton, Rachael L. and Cunha, Katia and Hasselquist, Sten and Holtzman, Jon A. and Jönsson, Henrik and Majewski, Steven R. and Shetrone, Matthew and Smith, Verne V. and Almeida, Andrés},
	month = sep,
	year = {2022},
	note = {ADS Bibcode: 2022ApJS..262...34H},
	pages = {34},
}

@article{griffith_abundance_2019,
	title = {Abundance {Ratios} in {GALAH} {DR2} and {Their} {Implications} for {Nucleosynthesis}},
	volume = {886},
	issn = {0004-637X},
	url = {https://ui.adsabs.harvard.edu/abs/2019ApJ...886...84G},
	doi = {10.3847/1538-4357/ab4b5d},
	urldate = {2026-04-06},
	journal = {\apj},
	publisher = {IOP},
	author = {Griffith, Emily and Johnson, Jennifer A. and Weinberg, David H.},
	month = dec,
	year = {2019},
	note = {ADS Bibcode: 2019ApJ...886...84G},
	pages = {84},
}

@article{hunter_matplotlib_2007,
	title = {Matplotlib: {A} {2D} {Graphics} {Environment}},
	volume = {9},
	issn = {1558-366X},
	shorttitle = {Matplotlib},
	url = {https://ieeexplore.ieee.org/document/4160265},
	doi = {10.1109/MCSE.2007.55},
	number = {3},
	urldate = {2024-05-20},
	journal = {Computing in Science \& Engineering},
	author = {Hunter, John D.},
	month = may,
	year = {2007},
	note = {Conference Name: Computing in Science \& Engineering},
	pages = {90--95},
}

@article{pedregosa_scikit-learn_2011,
	title = {Scikit-learn: {Machine} {Learning} in {Python}},
	volume = {12},
	issn = {1533-7928},
	shorttitle = {Scikit-learn},
	url = {http://jmlr.org/papers/v12/pedregosa11a.html},
	number = {85},
	urldate = {2024-05-20},
	journal = {Journal of Machine Learning Research},
	author = {Pedregosa, Fabian and Varoquaux, Gaël and Gramfort, Alexandre and Michel, Vincent and Thirion, Bertrand and Grisel, Olivier and Blondel, Mathieu and Prettenhofer, Peter and Weiss, Ron and Dubourg, Vincent and Vanderplas, Jake and Passos, Alexandre and Cournapeau, David and Brucher, Matthieu and Perrot, Matthieu and Duchesnay, Édouard},
	year = {2011},
	pages = {2825--2830},
}

@article{astropy_collaboration_astropy_2013,
	title = {Astropy: {A} community {Python} package for astronomy},
	volume = {558},
	copyright = {© ESO, 2013},
	issn = {0004-6361, 1432-0746},
	shorttitle = {Astropy},
	url = {https://www.aanda.org/articles/aa/abs/2013/10/aa22068-13/aa22068-13.html},
	doi = {10.1051/0004-6361/201322068},
	language = {en},
	urldate = {2024-05-20},
	journal = {\aap},
	publisher = {EDP Sciences},
	author = {{Astropy Collaboration} and Robitaille, Thomas P. and Tollerud, Erik J. and Greenfield, Perry and Droettboom, Michael and Bray, Erik and Aldcroft, Tom and Davis, Matt and Ginsburg, Adam and Price-Whelan, Adrian M. and Kerzendorf, Wolfgang E. and Conley, Alexander and Crighton, Neil and Barbary, Kyle and Muna, Demitri and Ferguson, Henry and Grollier, Frédéric and Parikh, Madhura M. and Nair, Prasanth H. and Günther, Hans M. and Deil, Christoph and Woillez, Julien and Conseil, Simon and Kramer, Roban and Turner, James E. H. and Singer, Leo and Fox, Ryan and Weaver, Benjamin A. and Zabalza, Victor and Edwards, Zachary I. and Bostroem, K. Azalee and Burke, D. J. and Casey, Andrew R. and Crawford, Steven M. and Dencheva, Nadia and Ely, Justin and Jenness, Tim and Labrie, Kathleen and Lim, Pey Lian and Pierfederici, Francesco and Pontzen, Andrew and Ptak, Andy and Refsdal, Brian and Servillat, Mathieu and Streicher, Ole},
	month = oct,
	year = {2013},
	pages = {A33},
}

@article{astropy_collaboration_astropy_2018,
	title = {The {Astropy} {Project}: {Building} an {Open}-science {Project} and {Status} of the v2.0 {Core} {Package}*},
	volume = {156},
	issn = {1538-3881},
	shorttitle = {The {Astropy} {Project}},
	url = {https://dx.doi.org/10.3847/1538-3881/aabc4f},
	doi = {10.3847/1538-3881/aabc4f},
	language = {en},
	number = {3},
	urldate = {2024-05-20},
	journal = {\aj},
	publisher = {The American Astronomical Society},
	author = {{Astropy Collaboration} and Price-Whelan, A. M. and Sipőcz, B. M. and Günther, H. M. and Lim, P. L. and Crawford, S. M. and Conseil, S. and Shupe, D. L. and Craig, M. W. and Dencheva, N. and Ginsburg, A. and VanderPlas, J. T. and Bradley, L. D. and Pérez-Suárez, D. and Val-Borro, M. de and Contributors), (Primary Paper and Aldcroft, T. L. and Cruz, K. L. and Robitaille, T. P. and Tollerud, E. J. and Committee), (Astropy Coordination and Ardelean, C. and Babej, T. and Bach, Y. P. and Bachetti, M. and Bakanov, A. V. and Bamford, S. P. and Barentsen, G. and Barmby, P. and Baumbach, A. and Berry, K. L. and Biscani, F. and Boquien, M. and Bostroem, K. A. and Bouma, L. G. and Brammer, G. B. and Bray, E. M. and Breytenbach, H. and Buddelmeijer, H. and Burke, D. J. and Calderone, G. and Rodríguez, J. L. Cano and Cara, M. and Cardoso, J. V. M. and Cheedella, S. and Copin, Y. and Corrales, L. and Crichton, D. and D’Avella, D. and Deil, C. and Depagne, É and Dietrich, J. P. and Donath, A. and Droettboom, M. and Earl, N. and Erben, T. and Fabbro, S. and Ferreira, L. A. and Finethy, T. and Fox, R. T. and Garrison, L. H. and Gibbons, S. L. J. and Goldstein, D. A. and Gommers, R. and Greco, J. P. and Greenfield, P. and Groener, A. M. and Grollier, F. and Hagen, A. and Hirst, P. and Homeier, D. and Horton, A. J. and Hosseinzadeh, G. and Hu, L. and Hunkeler, J. S. and Ivezić, Ž and Jain, A. and Jenness, T. and Kanarek, G. and Kendrew, S. and Kern, N. S. and Kerzendorf, W. E. and Khvalko, A. and King, J. and Kirkby, D. and Kulkarni, A. M. and Kumar, A. and Lee, A. and Lenz, D. and Littlefair, S. P. and Ma, Z. and Macleod, D. M. and Mastropietro, M. and McCully, C. and Montagnac, S. and Morris, B. M. and Mueller, M. and Mumford, S. J. and Muna, D. and Murphy, N. A. and Nelson, S. and Nguyen, G. H. and Ninan, J. P. and Nöthe, M. and Ogaz, S. and Oh, S. and Parejko, J. K. and Parley, N. and Pascual, S. and Patil, R. and Patil, A. A. and Plunkett, A. L. and Prochaska, J. X. and Rastogi, T. and Janga, V. Reddy and Sabater, J. and Sakurikar, P. and Seifert, M. and Sherbert, L. E. and Sherwood-Taylor, H. and Shih, A. Y. and Sick, J. and Silbiger, M. T. and Singanamalla, S. and Singer, L. P. and Sladen, P. H. and Sooley, K. A. and Sornarajah, S. and Streicher, O. and Teuben, P. and Thomas, S. W. and Tremblay, G. R. and Turner, J. E. H. and Terrón, V. and Kerkwijk, M. H. van and Vega, A. de la and Watkins, L. L. and Weaver, B. A. and Whitmore, J. B. and Woillez, J. and Zabalza, V. and Contributors), (Astropy},
	month = aug,
	year = {2018},
	pages = {123},
}

@article{harris_array_2020,
	title = {Array programming with {NumPy}},
	volume = {585},
	copyright = {2020 The Author(s)},
	issn = {1476-4687},
	url = {https://www.nature.com/articles/s41586-020-2649-2},
	doi = {10.1038/s41586-020-2649-2},
	language = {en},
	number = {7825},
	urldate = {2025-07-29},
	journal = {\nat},
	publisher = {Nature Publishing Group},
	author = {Harris, Charles R. and Millman, K. Jarrod and van der Walt, Stéfan J. and Gommers, Ralf and Virtanen, Pauli and Cournapeau, David and Wieser, Eric and Taylor, Julian and Berg, Sebastian and Smith, Nathaniel J. and Kern, Robert and Picus, Matti and Hoyer, Stephan and van Kerkwijk, Marten H. and Brett, Matthew and Haldane, Allan and del Río, Jaime Fernández and Wiebe, Mark and Peterson, Pearu and Gérard-Marchant, Pierre and Sheppard, Kevin and Reddy, Tyler and Weckesser, Warren and Abbasi, Hameer and Gohlke, Christoph and Oliphant, Travis E.},
	month = sep,
	year = {2020},
	pages = {357--362},
}

@article{virtanen_scipy_2020,
	title = {{SciPy} 1.0: fundamental algorithms for scientific computing in {Python}},
	volume = {17},
	copyright = {2020 The Author(s)},
	issn = {1548-7105},
	shorttitle = {{SciPy} 1.0},
	url = {https://www.nature.com/articles/s41592-019-0686-2},
	doi = {10.1038/s41592-019-0686-2},
	language = {en},
	number = {3},
	urldate = {2024-05-20},
	journal = {Nature Methods},
	publisher = {Nature Publishing Group},
	author = {Virtanen, Pauli and Gommers, Ralf and Oliphant, Travis E. and Haberland, Matt and Reddy, Tyler and Cournapeau, David and Burovski, Evgeni and Peterson, Pearu and Weckesser, Warren and Bright, Jonathan and van der Walt, Stéfan J. and Brett, Matthew and Wilson, Joshua and Millman, K. Jarrod and Mayorov, Nikolay and Nelson, Andrew R. J. and Jones, Eric and Kern, Robert and Larson, Eric and Carey, C. J. and Polat, İlhan and Feng, Yu and Moore, Eric W. and VanderPlas, Jake and Laxalde, Denis and Perktold, Josef and Cimrman, Robert and Henriksen, Ian and Quintero, E. A. and Harris, Charles R. and Archibald, Anne M. and Ribeiro, Antônio H. and Pedregosa, Fabian and van Mulbregt, Paul},
	month = mar,
	year = {2020},
	pages = {261--272},
}

@misc{reback_pandas-devpandas_2021,
	title = {pandas-dev/pandas: {Pandas} 1.2.4},
	shorttitle = {pandas-dev/pandas},
	url = {https://zenodo.org/record/4681666},
	doi = {10.5281/zenodo.4681666},
	urldate = {2021-05-11},
	publisher = {Zenodo},
	author = {Reback, Jeff and McKinney, Wes and jbrockmendel and Bossche, Joris Van den and Augspurger, Tom and Cloud, Phillip and Hawkins, Simon and gfyoung and Sinhrks and Roeschke, Matthew and Klein, Adam and Petersen, Terji and Tratner, Jeff and She, Chang and William Ayd and Shahar Naveh and patrick and Marc Garcia and Jeremy Schendel and Andy Hayden and Daniel Saxton and Vytautas Jancauskas and Marco Gorelli and Richard Shadrach and Ali McMaster and Pietro Battiston and Skipper Seabold and Kaiqi Dong and chris-b1 and h-vetinari},
	month = apr,
	year = {2021},
}

@article{astropy_collaboration_astropy_2022,
	title = {The {Astropy} {Project}: {Sustaining} and {Growing} a {Community}-oriented {Open}-source {Project} and the {Latest} {Major} {Release} (v5.0) of the {Core} {Package}*},
	volume = {935},
	issn = {0004-637X},
	shorttitle = {The {Astropy} {Project}},
	url = {https://dx.doi.org/10.3847/1538-4357/ac7c74},
	doi = {10.3847/1538-4357/ac7c74},
	language = {en},
	number = {2},
	urldate = {2024-05-20},
	journal = {\apj},
	publisher = {The American Astronomical Society},
	author = {{Astropy Collaboration} and Price-Whelan, Adrian M. and Lim, Pey Lian and Earl, Nicholas and Starkman, Nathaniel and Bradley, Larry and Shupe, David L. and Patil, Aarya A. and Corrales, Lia and Brasseur, C. E. and Nöthe, Maximilian and Donath, Axel and Tollerud, Erik and Morris, Brett M. and Ginsburg, Adam and Vaher, Eero and Weaver, Benjamin A. and Tocknell, James and Jamieson, William and Kerkwijk, Marten H. van and Robitaille, Thomas P. and Merry, Bruce and Bachetti, Matteo and Günther, H. Moritz and Authors, Paper and Aldcroft, Thomas L. and Alvarado-Montes, Jaime A. and Archibald, Anne M. and Bódi, Attila and Bapat, Shreyas and Barentsen, Geert and Bazán, Juanjo and Biswas, Manish and Boquien, Médéric and Burke, D. J. and Cara, Daria and Cara, Mihai and Conroy, Kyle E. and Conseil, Simon and Craig, Matthew W. and Cross, Robert M. and Cruz, Kelle L. and D’Eugenio, Francesco and Dencheva, Nadia and Devillepoix, Hadrien A. R. and Dietrich, Jörg P. and Eigenbrot, Arthur Davis and Erben, Thomas and Ferreira, Leonardo and Foreman-Mackey, Daniel and Fox, Ryan and Freij, Nabil and Garg, Suyog and Geda, Robel and Glattly, Lauren and Gondhalekar, Yash and Gordon, Karl D. and Grant, David and Greenfield, Perry and Groener, Austen M. and Guest, Steve and Gurovich, Sebastian and Handberg, Rasmus and Hart, Akeem and Hatfield-Dodds, Zac and Homeier, Derek and Hosseinzadeh, Griffin and Jenness, Tim and Jones, Craig K. and Joseph, Prajwel and Kalmbach, J. Bryce and Karamehmetoglu, Emir and Kałuszyński, Mikołaj and Kelley, Michael S. P. and Kern, Nicholas and Kerzendorf, Wolfgang E. and Koch, Eric W. and Kulumani, Shankar and Lee, Antony and Ly, Chun and Ma, Zhiyuan and MacBride, Conor and Maljaars, Jakob M. and Muna, Demitri and Murphy, N. A. and Norman, Henrik and O’Steen, Richard and Oman, Kyle A. and Pacifici, Camilla and Pascual, Sergio and Pascual-Granado, J. and Patil, Rohit R. and Perren, Gabriel I. and Pickering, Timothy E. and Rastogi, Tanuj and Roulston, Benjamin R. and Ryan, Daniel F. and Rykoff, Eli S. and Sabater, Jose and Sakurikar, Parikshit and Salgado, Jesús and Sanghi, Aniket and Saunders, Nicholas and Savchenko, Volodymyr and Schwardt, Ludwig and Seifert-Eckert, Michael and Shih, Albert Y. and Jain, Anany Shrey and Shukla, Gyanendra and Sick, Jonathan and Simpson, Chris and Singanamalla, Sudheesh and Singer, Leo P. and Singhal, Jaladh and Sinha, Manodeep and Sipőcz, Brigitta M. and Spitler, Lee R. and Stansby, David and Streicher, Ole and Šumak, Jani and Swinbank, John D. and Taranu, Dan S. and Tewary, Nikita and Tremblay, Grant R. and Val-Borro, Miguel de and Kooten, Samuel J. Van and Vasović, Zlatan and Verma, Shresth and Cardoso, José Vinícius de Miranda and Williams, Peter K. G. and Wilson, Tom J. and Winkel, Benjamin and Wood-Vasey, W. M. and Xue, Rui and Yoachim, Peter and Zhang, Chen and Zonca, Andrea and Contributors, Astropy Project},
	month = aug,
	year = {2022},
	pages = {167},
}

@misc{the_pandas_development_team_pandas-devpandas_2025,
	title = {pandas-dev/pandas: {Pandas}},
	shorttitle = {pandas-dev/pandas},
	url = {https://zenodo.org/records/15831829},
	doi = {10.5281/zenodo.15831829},
	urldate = {2025-07-29},
	publisher = {Zenodo},
	author = {{The pandas development team}},
	month = jul,
	year = {2025},
}

@article{mendez-delgado_gradients_2022,
	title = {Gradients of chemical abundances in the {Milky} {Way} from {H} {II} regions: distances derived from {Gaia} {EDR3} parallaxes and temperature inhomogeneities},
	volume = {510},
	issn = {0035-8711},
	shorttitle = {Gradients of chemical abundances in the {Milky} {Way} from {H} {II} regions},
	url = {https://ui.adsabs.harvard.edu/abs/2022MNRAS.510.4436M},
	doi = {10.1093/mnras/stab3782},
	urldate = {2025-05-29},
	journal = {\mnras},
	publisher = {OUP},
	author = {Méndez-Delgado, J. E. and Amayo, A. and Arellano-Córdova, K. Z. and Esteban, C. and García-Rojas, J. and Carigi, L. and Delgado-Inglada, G.},
	month = mar,
	year = {2022},
	note = {ADS Bibcode: 2022MNRAS.510.4436M},
	pages = {4436--4455},
}

@article{johnson_milky_2025,
	title = {The {Milky} {Way} {Radial} {Metallicity} {Gradient} as an {Equilibrium} {Phenomenon}: {Why} {Old} {Stars} {Are} {Metal} {Rich}},
	volume = {988},
	issn = {0004-637X},
	shorttitle = {The {Milky} {Way} {Radial} {Metallicity} {Gradient} as an {Equilibrium} {Phenomenon}},
	url = {https://ui.adsabs.harvard.edu/abs/2025ApJ...988....8J},
	doi = {10.3847/1538-4357/addbe5},
	urldate = {2025-12-02},
	journal = {\apj},
	publisher = {IOP},
	author = {Johnson, James W. and Weinberg, David H. and Blanc, Guillermo A. and Bonaca, Ana and Rudie, Gwen C. and Lu, Yuxi (Lucy) and Reichardt Chu, Bronwyn and Griffith, Emily J. and Sit, Tawny and Johnson, Jennifer A. and Dubay, Liam O. and Weller, Miqaela K. and Boyea, Daniel A. and Bird, Jonathan C.},
	month = jul,
	year = {2025},
	note = {ADS Bibcode: 2025ApJ...988....8J},
	pages = {8},
}

@article{molero_modelling_2025,
	title = {Modelling chemical clocks: {Theoretical} evidences of the space and time evolution of [s/α] in the {Galactic} disc with {Gaia}-{ESO} survey},
	volume = {694},
	issn = {0004-6361},
	shorttitle = {Modelling chemical clocks},
	url = {https://ui.adsabs.harvard.edu/abs/2025A&A...694A.274M},
	doi = {10.1051/0004-6361/202453466},
	urldate = {2026-03-09},
	journal = {\aap},
	publisher = {EDP},
	author = {Molero, M. and Magrini, L. and Palla, M. and Cescutti, G. and Viscasillas Vázquez, C. and Casali, G. and Spitoni, E. and Matteucci, F. and Randich, S.},
	month = feb,
	year = {2025},
	note = {ADS Bibcode: 2025A\&A...694A.274M},
	pages = {A274},
}

@article{molero_constraining_2025,
	title = {Constraining r-process nucleosynthesis with multi-objective {Galactic} chemical evolution models},
	url = {https://ui.adsabs.harvard.edu/abs/2025arXiv251113372M/abstract},
	doi = {10.48550/arXiv.2511.13372},
	language = {en},
	urldate = {2026-03-09},
	journal = {eprint arXiv:2511.13372},
	author = {Molero, M. and Arcones, A. and Montes, F. and Hansen, C. J.},
	month = nov,
	year = {2025},
	pages = {arXiv:2511.13372},
}

@article{bailer-jones_estimating_2021,
	title = {Estimating {Distances} from {Parallaxes}. {V}. {Geometric} and {Photogeometric} {Distances} to 1.47 {Billion} {Stars} in {Gaia} {Early} {Data} {Release} 3},
	volume = {161},
	issn = {0004-6256},
	url = {https://ui.adsabs.harvard.edu/abs/2021AJ....161..147B},
	doi = {10.3847/1538-3881/abd806},
	urldate = {2025-01-31},
	journal = {\aj},
	publisher = {IOP},
	author = {Bailer-Jones, C. A. L. and Rybizki, J. and Fouesneau, M. and Demleitner, M. and Andrae, R.},
	month = mar,
	year = {2021},
	note = {ADS Bibcode: 2021AJ....161..147B},
	pages = {147},
}

@article{zhang_parameters_2023,
	title = {Parameters of 220 million stars from {Gaia} {BP}/{RP} spectra},
	volume = {524},
	issn = {0035-8711},
	url = {https://ui.adsabs.harvard.edu/search/fq=%7B!type%3Daqp%20v%3D%24fq_database%7D&fq_database=(database%3Aastronomy%20OR%20database%3Aphysics)&q=author%3A%22%5Ezhang%22%2C%20author%3A%22green%22%2C%20author%3A%22rix%22%2C%20year%3A2023&sort=date%20desc%2C%20bibcode%20desc&p_=0},
	doi = {10.1093/mnras/stad1941},
	language = {en},
	number = {2},
	urldate = {2026-03-03},
	journal = {\mnras, Volume 524, Issue 2, pp.1855-1884},
	author = {Zhang, Xiangyu and Green, Gregory M. and Rix, Hans-Walter},
	month = sep,
	year = {2023},
	pages = {1855},
}

@misc{pakstiene_calibration_2026,
	title = {Calibration of the [{C}/{N}] and [{Y}/{Mg}] chemical clocks with asteroseismic ages from the {TESS} space mission},
	url = {http://arxiv.org/abs/2602.21413},
	doi = {10.48550/arXiv.2602.21413},
	urldate = {2026-02-26},
	publisher = {arXiv},
	author = {Pakštienė, E. and Tautvaišienė, G. and Bagdonas, V. and Kjeldsen, H. and Winther, M. L. and Drazdauskas, A. and Vázquez, C. Viscasillas and Chorniy, Y. and Mikolaitis, Š and Minkevičiūtė, R. and Stonkutė, E.},
	month = feb,
	year = {2026},
	note = {arXiv:2602.21413 [astro-ph]},
}

@article{bird_inside_2021,
	title = {Inside out and upside-down: {The} roles of gas cooling and dynamical heating in shaping the stellar age-velocity relation},
	volume = {503},
	issn = {0035-8711},
	shorttitle = {Inside out and upside-down},
	url = {https://ui.adsabs.harvard.edu/abs/2021MNRAS.503.1815B},
	doi = {10.1093/mnras/stab289},
	urldate = {2026-02-20},
	journal = {\mnras},
	publisher = {OUP},
	author = {Bird, Jonathan C. and Loebman, Sarah R. and Weinberg, David H. and Brooks, Alyson M. and Quinn, Thomas R. and Christensen, Charlotte R.},
	month = may,
	year = {2021},
	note = {ADS Bibcode: 2021MNRAS.503.1815B},
	pages = {1815--1827},
}

@article{razera_abundance_2022,
	title = {Abundance analysis of {APOGEE} spectra for 58 metal-poor stars from the bulge spheroid},
	volume = {517},
	issn = {0035-8711},
	url = {https://ui.adsabs.harvard.edu/abs/2022MNRAS.517.4590R},
	doi = {10.1093/mnras/stac2136},
	urldate = {2026-02-18},
	journal = {\mnras},
	publisher = {OUP},
	author = {Razera, R. and Barbuy, B. and Moura, T. C. and Ernandes, H. and Pérez-Villegas, A. and Souza, S. O. and Chiappini, C. and Queiroz, A. B. A. and Anders, F. and Fernández-Trincado, J. G. and Friaça, A. C. S. and Cunha, K. and Smith, V. V. and Santiago, B. X. and Schiavon, R. P. and Valentini, M. and Minniti, D. and Schultheis, M. and Geisler, D. and Sobeck, J. and Placco, V. M. and Zoccali, M.},
	month = dec,
	year = {2022},
	note = {ADS Bibcode: 2022MNRAS.517.4590R},
	pages = {4590--4606},
}

@ARTICLE{ernandes_abundances_2026,
       author = {{Ernandes}, H. and {Barbuy}, B. and {Chiappini}, C. and {Feltzing}, S. and {P{\'e}rez-Villegas}, A. and {Fria{\c{c}}a}, A.~C.~S. and {Souza}, S.~O. and {Nunes}, R.~P. and {Queiroz}, A.~B.~A. and {Fern{\'a}ndez-Trincado}, J.~G. and {Rocha de Abreu}, A.~L. and {Plotnikova}, A.},
        title = "{Abundances in 78 metal-rich bulge spheroid stars from APOGEE}",
      journal = {\aap},
         year = 2026,
        month = mar,
       volume = {707},
          eid = {A328},
        pages = {A328},
          doi = {10.1051/0004-6361/202558789},
archivePrefix = {arXiv},
       eprint = {2602.12415},
 primaryClass = {astro-ph.GA},
       adsurl = {https://ui.adsabs.harvard.edu/abs/2026A&A...707A.328E}
}

@article{price-whelan_gala_2017,
	title = {Gala: {A} {Python} package for galactic dynamics},
	volume = {2},
	shorttitle = {Gala},
	url = {https://ui.adsabs.harvard.edu/abs/2017JOSS....2..388P},
	doi = {10.21105/joss.00388},
	urldate = {2026-02-12},
	journal = {The Journal of Open Source Software},
	author = {Price-Whelan, Adrian M.},
	month = oct,
	year = {2017},
	note = {ADS Bibcode: 2017JOSS....2..388P},
	pages = {388},
}

@article{hunt_multiple_2022,
	title = {Multiple phase spirals suggest multiple origins in {Gaia} {DR3}},
	volume = {516},
	issn = {0035-8711},
	url = {https://ui.adsabs.harvard.edu/abs/2022MNRAS.516L...7H},
	doi = {10.1093/mnrasl/slac082},
	urldate = {2026-02-12},
	journal = {\mnras},
	publisher = {OUP},
	author = {Hunt, Jason A. S. and Price-Whelan, Adrian M. and Johnston, Kathryn V. and Darragh-Ford, Elise},
	month = oct,
	year = {2022},
	note = {ADS Bibcode: 2022MNRAS.516L...7H},
	pages = {L7--L11},
}

@article{hawkins_using_2015,
	title = {Using chemical tagging to redefine the interface of the {Galactic} disc and halo},
	volume = {453},
	issn = {0035-8711},
	url = {https://ui.adsabs.harvard.edu/abs/2015MNRAS.453..758H},
	doi = {10.1093/mnras/stv1586},
	urldate = {2026-02-09},
	journal = {\mnras},
	publisher = {OUP},
	author = {Hawkins, K. and Jofré, P. and Masseron, T. and Gilmore, G.},
	month = oct,
	year = {2015},
	note = {ADS Bibcode: 2015MNRAS.453..758H},
	pages = {758--774},
}

@article{casali_tracing_2025,
	title = {Tracing the {Milky} {Way}: calibrating chemical ages with high-precision {Kepler} data},
	volume = {541},
	issn = {0035-8711},
	shorttitle = {Tracing the {Milky} {Way}},
	url = {https://ui.adsabs.harvard.edu/abs/2025MNRAS.541.2631C},
	doi = {10.1093/mnras/staf1047},
	urldate = {2026-02-05},
	journal = {\mnras},
	publisher = {OUP},
	author = {Casali, G. and Montalbán, J. and Miglio, A. and Casagrande, L. and Magrini, L. and Chiappini, C. and Bragaglia, A. and Matteuzzi, M. and Brogaard, K. and Stokholm, A. and Grisoni, V. and Tailo, M. and Willett, E.},
	month = aug,
	year = {2025},
	note = {ADS Bibcode: 2025MNRAS.541.2631C},
	pages = {2631--2650},
}

@article{otto_open_2026,
	title = {The {Open} {Cluster} {Chemical} {Abundances} and {Mapping} {Survey}. {VIII}. {Galactic} {Chemical} {Gradient} and {Azimuthal} {Analysis} from {SDSS}/{MWM} {DR19}},
	volume = {171},
	issn = {0004-6256},
	url = {https://ui.adsabs.harvard.edu/abs/2026AJ....171...91O},
	doi = {10.3847/1538-3881/ae28d8},
	urldate = {2026-01-30},
	journal = {\aj},
	publisher = {IOP},
	author = {Otto, Jonah M. and Frinchaboy, Peter M. and Myers, Natalie R. and Johnson, James W. and Donor, John and Hossain, Ahabar and Mészáros, Szabolcs and Wallace, Hailey and Cunha, Katia and Bhattarai, Binod and Sinha, Amaya and Zasowski, Gail and Loebman, Sarah R. and Wiggins, Alessa I. and Price-Whelan, Adrian M. and Spoo, Taylor and Souto, Diogo and Bizyaev, Dmitry and Pan, Kaike and Saydjari, Andrew K. and Song, Ying-Yi},
	month = feb,
	year = {2026},
	note = {ADS Bibcode: 2026AJ....171...91O},
	pages = {91},
}

@article{hasselquist_identification_2016,
	title = {Identification of {Neodymium} in the {Apogee} {H}-{Band} {Spectra}},
	volume = {833},
	issn = {0004-637X},
	url = {https://ui.adsabs.harvard.edu/abs/2016ApJ...833...81H},
	doi = {10.3847/1538-4357/833/1/81},
	urldate = {2026-01-28},
	journal = {\apj},
	publisher = {IOP},
	author = {Hasselquist, Sten and Shetrone, Matthew and Cunha, Katia and Smith, Verne V. and Holtzman, Jon and Lawler, J. E. and Allende Prieto, Carlos and Beers, Timothy C. and Chojnowski, Drew and Fernández-Trincado, J. G. and García-Hernández, D. A. and Hearty, Fred R. and Majewski, Steven R. and Pereira, C. B. and Placco, Vinicius M. and Villanova, Sandro and Zamora, Olga},
	month = dec,
	year = {2016},
	note = {ADS Bibcode: 2016ApJ...833...81H},
	pages = {81},
}

@misc{hubeny_tlusty_2021,
	title = {{TLUSTY} and {SYNSPEC} {Users}'s {Guide} {IV}: {Upgraded} {Versions} 208 and 54},
	shorttitle = {{TLUSTY} and {SYNSPEC} {Users}'s {Guide} {IV}},
	url = {https://ui.adsabs.harvard.edu/abs/2021arXiv210402829H},
	doi = {10.48550/arXiv.2104.02829},
	urldate = {2026-01-28},
	publisher = {arXiv},
	author = {Hubeny, Ivan and Allende Prieto, Carlos and Osorio, Yeisson and Lanz, Thierry},
	month = apr,
	year = {2021},
	note = {ADS Bibcode: 2021arXiv210402829H},
}

@article{abdurrouf_seventeenth_2022,
	title = {The {Seventeenth} {Data} {Release} of the {Sloan} {Digital} {Sky} {Surveys}: {Complete} {Release} of {MaNGA}, {MaStar}, and {APOGEE}-2 {Data}},
	volume = {259},
	issn = {0067-0049},
	shorttitle = {The {Seventeenth} {Data} {Release} of the {Sloan} {Digital} {Sky} {Surveys}},
	url = {https://ui.adsabs.harvard.edu/abs/2022ApJS..259...35A},
	doi = {10.3847/1538-4365/ac4414},
	urldate = {2022-10-15},
	journal = {\apjs},
	author = {{Abdurro'uf} and Accetta, Katherine and Aerts, Conny and Silva Aguirre, Víctor and Ahumada, Romina and Ajgaonkar, Nikhil and Filiz Ak, N. and Alam, Shadab and Allende Prieto, Carlos and Almeida, Andrés and Anders, Friedrich and Anderson, Scott F. and Andrews, Brett H. and Anguiano, Borja and Aquino-Ortíz, Erik and Aragón-Salamanca, Alfonso and Argudo-Fernández, Maria and Ata, Metin and Aubert, Marie and Avila-Reese, Vladimir and Badenes, Carles and Barbá, Rodolfo H. and Barger, Kat and Barrera-Ballesteros, Jorge K. and Beaton, Rachael L. and Beers, Timothy C. and Belfiore, Francesco and Bender, Chad F. and Bernardi, Mariangela and Bershady, Matthew A. and Beutler, Florian and Bidin, Christian Moni and Bird, Jonathan C. and Bizyaev, Dmitry and Blanc, Guillermo A. and Blanton, Michael R. and Boardman, Nicholas Fraser and Bolton, Adam S. and Boquien, Médéric and Borissova, Jura and Bovy, Jo and Brandt, W. N. and Brown, Jordan and Brownstein, Joel R. and Brusa, Marcella and Buchner, Johannes and Bundy, Kevin and Burchett, Joseph N. and Bureau, Martin and Burgasser, Adam and Cabang, Tuesday K. and Campbell, Stephanie and Cappellari, Michele and Carlberg, Joleen K. and Wanderley, Fábio Carneiro and Carrera, Ricardo and Cash, Jennifer and Chen, Yan-Ping and Chen, Wei-Huai and Cherinka, Brian and Chiappini, Cristina and Choi, Peter Doohyun and Chojnowski, S. Drew and Chung, Haeun and Clerc, Nicolas and Cohen, Roger E. and Comerford, Julia M. and Comparat, Johan and da Costa, Luiz and Covey, Kevin and Crane, Jeffrey D. and Cruz-Gonzalez, Irene and Culhane, Connor and Cunha, Katia and Dai, Y. Sophia and Damke, Guillermo and Darling, Jeremy and Davidson, Jr., James W. and Davies, Roger and Dawson, Kyle and De Lee, Nathan and Diamond-Stanic, Aleksandar M. and Cano-Díaz, Mariana and Sánchez, Helena Domínguez and Donor, John and Duckworth, Chris and Dwelly, Tom and Eisenstein, Daniel J. and Elsworth, Yvonne P. and Emsellem, Eric and Eracleous, Mike and Escoffier, Stephanie and Fan, Xiaohui and Farr, Emily and Feng, Shuai and Fernández-Trincado, José G. and Feuillet, Diane and Filipp, Andreas and Fillingham, Sean P. and Frinchaboy, Peter M. and Fromenteau, Sebastien and Galbany, Lluís and García, Rafael A. and García-Hernández, D. A. and Ge, Junqiang and Geisler, Doug and Gelfand, Joseph and Géron, Tobias and Gibson, Benjamin J. and Goddy, Julian and Godoy-Rivera, Diego and Grabowski, Kathleen and Green, Paul J. and Greener, Michael and Grier, Catherine J. and Griffith, Emily and Guo, Hong and Guy, Julien and Hadjara, Massinissa and Harding, Paul and Hasselquist, Sten and Hayes, Christian R. and Hearty, Fred and Hernández, Jesús and Hill, Lewis and Hogg, David W. and Holtzman, Jon A. and Horta, Danny and Hsieh, Bau-Ching and Hsu, Chin-Hao and Hsu, Yun-Hsin and Huber, Daniel and Huertas-Company, Marc and Hutchinson, Brian and Hwang, Ho Seong and Ibarra-Medel, Héctor J. and Chitham, Jacob Ider and Ilha, Gabriele S. and Imig, Julie and Jaekle, Will and Jayasinghe, Tharindu and Ji, Xihan and Johnson, Jennifer A. and Jones, Amy and Jönsson, Henrik and Katkov, Ivan and Khalatyan, Dr., Arman and Kinemuchi, Karen and Kisku, Shobhit and Knapen, Johan H. and Kneib, Jean-Paul and Kollmeier, Juna A. and Kong, Miranda and Kounkel, Marina and Kreckel, Kathryn and Krishnarao, Dhanesh and Lacerna, Ivan and Lane, Richard R. and Langgin, Rachel and Lavender, Ramon and Law, David R. and Lazarz, Daniel and Leung, Henry W. and Leung, Ho-Hin and Lewis, Hannah M. and Li, Cheng and Li, Ran and Lian, Jianhui and Liang, Fu-Heng and Lin, Lihwai and Lin, Yen-Ting and Lin, Sicheng and Lintott, Chris and Long, Dan and Longa-Peña, Penélope and López-Cobá, Carlos and Lu, Shengdong and Lundgren, Britt F. and Luo, Yuanze and Mackereth, J. Ted and de la Macorra, Axel and Mahadevan, Suvrath and Majewski, Steven R. and Manchado, Arturo and Mandeville, Travis and Maraston, Claudia and Margalef-Bentabol, Berta and Masseron, Thomas and Masters, Karen L. and Mathur, Savita and McDermid, Richard M. and Mckay, Myles and Merloni, Andrea and Merrifield, Michael and Meszaros, Szabolcs and Miglio, Andrea and Di Mille, Francesco and Minniti, Dante and Minsley, Rebecca and Monachesi, Antonela and Moon, Jeongin and Mosser, Benoit and Mulchaey, John and Muna, Demitri and Muñoz, Ricardo R. and Myers, Adam D. and Myers, Natalie and Nadathur, Seshadri and Nair, Preethi and Nandra, Kirpal and Neumann, Justus and Newman, Jeffrey A. and Nidever, David L. and Nikakhtar, Farnik and Nitschelm, Christian and O'Connell, Julia E. and Garma-Oehmichen, Luis and Luan Souza de Oliveira, Gabriel and Olney, Richard and Oravetz, Daniel and Ortigoza-Urdaneta, Mario and Osorio, Yeisson and Otter, Justin and Pace, Zachary J. and Padilla, Nelson and Pan, Kaike and Pan, Hsi-An and Parikh, Taniya and Parker, James and Peirani, Sebastien and Peña Ramírez, Karla and Penny, Samantha and Percival, Will J. and Perez-Fournon, Ismael and Pinsonneault, Marc and Poidevin, Frédérick and Poovelil, Vijith Jacob and Price-Whelan, Adrian M. and Bárbara de Andrade Queiroz, Anna and Raddick, M. Jordan and Ray, Amy and Rembold, Sandro Barboza and Riddle, Nicole and Riffel, Rogemar A. and Riffel, Rogério and Rix, Hans-Walter and Robin, Annie C. and Rodríguez-Puebla, Aldo and Roman-Lopes, Alexandre and Román-Zúñiga, Carlos and Rose, Benjamin and Ross, Ashley J. and Rossi, Graziano and Rubin, Kate H. R. and Salvato, Mara and Sánchez, Sebástian F. and Sánchez-Gallego, José R. and Sanderson, Robyn and Santana Rojas, Felipe Antonio and Sarceno, Edgar and Sarmiento, Regina and Sayres, Conor and Sazonova, Elizaveta and Schaefer, Adam L. and Schiavon, Ricardo and Schlegel, David J. and Schneider, Donald P. and Schultheis, Mathias and Schwope, Axel and Serenelli, Aldo and Serna, Javier and Shao, Zhengyi and Shapiro, Griffin and Sharma, Anubhav and Shen, Yue and Shetrone, Matthew and Shu, Yiping and Simon, Joshua D. and Skrutskie, M. F. and Smethurst, Rebecca and Smith, Verne and Sobeck, Jennifer and Spoo, Taylor and Sprague, Dani and Stark, David V. and Stassun, Keivan G. and Steinmetz, Matthias and Stello, Dennis and Stone-Martinez, Alexander and Storchi-Bergmann, Thaisa and Stringfellow, Guy S. and Stutz, Amelia and Su, Yung-Chau and Taghizadeh-Popp, Manuchehr and Talbot, Michael S. and Tayar, Jamie and Telles, Eduardo and Teske, Johanna and Thakar, Ani and Theissen, Christopher and Tkachenko, Andrew and Thomas, Daniel and Tojeiro, Rita and Hernandez Toledo, Hector and Troup, Nicholas W. and Trump, Jonathan R. and Trussler, James and Turner, Jacqueline and Tuttle, Sarah and Unda-Sanzana, Eduardo and Vázquez-Mata, José Antonio and Valentini, Marica and Valenzuela, Octavio and Vargas-González, Jaime and Vargas-Magaña, Mariana and Alfaro, Pablo Vera and Villanova, Sandro and Vincenzo, Fiorenzo and Wake, David and Warfield, Jack T. and Washington, Jessica Diane and Weaver, Benjamin Alan and Weijmans, Anne-Marie and Weinberg, David H. and Weiss, Achim and Westfall, Kyle B. and Wild, Vivienne and Wilde, Matthew C. and Wilson, John C. and Wilson, Robert F. and Wilson, Mikayla and Wolf, Julien and Wood-Vasey, W. M. and Yan, Renbin and Zamora, Olga and Zasowski, Gail and Zhang, Kai and Zhao, Cheng and Zheng, Zheng and Zheng, Zheng and Zhu, Kai},
	month = apr,
	year = {2022},
	note = {ADS Bibcode: 2022ApJS..259...35A},
	pages = {35},
}

@misc{sdss_collaboration_nineteenth_2025,
	title = {The {Nineteenth} {Data} {Release} of the {Sloan} {Digital} {Sky} {Survey}},
	url = {https://ui.adsabs.harvard.edu/abs/2025arXiv250707093S},
	doi = {10.48550/arXiv.2507.07093},
	urldate = {2026-01-28},
	publisher = {arXiv},
	author = {{SDSS Collaboration} and Adamane Pallathadka, Gautham and Aghakhanloo, Mojgan and Aird, James and Almeida, Andrés and Amrita, Singh and Anders, Friedrich and Anderson, Scott F. and Arseneau, Stefan and González Avila, Consuelo and Aviram, Shir and Aydar, Catarina and Badenes, Carles and Barrera-Ballesteros, Jorge K. and Bauer, Franz E. and Behmard, Aida and Berg, Michelle and Besser, F. and Moni Bidin, Christian and Bizyaev, Dmitry and Blanc, Guillermo and Blanton, Michael R. and Bovy, Jo and Brandt, William Nielsen and Brownstein, Joel R. and Buchner, Johannes and Bulbul, Esra and Burchett, Joseph N. and Carigi, Leticia and Carlberg, Joleen K. and Casey, Andrew R. and Chakraborty, Priyanka and Chanamé, Julio and Chandra, Vedant and Chiappini, Cristina and Chilingarian, Igor and Comparat, Johan and Covey, Kevin and Crumpler, Nicole and Cunha, Katia and D'Onghia, Elena and Dai, Xinyu and Darling, Jeremy and Davis, Megan and De Lee, Nathan and Deacon, Niall and Méndez Delgado, José Eduardo and Demasi, Sebastian and Demianenko, Mariia and Demke, Delvin and Donor, John and Drory, Niv and Villa Durango, Monica Alejandra and Dwelly, Tom and Egorov, Oleg and Egorova, Evgeniya and El-Badry, Kareem and Eracleous, Mike and Fan, Xiaohui and Farr, Emily and Finkbeiner, Douglas P. and Fries, Logan and Frinchaboy, Peter and Gentile Fusillo, Nicola Pietro and Serrano Félix, Luis Daniel and Gaensicke, Boris and Galligan, Emma and García, Pablo and Gelfand, Joseph and Grabowski, Katie and Grebel, Eva and Green, Paul J and Greve, Hannah and Grier, Catherine and Griffith, Emily and Guetzoyan, Paloma and Gupta, Pramod and Hackshaw, Zoe and Hall, Patrick B. and Hawkins, Keith and Hegedűs, Viola and Hekker, Saskia and Herbst, T. M. and Hermes, J. J. and Hernández-García, Lorena and Hiremath, Pranavi and Hogg, David W and Holtzman, Jon and Horne, Keith and Horta, Danny and Huang, Yang and Hutchinson, Brian and Häberle, Maximilian and Ibarra-Medel, Hector Javier and Ji, Alexander P. and Jofre, Paula and Johnson, James W. and Johnson, Jennifer and Johnston, Evelyn J. and Kaldor, Mary and Katkov, Ivan and Khalatyan, Arman and Khoperskov, Sergey and Klessen, Ralf and Kluge, Matthias and Koekemoer, Anton M. and Kollmeier, Juna A. and Kounkel, Marina and Kreckel, Kathryn and Krishnarao, Dhanesh and Krumpe, Mirko and Lacerna, Ivan and Laporte, Chervin and Lepine, Sebastien and Li, Jing and Liang, Fu-Heng and Limberg, Guilherme and Liu, Xin and Loebman, Sarah and Long, Knox and Lu, Yuxi and Lucey, Madeline and Lugo-Aranda, Alejandra Z. and Martínez Martinez-Aldama, Mary Loli and McKinnon, Kevin and Medan, Ilija and Merloni, Andrea and Morrison, Sean and Myers, Natalie and Mészáros, Szabolcs and Müller-Horn, Johanna and Nepal, Samir and Ness, Melissa and Nidever, David and Nitschelm, Christian and Oravetz, Audrey and Otto, Jonah and Pan, Kaike and Pérez Paolino, Facundo and Negrete Peñaloza, Castalia Alenka and Pinsonneault, Marc and Taghizadeh Popp, Manuchehr and Price-Whelan, Adrian and Pulatova, Nadiia and Queiroz, Anna Barbara and Raddick, Jordan and Rankine, Amy and Rix, Hans-Walter and Román-Zúñiga, Carlos and Fernández Rosso, Daniela and Runnoe, Jessie and Mahmud Saad, Serat and Salvato, Mara and Sanchez, Sebastian F. and Sattler, Natascha and Saydjari, Andrew and Sayres, Conor and Schlaufman, Kevin and Schneider, Donald P. and Schwope, Axel and Seaton, Lucas M. and Seeburger, Rhys and Serna, Javier and Sharma, Sanjib and Shen, Yue and Sinha, Amaya and Sizemore, Brian and Sniegowska, Marzena and Song, Yingyi and Souto, Diogo and Stassun, Keivan and Steinmetz, Matthias and Stone, Zachary and Stone-Martinez, Alexander and Stringfellow, Guy S. and Mata Sánchez, Aurora and Sánchez-Gallego, José and Tan, Jonathan and Tayar, Jamie and Thai, Riley and Thakar, Ani and Thibodeaux, Pierre and Ting, Yuan-Sen and Tkachenko, Andrew and Trakhtenbrot, Benny and Fernandez Trincado, Jose G. and Troup, Nicholas and Trump, Jonathan R. and Ulloa, Natalie and Van der Marel, Roeland P. and Vera, Pablo and Villanova, Sandro and Villaseñor, Jaime and Wang, Ji and Way, Zachary and Weijmans, Anne-Marie and Wheeler, Adam and Wilson, John C. and Wofford, Aida and Wong, Tony},
	month = jul,
	year = {2025},
	note = {ADS Bibcode: 2025arXiv250707093S},
}

@article{bowen_optical_1973,
	title = {The optical design of the 40-in. telescope and of the {Irénée} {DuPont} telescope at {Las} {Campanas} {Observatory}, {Chile}.},
	volume = {12},
	issn = {0003-6935},
	url = {https://ui.adsabs.harvard.edu/abs/1973ApOpt..12.1430B},
	doi = {10.1364/AO.12.001430},
	urldate = {2023-06-26},
	journal = {Applied Optics},
	author = {Bowen, I. S. and Vaughan, Jr., A. H.},
	month = jan,
	year = {1973},
	note = {ADS Bibcode: 1973ApOpt..12.1430B},
	pages = {1430--1434},
}
\bibliographystyle{aasjournalv7}

\end{document}